\documentclass[twocolumn,superscriptaddress,floatfix,prx]{revtex4-1}
\usepackage{graphicx,amsfonts,amssymb,amsmath,enumerate}
\usepackage{color}
\usepackage{bbm}

\newcommand{\beq}{\begin{equation}}
\newcommand{\eeq}{\end{equation}}
\newcommand{\bea}{\begin{eqnarray}}
\newcommand{\eea}{\end{eqnarray}}

\newcommand{\cA}{{\cal A}}

\newcommand{\sign}{{\rm sign}}
\newcommand{\mi} {\mathbbm{1}}

\begin{document}
\title{Floquet dynamics of classical and quantum cavity fields}

\author{Ivar Martin}
\affiliation{Materials Science Division, Argonne National Laboratory, Argonne, Illinois 60439, USA}

\date{\today}

\begin{abstract}

We show that the time-dependence of electromagnetic field in a periodically modulated cavity can be effectively analyzed using a {\em Floquet map.} The map relates the field states separated by one period of the drive; iterative application of the map allows to determine field configuration after arbitrary number of drive periods.  For resonant and near-resonant drives, the map has stable and unstable fixed points, which are the loci of infinite energy concentration in the long time limit. The Floquet map method can be applied both to classical and quantum massless field problems, including the dynamical Casimir effect. The stroboscopic time evolution implemented by the map can be interpreted in terms of the wave propagation in a curved space, with the fixed points of the map corresponding to the black hole and white hole horizons. More practically, the map can be used to design protocols for signal compression/decompression, cooling, and sensing.
\end{abstract}

\maketitle

\section{Introduction}

Periodically driven systems lie between static and truly time-dependent systems.  
Though nominally only slightly more complex to specify than their time-independent counterparts, their behaviors can be qualitatively different. A notable classical example is the onset of chaos in a kicked rotor \cite{Chirikov1979} and in a driven nonlinear oscillator \cite{ NLO83}.  In quantum systems, periodic driving can lead to  new states of matter, impossible in equilibrium \cite{Oka09, Kitagawa2011, AFAI-1, AFAI-2, vonKeyserlingk2016a, Yao17}. 
More broadly, a digital computer can be seen as a physical system that periodically applies a specific set of rules to its current state, thus recomputing the input for the next clock cycle \cite{Landauer, Bennett, minsky1967computation, wolfram2002new}, producing a combinatorial variety of outputs.

A periodically driven closed physical system breaks the continuous time translation symmetry, which characterizes stationary systems. However, it preserves the discrete time translation symmetry: a time shift by an integer number of drive periods is a good symmetry, which can be used to simplify  analysis. 
In this work we will assume that the drive is supplied by an external clock, which breaks the continuous time translational symmetry explicitly. However, this is not a necessary assumption, since  -- as long as there is an energy source -- there are many mechanism for generating oscillations  (e.g., a wound up spring in a mechanical clock or dc current in a quartz clock or a semiconductor laser) \footnote{
It should be mentioned that recently it has been speculated that the continuous time-translational symmetry can break spontaneously even in the ground states of quantum systems \cite{Wilcz}. It has been subsequently shown to be impossible under rather general conditions \cite{Bruno, Oshi}. In contrast, it appears that the {\em discrete} time translation symmetry of {\em driven} quantum systems can indeed further break spontaneously, for instance making the system evolve with twice the period of the drive \cite{Sondhi2016, Nayak2016, Yao17, choi2017observation, zhang2017observation}.}. 

Continuous time translation symmetry leads to energy conservation; on the other hand, the discrete time translation symmetry only conserves {\em quasienergy}, a bounded quantity. 
Since the quasienergy is defined only modulo the quantum of drive frequency,  the  {\em instantaneous} energy of a driven system becomes unbounded, leading to a richer dynamics not limited to the constant energy hypersurface of the phase space.

Just like the energy eigenstates provide a natural basis for static quantum systems, the quasi-energy eigenstates -- or Floquet eigenstates -- provide a convenient basis for periodically driven systems. They are the eigenstates of the evolution operator over the single period of the drive and hence their time evolution is particularly simple: after a period of drive they come back to themselves, up to a phase that defines quasienergy. An arbitrary initial state of a driven system can be expanded in terms of the Floquet states, which completely determines its time evolution.

For finite-dimensional Hamiltonians this is a very efficient approach that revealed possibility of such exotic quantum states as the Floquet topological insulators and time crystals. On the other hand, for truly (non-factorizable) infinite dimensional systems, the application of Floquet technology is hampered by the need to diagonalize infinite matrices. 
Even  the ``simple'' case of a single quantum parametrically driven oscillator, albeit solvable, requires application of special techniques \cite{Selez95}.

In this paper we demonstrate how a Floquet-like approach can be applied to solve the problem of a periodically driven one-dimensional electromagnetic cavity, e.g. a cavity whose length is periodically modulated. A static cavity contains an infinite number of eigenmodes, each a harmonic oscillator, making the problem equivalent to an infinite number of parametrically driven harmonic  oscillators.
Yet, as we will show, it is possible to construct a {\em map} that relates the electromagnetic field $ \cA(x,t_0)$  at some initial time $t_0$ to the field one period of drive later, $\cA(x, t_0+T)$.  The same map can be applied repeatedly to generate the field configuration at any other discrete time, $\cA(x, t_0+nT)$.  The map function plays a role similar to the Floquet evolution operator mentioned earlier. For that reason, we will refer to the map as the  {\em Floquet map}. 

The Floquet map reveals some striking features of a classical field in a driven cavity. For instance, it shows that if the cavity is driven close to its $q^{th}$ eigenfrequency, any initial field configuration becomes exponentially concentrated and amplified around $q$ {\em fixed-point trajectories}, one for each stable fixed point of the stroboscopic map function. This feature offers a  possibility of creating ultrashort and intense pulses (related to mode-locking in lasers), or for concentrating and removing thermal or any other electromagnetic energy from the cavity (``parametric cooling").

The Floquet map function is closely related to the conformal transformation function that has been employed to relate a driven cavity problem to a stationary one \cite{Moore70} (reviewed in Section \ref{sec:Moore}); it is known explicitly only in few special cases \cite{Moore70, Law94}. Various approximate and numerical techniques have been proposed to construct the conformal transformation function \cite{Fulling76, Dalvit98}. However, they didn't take advantage of the   Floquet structure of the problem and tried to solve for it directly, apparently running into difficulties in the long time limit. 

The knowledge of the Floquet map, and hence also the conformal map function,  allows to solve the problem of the quantum cavity field. In particular, one can find how the vacuum energy density responds to driving. It has been noticed previously that starting from vacuum, as a result of driving, the energy density develops  exponentially  sharp peaks \cite{Law94}. As we will show, these peaks correspond precisely to the stable fixed points of the Floquet map.  In the case of weak driving, the  Floquet map can be explicitly calculated for arbitrary number of drive periods (Section \ref{sec:Weak}). That allows to show analytically both the emergence of the peaks, and that the vacuum energy density between the peaks generically drops to a value {\em below} the static Casimir energy density  (Section \ref{sec:Casi}). 

The Floquet map allows to efficiently evolve any initial field configuration over any integer number of drive periods. If a field configuration at some other intermediate time is desired, it can also be found, by first applying the Floquet map for ``coarse approach", and then evolving the wave equation directly over the time interval shorter than the drive period.

Moreover, it is easy to concatenate multiple periodic drive protocols, using output of one as an input for the following one. A special case of this is a time-reversal protocol (Section \ref{sec:TR}). The natural time evolution for resonant or near resonant driving, independent of the initial conditions, is for the radiation to be compressed to exponentially short pulses. However, there is no information loss: for any drive protocol, it is possible to construct  a complementary  protocol that inverts the discrete time evolution. This could be of interest for signal compression and decompression (Section \ref{sec:CompDecomp}).

A  consequence of the fixed point evolution is that there {\em appears to be} loss of unitarity in the long time limit -- any initial state becomes infinitely compressed into a set of (stable fixed) points. Remarkably, propagating back in time has the same effect, except that the role of the stable fixed points is taken by the unstable fixed points. That is, if a system has experienced periodic drive starting from $t = -\infty$ and has a nontrivial field configuration at time time $t = 0$, then it necessarily originated from an unstable fixed point(s) at $t = -\infty$ (``bang") and is destined to end up in a stable fixed point(s) at time $t = \infty$ (``crunch").  This analogy with the black and white hole dynamics can be used to construct a stroboscopic toy model of General Relativity, with the speed of light and singularity structure controlled by the drive amplitude and detuning from the cavity resonances (Section \ref{sec:GR}).

Besides a cavity with moving mirrors, essentially the same phenomena can be obtained in fixed mirror cavities or circular optical fiber resonators by modulating the optical properties of the medium. Spatio-temporally modulating the refractive index changes the {\em optical length}, which is analogous to changing the actual length of the cavity \cite{ Yablo1989, DodonovM93, Leonhardt2006} (Appendix \ref{sec:nxt}). Moreover,  besides the electromagnetic waves, other waves with weak dispersion (dependence of propagation speed on wavelength) can be used as well. Among them, phonons in crystals or Bose condensates, or antiferromagnons. The nonlinearities present in their dispersions can lead to additional phenomena, such as shock waves and caustics.

\section{Model: 1D optical cavity with a moving mirror.}

In this section we formulate the model of 1D cavity with moving mirror, present the formal solution in terms of the conformal transformation originally constructed by G. Moore \cite{Moore70}, and give its interpretation in terms of the vector potential transport along the light rays (light world lines = null lines). We then recast the problem of a linear resonator cavity into ring cavity of twice the size with only one direction of light propagation, and finally introduce the indicator basis for the dynamical problem which plays the same role as the eigenbasis of the stationary problem.

\subsection{Statement of the problem}

Consider an electromagnetic linear cavity defined by two mirrors, one stationary at $x = 0$, and  the  other following a trajectory $x = z(t)$.  This may be a linear strip-line resonator or a cavity bounded by infinite parallel mirrors. In the latter case, we will not concern ourselves, however, with the wave propagation in the direction parallel to the mirrors, as the wave-vector in that direction is preserved and can be included straightforwardly  (if mirrors are finite, then the role of the in-plane momentum is played by the mode index).

We will work in the Coulomb (transverse) gauge. The wave equation that describes the evolution of the vector potential  $\cA(t,x)$ between the mirrors is 
\beq
\frac{\partial ^2 \cA}{\partial t^2} - \frac{\partial ^2 \cA}{\partial x^2} = 0.\label{eq:WE}
\eeq
(we use the units where the speed of light is $c = 1$).
For a perfect mirror, either static or moving, the correct boundary conditions  are $\cA(t,0) = 0$ and $\cA(t, z(t)) = 0$. 
They correspond to the vanishing surface electric field in the frame where the mirror is stationary. Indeed, electric field in the frame moving with velocity $v$ is related to the electric and magnetic fields in the lab frame via $E'_z= \gamma (E_z - v B_y) = \gamma (\partial_t \cA + \dot z \partial_x \cA)$, where $\gamma = 1/\sqrt{1 - v^2}$. It is proportional to the directional derivative of the ($z$ component of) vector potential along the mirror trajectory. Hence if $\cA$ is set to 0 (or any other constant) on the mirrors, the zero electric field boundary conditions is automatically satisfied \cite{Moore70}. 

The goal is to solve for $\cA(t,x)$ starting from any initial field configuration. For a stationary resonator, the problem is trivially solved by expanding the initial conditions in terms of the eigenmodes of the cavity. When the cavity itself is time-dependent, the eigenmodes can only be defined for instantaneous mirror positions and thus no longer provide a good expansion basis. An alternative method is needed.

\subsection{Solution by conformal mapping}\label{sec:Moore}
It was shown by Moore \cite{Moore70} that the 1D problem with a moving mirror can be mapped onto a stationary problem with fixed mirrors by a conformal transformation,
\bea
t+x = F(s+w),\\
t-x = F(s - w).
\eea
It has the property $dt^2 - dx^2 = F'(s+w)F'(s-w)(ds^2 - dw^2)$, and preserves the form of the wave equation,
\beq
\frac{\partial ^2 \cA}{\partial s^2} - \frac{\partial ^2 \cA}{\partial w^2} = 0.\label{eq:WE2}
\eeq
The function $F$ can be chosen to map, e.g., $(t, 0)$ to $(s,0)$ and $(t,z(t))$ to $(s,1)$. The transformation is defined then by the conditions,
\bea
t &=& F(s),\\
t+z(t) &=& F(s +1),\\
t-z(t) &=& F(s - 1).
\eea
Introducing an inverse function $R = F^{-1}$, the last two equations are equivalent to  
\beq
R(t+z(t) ) - R(t-z(t) ) = 2,\label{eq:R}
\eeq
which, when solved, can be used to relate any solution of static problem to a  solution in the time-dependent problem. 
A general solution of Eq. (\ref{eq:WE2}) has the form 
$$\cA(s,w) = \cA_+(s+w) + \cA_-(s-w).$$ 
The zero boundary conditions at $w = 0,1$ imply  $\cA_+(s) =- \cA_-(s)$, and $\cA_+(s) = \cA_+(s+2)$.
Therefore a general solution to the original problem is
\bea
&A(s,w)\to\nonumber\\
& \cA_+\left ({ R(t-x)}\right)  - \cA_+\left ({R(t+x) }\right)  \equiv \tilde \cA(t,x). \label{eq:Gen}
\eea 
That this is a legitimate solution is clear -- as an additive function of $(x\pm t)$, it automatically satisfies the wave equation (\ref{eq:WE}) everywhere inside the cavity. The zero boundary conditions are satisfied trivially at $x = 0$ and at $x = z(t)$ from Eq. (\ref{eq:R}) and the periodic property of $A_+$.

For example, take an eigen-mode in the static problem. It corresponds to the following solution of the original problem
\bea
\cA_n(s,w) &=& e^{-i n\pi s}\sin (n\pi w)  \nonumber\\
&=& \frac{e^{-i n\pi R(t+x)} - e^{-i n\pi R(t-x)}}{2i} \equiv \tilde \cA_n(t,x).
\eea

If the mirror cannot move superluminally ($|\dot z| < 1$) \footnote{In principle, it is not impossible to``move'' the mirror superluminally, since no information needs to be transferred when the reflective properties of the medium are changed. A classic example of the same phenomenon is the ``propagation'' of the scissor cut}, then it is sufficient to define the function $R(x)$ on the interval $[-z(0), z(0))$. From that interval, $R(x)$ can be bootstrapped using Eq. (\ref{eq:R}) to all other $x$ (but only if mirror is subluminal!).  For instance, in the previous example, if we choose $R(x) = x/z(0)$ on $x \in [-z(0), z(0)]$, then $\cA_n(t,x)$ will clearly correspond to the cavity being initialized in the $n^{th}$ eignemode at $t = 0$.

\subsection{Interpretation of the Moore formula.}\label{sec:MooreInterp}
The structure of the Moore formula (\ref{eq:Gen}) reflects the fact that  the solution is being transported along the left and right moving rays, $t \pm x = {\rm const}$. This is indeed expected since, except for the reflection at the mirrors, the light propagates freely in vacuum, with the value of $\cA$ being constant on the world (null) lines. What happens at the mirror? Right before encountering the mirror, the world line is causally unaware of its existence. After reflection it reverses direction and begins to move away from the mirror. At the mirror, by the boundary conditions, the total (incoming plus reflected) vector potential vanishes. Thus we conclude that the vector potential simply flips sign upon reflection, but preserves magnitude as the world line bounces between the mirrors, Fig. \ref{fig:circle}a. That prescription is independent of whether the mirror is moving or static. For moving mirror, same conclusion is reached from the Lorenz transformation properties of the vector potential: being transverse, $\cA$ is invariant with respect to a boost in the direction of propagation, both before and after the reflection.

This  leads to the following approach to solving the initial value problem with moving mirrors. 
Suppose the initial conditions are specified at $t = 0$, represented as a sum of left and right moving components, $\cA_0(x) = \cA_{0+}(x)+ \cA_{0-}(x)$, consistent with the boundary conditions. To find $\cA_+(t, x)$ and  $\cA_-(t, x)$ at a different time, one should construct the null lines from $(x,t)$ back to $t = 0$ line. These trajectories are $x^\pm(t)$, satisfying the ``final" conditions $x^\pm(t) = x$ and $\dot x^\pm(t) = \pm 1$. Then, if  at initial time  $\dot x^\pm(0) = \pm 1$, then $\cA(t,x) = \cA_+(x^+(0))+ \cA_-(x^-(0))$; otherwise ($\dot x^\pm(0) = \mp 1$), we get $\cA(t,x) = -\cA_-(x^+(0))- \cA_+(x^-(0))$. 

An equivalent point of view is that the time evolution is a time-dependent map of a set of points $x_i(t)$ from $t = 0$ to any other time. As the points propagate, they carry the associated initial values of $\cA_{0\pm}(x_i)$ that determine the field configuration at any other time.

\subsection{Unfolding cavity into a ring}\label{sec:unfold}

Instead of keeping track of the right and left moving waves in the cavity, it is convenient to unfold the cavity of length $z(t)$ into a ring of twice the circumference, $2z(t)$, with only one direction of propagation. What used to be a right moving wave (``right mover") at point $x\in (0,z(t))$ maps onto a point $x$ in the ring, while the left mover  at the same point maps onto point $-x$, {\em with the same direction of propagation along the ring}, Figure \ref{fig:circle}. The ring construction makes explicit the cyclic nature of the wave propagation in the cavity: The left movers after reflecting from $x = 0$ boundary become right movers, and right movers after reflecting from the $x = z(t)$ appear as the left movers at $x = -z(t)$, the two points being identified on a ring. Upon reflection, the vector potential that they carry, $\cA_{0}$, flips sign; a full trajectory around the ring contains two reflections and hence two sign  changes.
 This corresponds to a chiral wave equation
\beq
\frac{\partial \cA}{\partial t} - \frac{\partial \cA}{\partial x} = 0,\label{eq:WEc}
\eeq
defined on a ring with a time-dependent circumference $2z(t)$. 
As a simple first order PDE it can be solved by the method of characteristics. The characteristics are the null lines that ``transport'' the vector potential $\cA$, with no change, expect for sign flips at points $(t,0)$ and $(t,z(t))$.

A general solution of the wave equation, Eq. (\ref{eq:Gen}), rewritten on a ring becomes
\bea
\tilde \cA(t,x) &=&  \cA_+\left ({R(t-x_s) }\right)  - \cA_+\left ({ R(t+x_s)}\right) |_{x_s\in{\rm seg}} \nonumber\\
&\to& \cA_+\left ({R(t-x) }\right) \sign(x)|_{x\in{\rm ring}}\label{eq:GenCh}
\eea
where $x_s$ is the positive coordinate in the original cavity, while $x$  lives on a ring and can be either positive (right mover) or negative (left mover). As was shown in Section \ref{sec:Moore}, $\cA_+(s)$ is an arbitrary function of $x$ with period $2$.  
An initial value problem with the initial vector potential $\cA_0(x)$ defined on the ring at $t = 0$, $[-z(0),z(0))$, is therefore solved by (\ref{eq:GenCh}), by matching to the initial conditions, 
\bea
\cA_+(s) &=& \cA_0(-z(0)s) \sign(s), \quad s\in [-1,1);\\
\eea
and constructing the inverse conformal function starting with 
\beq
R(x) = x, \quad x \in [-1,1),
\eeq
and bootstrapping with 
\beq
R(t+z(t)) = R(t-z(t)) +2.
\eeq
Notice that while $\cA_0(x)$ is in general discontinuous at $x = 0, z(0)$, the function $\cA_+(s)$ is continuos.

\subsection{Indicator basis}\label{sec:ib}
We will now construct the modes (basis) for  the dynamical problem in terms of which an arbitrary initial conditions can be expanded, and whose dynamics fully determines the time evolution. A convenient choice is the indicator basis. At $t= 0$ it is labeled by the position $x_0$ in $[-z(0), z(0))$,
\beq
\cA_{x_0}(0,x) = \mi_\delta(x - x_0),
\eeq
where, $\mi_\delta(x) = 1$ if $|x| < \delta/2$, and zero otherwise. If $x_0$ are  spaced by $\delta$, and $\delta \to 0$, any continuous function $\cA_0(x)$  can be expanded in this basis.
The time evolution is governed by the light-ray propagation (see Section \ref{sec:MooreInterp}),
\beq
\cA_{x_0}(t,x) = \mi_\delta(x - x_{x_0}(t)).
\eeq
Here $x_{x_0}(t)$ is the null line originating at $x_0$, $x_{x_0}(0) = x_0$. Hence the initial value problem is solved by
\beq
A(t,x) = \sum_{x_0}  c_{x_0} \cA_{x_0}(t,x),
\eeq
where 
\beq
c_{x_0} = \frac{1}{\delta}\int dx \cA_{x_0}(0,x) \cA_{0}(x).
\eeq

\section{Periodic drive: stroboscopic evolution and the Floquet map}\label{sec:FMap}
As we saw in Section \ref{sec:ib}, the solution of the dynamical problem simply reduces to finding the null line trajectories, 
$x_{x_0}(t)$.  This is the case in general, regardless of whether the drive $z(t)$ is periodic or not. If the drive is periodic, the problem simplifies dramatically.
In this section we show that the dynamics of the electromagnetic field in a periodically driven cavity is captured by the stroboscopic observation synchronized with the drive. 

The stroboscopic evolution is completely determined by a map that relates positions $x_0 = x_{x_0}(0)$ and  $x_{x_0}(T)$ one period later. If the map has fixed points, the long-time evolution leads to collapse of the field to a discrete set of fixed point trajectories. The map is invertible, which will allow us to construct a discrete time-reversal by simply changing the modulation protocol $z(t)$. For weak near-resonant drives we will calculate explicitly both the single period and multi-period map function.

\subsection {Floquet map} 
The chiral wave equation on a ring (\ref{eq:WEc}) is the first order in time, and hence the field evolution does not have ``memory" and only depends on the current and future cavity configurations.  In this sense, the dynamics is ``Markovian".
After each period of modulation the cavity returns to its initial state. Therefore, the field evolution over single period is a sufficient building block to construct evolution over any number of periods. Using the indicator basis (Section \ref{sec:ib}), the problem reduces to finding the map relating the null line spatial coordinate $x_0$ at $t = 0$ to the one at  $t = T$, $x_1 = x_{x_0}(T)$.

Let us introduce map function, $f(x): x_{t_0} \to x_{t_0+T}$, or using discrete-time indexing,
$x^{n+1} = f(x^n)$.  There is a single null line originating from any point on the ring that implements the map (as opposed to the original cavity, which had two separate null lines corresponding to left and right movers). The nested application of the map function corresponds to the evolution over multiple periods, $f^{(p)}(x): x_{t_0} \to x_{t_0+pT}$. It is convenient to define also the inverse map function, $g = f^{-1}$. Indeed, if $f$ corresponds to propagation by single period forward in time, $g$  propagates back in time by one period. 
The multi-period inverse map function implements the discrete version of function $R(x)$ from Eq. (\ref{eq:R}). Taking the reference time  to zero, $t_0 = 0$,
\beq
g^{(p)}(x) = z(0)\mod(R(pT - x), 2).\label{eq:gR}
\eeq 

A discrete maps is characterized by its {\em fixed points}. As the name suggests, a fixed point  remains invariant under the action of the map, $x_0 = f(x_0)$. The fixed points can be either stable or unstable, depending on whether a deviation from the fixed point, $\delta$, grows  or decays. To linear order, $\delta^n = f'(x_0)\delta^{n-1}$. Thus, stable fixed points have $|f'(x_0)|< 1$; for $|f'(x_0)| > 1$ they are unstable.  
There can also be higher (than 1) period fixed points, defined as $x_0 = f^{(p)}(x_0)$ [for instance, $f^{(2)}(x) = f(f(x))$].  There are necessarily at least $p$ of those, since if $x_0$ is a period $p$ fixed point, so is $f(x_0)$.
As we will see, the period $p$ fixed points emerge when the cavity is driven near the $p^{\rm th}$ resonance mode of the cavity. The stable fixed points correspond to the (stroboscopic) points where the cavity energy is becoming infinitely concentrated in the long time limit (``black holes"), and the unstable fixed points to the points from which the energy is being repelled (``white holes"). 
It is important to note, that the map is static only stroboscopically, starting from some seed time (or, equivalently, mirror oscillation phase). Observed over continuous time, the fixed points actually travel inside the cavity at the speed of light  tracing out {\em fixed-point trajectories}.

\begin{figure}[htb]
     \includegraphics[width=.3\columnwidth]{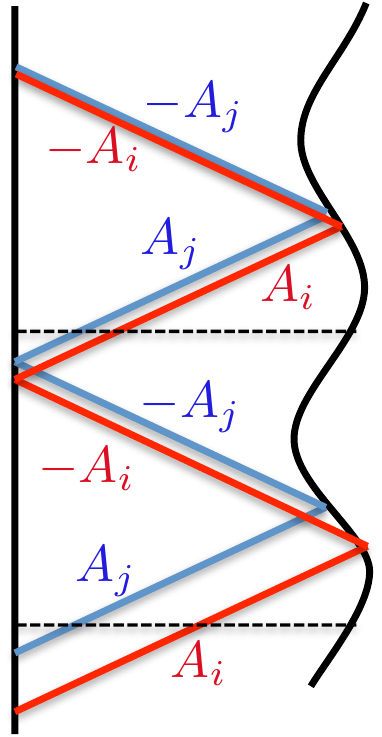}  \hspace{15 mm}
     \includegraphics[width=.5\columnwidth]{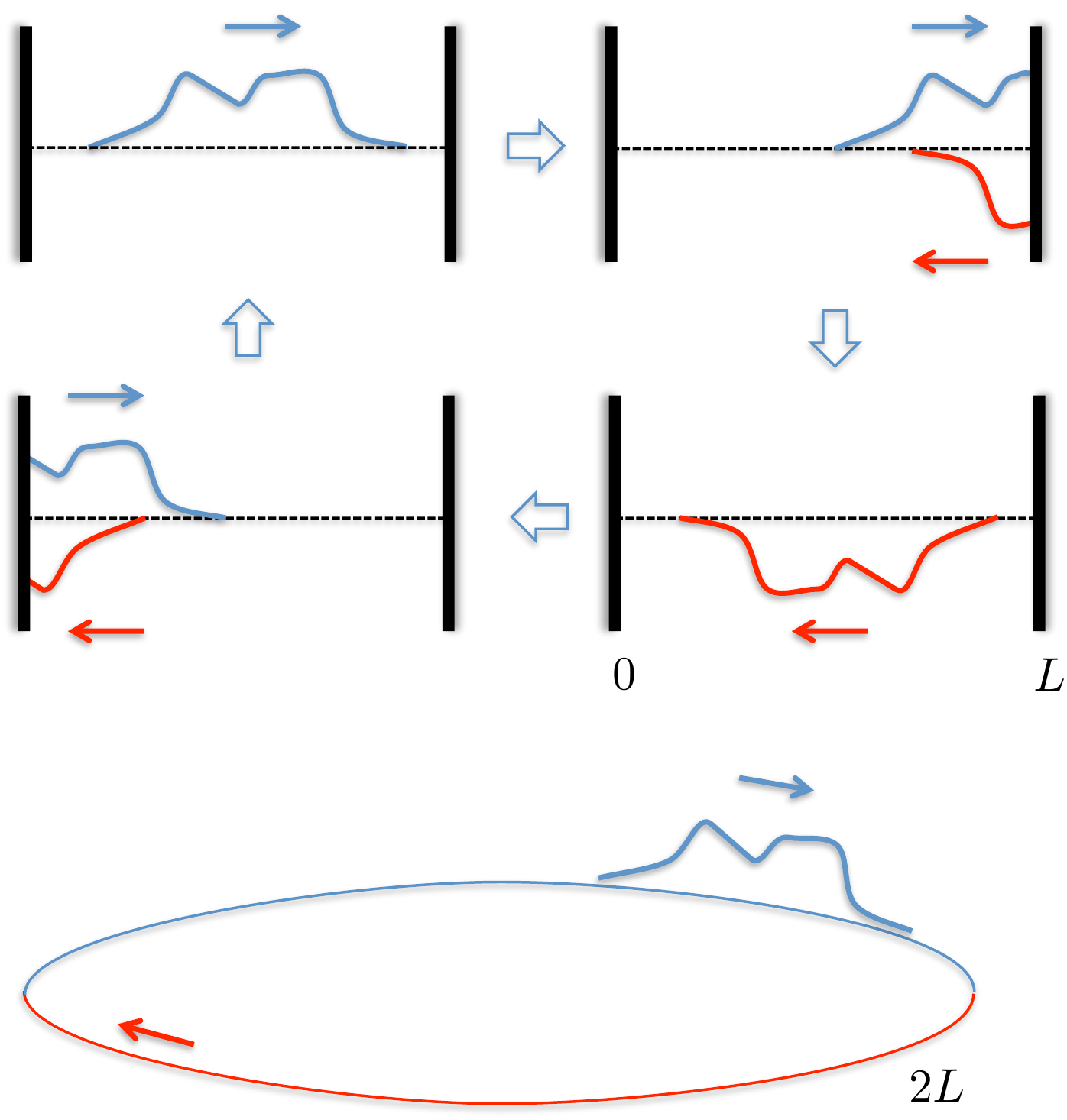}
	\caption{(a) World lines of waves and mirrors in a 1D cavity with modulated length. As the electromagnetic wave propagates in thecavity, the vector potential is transported along the null lines, only changing sign upon reflection from mirrors. A field configuration at any reference time (horizontal dashed line) thus maps onto a field configuration at any other time by following the null lines. (b) A linear cavity of length $z(t)$ can be ``unfolded" into a ring of twice the circumference, $2z(t)$, with only one direction of propagation. What used to be a right moving wave (``right mover") at point $x\in (0,z(t))$ maps onto a point $x$ in the ring, while the left mover  at the same point maps onto point $-x$, {\em with the same direction of propagation along the ring}. 
}
	\label{fig:circle}
\end{figure}

\subsection{Fixed points and their evolutions with parameters.}\label{sec:FPevo}

The simplest example of a situation that has fixed points is a 1D cavity driven  at the lowest frequency resonance, e.g., $$z(t) = L_0 + A\sin \Omega t$$ with $L_0 = cT/2 = \pi c/\Omega $. ($2L_0$ will  denote the distance light travels during the period of modulation; will use $L$ to denote the average cavity length, that can be different from $L_0$.)  Then, trajectories that encounter the moving wall at the ``neutral" position,  $z(t) = L_0$ (which happens in this specific example at times $t = nT/2$, $n$ integer), will keep coming back to the mirror after the mirror period $T$. Observed stroboscopically, they  correspond to fixed-points. There are two distinct trajectories of this kind: one with $\dot z(t) > 0$ at the time of the encounter ($t = nT$), and the other with $\dot z(t) < 0$ (encounters at $t = (n+1/2)T$), that produce negative and positive Doppler shift for the cavity light, respectively. The Doppler shift of the frequency or wavelength describes the contraction or expansion of the incoming wave (e.g., internode distance). Therefore, there is direct correspondence between the sign of the Doppler shift on the fixed-point trajectory and the stability of the corresponding fixed point: positive frequency shift corresponds to stable fixed points, and vice versa.

The fixed points persist also away from the perfect resonance conditions, for any smooth function $z(t)$ with period $T$ as long as $\min z(t) < L_0 < \max z(t)$. Geometrically, the fixed points trajectories pass through (reflect from) the intersections of $x = z(t)$ and $x = L_0$.   
As the function $z(t)$ changes, the fixed points can appear and disappear, typically in pairs of stable and unstable fixed points.  The  happens when   $x = L_0$ line {\em touches} $z(t)$. 

When fixed points are preset, in the long time limit, the light inside the cavity gets infinitely compressed into ultra short pulses that travel between the mirrors with the period of the drive. Outside the fixed point regime there is no infinite compression, and observed stroboscopically the ray position keeps shifting with every period.

In addition to the period-one fixed points, there can also be higher period fixed points. These occur when during the round trip through the cavity, the moving mirror performs several oscillations; they correspond to the cavity driven near its higher order resonance $p >1$. Similar to the $p = 1$ case, the location of the fixed point trajectories is defined by the intersection of $x = z(t)$ and $x = pL_0$ lines. When the fixed points exist, there are at least $2p$ of them, one stable and one unstable for every period of the drive.  For each fixed point, there is a corresponding fixed point trajectory. A particular consequence of having multiple stable fixed points is that under  driving,  any initially uniform (e.g. eigen-mode) of the cavity becomes split into $p$ equally spaced pulses. 

\subsection{Direct and inverse map functions}

In this section we construct the equations that the single-period direct ($f$) and inverse ($g$) map functions satisfy.
Suppose we are interested in the map function starting at time $t_0$,  $f: x(t_0)\to x(t_0 + T)$, or $x_0 \to x_1$ for short. For concreteness will assume that during the period of modulation, the ray encounters the moving wall precisely once; that is, the modulation frequency $\Omega$ is near the fundamental resonance (higher resonances can be treated analogously). The time of encounter with the moving mirror is determined by 
\bea
t_m &=& t_0 + \frac{z(t_m) - x_0}c \\
&=& t_0 + T - \frac{z(t_m)  + x_1}c.\label{eq:tm}
\eea
The total spatial length of the trajectory over the period of drive is 
\beq 
2L_0 = 2 z(t_m) - x_0 + x_1.\label{eq:path}
\eeq 
The above two equations specify the map $x_0 \to x_1$ implicitly. 
If the nonlinear equation (\ref{eq:tm}) is solvable analytically, the map can be constructed explicitly. Otherwise we can resort to numerics or perturbation theory.  Due to the assumed periodicity of $z(t)$, the identical map relates any $x_n$ and $x_{n+1}$. Somewhat more explicitly, the direct and the inverse maps are given by
\bea
f(x_{0}) = x_1 &=& -2 [z(t_m) - L_0]  + x_0,\nonumber\\
g(x_1) = x_0&=& 2 [z(t_m) - L_0]  + x_{1}.\label{eq:dirinv}
\eea 

Maps of this kind, connecting two points on a circle, have been first studied by Andrey Kolmogorov in relation to the dynamics of kicked spinning rotors and are known to have a number of interesting properties.

\subsection{Symmetry between the direct and inverse maps. Discrete time reversal.}\label{sec:TR}
Even though in general it is impossible to solve Eqs. (\ref{eq:tm}, \ref{eq:dirinv}) analytically, there is a useful relationship between the direct and inverse maps that allows to construct a practical time reversal protocol. Such a protocol can be used to compress a signal to a very short duration and subsequently to perfectly reconstruct the original. 

For concreteness let us stay close to the fundamental resonance. In this case, $z(t) \approx  L_0.$   From Eq. (\ref{eq:tm}),
\bea
z(t_m) = z(t_0 - [{x_0 - z(t_m)}]/c)\label{eq:ztm}\\
= z(t_0  - [x_1+ z(t_m) ]/c).\nonumber
\eea
Now, applying the transformation $z(t) \to 2L_0 - z(t)$ is equivalent to swapping $x_0$ and $x_1$ in both (\ref{eq:dirinv}) and (\ref{eq:tm}). The modified modulation protocol makes the discrete-time sequence $x_n$ run backwards in time! 
For example, driving the mirror with some periodic protocol $ z(t)$ up to some time $t^*$ and then changing it to $2L_0 - z(t)$ for $t > t^*$ is equivalent to doing a time reversal operation relative to time $t^*$  
\bea
Z_{TR}(t) = \left\{
\begin{array}{ll}
  z(t),& \quad t < t^*   \\
  2L_0 - z(t ),&  \quad t \ge t^* .
\end{array}
\right.\label{eq:TR}
\eea
Naturally, this can only be  safely done at times when  $2L_0 - z(t^*) = z(t^*)$ -- to avoid discontinuity in the trajectory $Z(t)$ at $t^*$. Also, for a perfect transformation one has to avoid the situation when any part of the wave packet arrives at the moving mirror at time $t^*$.  Notice that this discrete time reversal is not the same as the trivial -- continuous -- time reversal $z(t) \to z(-t)$ and $x \to -x$, as it does not require the time reversal of the light world lines.

If the cavity is driven at a higher harmonic, $z(t) \approx p L_0$,  the corresponding time-reversal transformation that undoes the higher-period map $f^{(p)}$ is
\bea
Z^{(p)}_{TR}(t) = \left\{
\begin{array}{ll}
  z(t),& \quad t < t^*   \\
  2pL_0 - z(t ),&  \quad t \ge t^* .
\end{array}
\right.\label{eq:TRp}
\eea

\section{Weak modulation: Perturbative solution}\label{sec:Weak}

In this section we will explicitly construct the single and multiperiod Floquet map functions for a weakly modulated mirror, for any $z(t)$.

When the drive amplitude is small, $[\max z(t) - \min z(t)]/2\ll  [\max z(t) + \min z(t)]/2 \approx L_0$, to the lowest order approximation, the solution of (\ref{eq:ztm}) is $t_m^0  = t_0 + [L_0- x_0]/c $. Hence, the direct map is approximately
\beq
f(x_0) = x_1 =  2L_0 - 2z\left(t_0 + \frac{L_0- x_0}{c}\right)  + x_0. \label{eq:map}
\eeq
Due to its simple explicit form, the result of the iterative map application map (both forward and back in time) can be easily computed.  (Note that through Eq. (\ref{eq:gR}), the iterated map function $g^{(p)}$ also determines the Moore conformal transformation function $R$.)

\subsubsection{Long-time evolution}
For a known map function, $x_{n+1} = f(x_n)$ can be iterated to determine  $x_{n+q}$ for an arbitrary integer $q$. Despite being explicit this is not always a convenient procedure, for instance if we are interested in the limit of large $q$.  This is particularly true when the drive strength is weak, and every application of the map function has only a small effect. Fortunately, this is precisely the limit in which an explicit evaluation of the iterated map is possible, in the particularly interesting case of the nearly resonant drive. 

For concreteness, consider the case of drive near the fundamental resonance,  $L \approx L_0 = c T/2$.
Under the specified conditions the relative change of $x_n$  after one application of the map function  is small, $|x_{n+1} - x_n | \ll x_n$. Thus, the iteration step $n$ can be approximately treated as a continuous variable, transforming the 
 map equation (\ref{eq:map}) into an ODE,
\beq
\frac{dx}{dn} = 2L_0 - 2z\left(t_0 + \frac{L- x}{c}\right),  \label{eq:mapc}
\eeq
or in the integral form 
\beq
\int_{x_0}^{x_n}\frac{dx}{2L_0 - 2z\left(t_0 + \frac{L- x}{c}\right)} = n. \label{eq:xt}
\eeq
There are two qualitatively distinct regimes that this expression covers, depending on whether the denominator under the integral has roots as a function of $x$ or not. The roots correspond to the fixed point of the iteration. If there are no roots, then as a function of time $x$ keeps drifting with an approximate rate $dx/dn \approx 2(L_0 - L)$.

\subsubsection{Weak harmonic modulation} \label{sec:whm}
Here we present the explicit results for an important case of weak harmonic modulation,
\beq
z(t) = L + A \sin(\Omega t). \label{eq:ex1}
\eeq
Suppose that the modulation is at the fundamental resonance, $L= L_0$, $|A| \ll L_0 $, which makes the conditions of (\ref{eq:mapc}) satisfied (the near-resonant case $L\ne L_0$ can also be solved analytically).
 
 After conveniently choosing the reference time $t_0$, the map ODE is
 \beq
 \frac {dx}{dn} = 2A\sin\frac{\pi x}{L_0},\label{eq:mapr}
 \eeq
or, introducing rescaled variables, $\tilde x = \pi x/L_0$ and $\tilde A = \pi A/L_0$, 
 \beq
 \frac {d\tilde x}{dn} = 2\tilde A\sin\tilde  x.\label{eq:mapr}
 \eeq
 
For $\tilde A>0$, it has an unstable fixed point at $\tilde x_u = 0$ and a stable one at $\tilde x_s  = \pi$.  It is clear from (\ref{eq:mapr}) that if the initial $\tilde x_0$ is inside the open interval $(0,\pi)$ the evolution makes it gradually move toward the stable fixed point [similarly for the interval $(\pi, 2\pi)$]. The equation (\ref{eq:mapr}) can be  integrated, relating the initial and the final values of $\tilde x$,
 \beq
 \frac{\tan \tilde x_n/2}{ \tan \tilde  x_0/2}= e^{ 2\tilde  An}. \label{eq:xnxo}
 \eeq
In the long time ($n\gg 1$) limit, this gives convergence to the stable fixed point, unless the initial value $\tilde x_0$ is precisely at the unstable fixed point. Indeed, let us define $\delta_n\equiv \pi - x_n \ll 1$.  Then, if $\delta_0 \equiv \tilde x_0 \ll 1$ (start near unstable fixed point) we find $\delta_n\equiv \pi - \tilde x_n \approx  4/\delta_0 \exp (-2\tilde  An)$. If $\delta_0 \equiv \pi - \tilde x_0 \ll 1$ then $\delta_n\approx  \delta_0 \exp (-2\tilde  An)$. The limiting cases can be understood by realizing that in the long time limit, most of the iteration time spent in the vicinity of the fixed points, either stable or unstable. For instance, starting near the unstable fixed point, it takes $n_u\sim (\log 1/\delta_0)/f'_u$ iterations to reach $\tilde x$  of order 1. From there, the remaining $n_s=n-n_u$ iterations go from $1$ to $\delta_n$ near the stable fixed point, $\delta_n\sim (f'_s)^{n_s}$. Recalling now that $f'_u\approx 1+ 2\tilde A$, and $f'_s\approx 1- 2\tilde A$ we recover one of the limiting expressions above.

Finally, the multi-period map function is 
\bea
 \tilde x_n = 2\arctan(e^{2\tilde A n}\tan\frac{\tilde x_0}2).
\label{eq:g}
 \eea

The case of a drive at higher cavity resonance can be treated analogously. For a closer parallel, let us keep the cavity size fixed at $L_0$ and increase the drive frequency, $\Omega = q\Omega_0 = q \pi c/L_0$, while keeping the ``sampling frequency" still at $\Omega_0$. This way one period of sampling corresponds to $q$ periods of drive. 
The new ``continuous time'' evolution equation is
 \beq
 \frac {d\tilde x}{dn} = 2\tilde A\sin q \tilde  x,\label{eq:maprq}
 \eeq 
 which obviously has a solution
\bea
 \tilde g(\tilde x_n) &=& (2/q)\arctan(e^{-2q n\tilde A }\tan\frac{q\tilde x_n}2).
\label{eq:gq}
 \eea
It has $q$ stable and $q$ unstable fixed points, separated by distance $\pi/q$. Note that $qn$ that appears in the exponential counts the true number of periods of the drive, and not of the ``sampling."

\section{Energy density and energy}\label{sec:En}

Fixed points of the stroboscopic cavity field evolution described by the Floquet map correspond to the strong concentration of the field energy. In this section we examine this effect in detail for the case of classical fields. We find that the peaks become exponentially sharer and more intense with time. Between the peaks, the energy density becomes exponentially suppressed. The number of peaks is given by the order of the driven resonance, $q = 1$ for the fundamental and $q> 1$ for the higher modes. Such strong effects promise tantalizing opportunities for creation of high intensity pulses of energy, and also for ``sweeping" the cavities -- concentrating and removing unwanted energy, which is equivalent to cooling (Section \ref{sec:sweep}). In Section \ref{sec:Casi} we will repeat the analysis for the quantum fields, in particular examine how the vacuum energy is affected by pumping.

\subsection{Qualitative picture}
In the resonant or near-resonant retime, after the fixed point dynamics is established, every reflection of the pulse from the moving mirror leads to a multiplicative Doppler shift, $k \to k (c+v)/(c-v)$, where $k$ is the incoming wavevector and $v$ is the mirror velocity at the intersection with the fixed point trajectory (world line). 
Coarse-graining over the fixed point trajectory period,  $T$ (approximately the fundamental mode period for small amplitudes of modulation)  
\beq
k(t) \sim k_0 \left(\frac {c+v}{c-v}\right)^{t/T}\approx k_0 e^{\frac{2v t}{cT}}\label{eq:kDop}
\eeq
(we assumed here that $|v| \ll c$).
The energy density of a wave scales as  $k^2\cA^2$, and the total energy as $k\cA^2$. Since the vector potential remains constant on the null lines (up to the sign flips after every reflection), these quantities grow exponentially with time as $e^{\frac{4v t}{cT}}$ and $e^{\frac{2v t}{cT}}$, respectively (and decay exponentially for the unstable fixed point trajectories).  

The wave vector behavior in Eq. (\ref{eq:kDop}) is identical to the one in the uniformly accelerating frame. This is not a coincidence (for details see Appendix \ref{sec:unruh}).

\subsection{Quantitative picture}\label{sec:quali}
We now quantify the above qualitative arguments.
Suppose the cavity is initialized in the state with the vector potential $A_0(x)$, defined on the initial circle $[-z(0),z(0))$, Fig. \ref{fig:circle} which takes care of the left and right movers, and thus is equivalent do defining the initial values and derivates of $\cA$ on the segment $[0,z(0)]$ needed for the solution of the wave equation (\ref{eq:WE}). 
The value of the vector potential at any later space-time point $(t,x_t)$ can be obtained by transporting back to time $t = 0$, to determine the corresponding value of $x_0$ and reading off $\cA(x_0)$, as was discussed in Section \ref{sec:ib}. Let us introduce for convenience the inverse map function, such that $x_0 = g^{(t)}(x_t)$, which extends the discrete inverse map function $g^{(p)}$ (see, e.g., Section \ref{sec:whm}) to all times.
Then, the general solution of the wave equation is
\beq
\cA(x_t, t) = \cA_0(g^{(t)}(x_t)).
\eeq
The cavity energy is the integral of the energy density  (00 component of the energy-stress tensor $T$),
\bea
E(t) &=& \int dx_t T_{00}(x_t,t)\\
 &=& \frac12\int dx_t \left[\left(\frac{\partial\cA(x_t, t)}{\partial x_t}\right)^2 + \left(\frac{\partial\cA(x_t, t)}{\partial t}\right)^2\right]\nonumber\\
 &=& \int dx_t\left[\frac{\partial  g^{(t)}(x_t)}{\partial x_t}\right]^2\left[{\cA'_0(g^{(t)}(x_t)})\right]^2\label{eq:edxt}\\
 &=& \int dx_0\frac{\partial  g^{(t)}(x_t)}{\partial x_t}\left[\cA'_0(x_0)\right]^2\nonumber\\
 &=& \int dx_0\frac{1}{\partial f^{(t)}(x_0)/\partial x_0}\left[\cA'_0(x_0)\right]^2\label{eq:edx0}
\eea
(we used here $|\partial_x g^{(t)}(x)| = |\partial_t g^{(t)}(x)|$ to eliminate time derivatives).
Note that $\propto |\cA'_0|^2$ is the initial energy density.
 For instance, if the  initial  energy density is uniform, then as a function of time it becomes proportional to  $({\partial  g^{(t)}(x_t)}/{\partial x_t})^2$. 
 
 \subsection{Weak harmonic modulation}\label{sec:EC}
Let us go back to the example of harmonically modulated cavity of Eq. (\ref{eq:ex1}).
In the case of weak modulation on the fundamental resonance, from Eq. (\ref{eq:xnxo}) the inverse map function is 
\bea
 \tilde x_0 = \tilde g(\tilde x_n) &=& 2\arctan(e^{-2\tilde A n}\tan\frac{\tilde x_n}2),
\label{eq:g}
 \eea
and its derivative is 
\bea
 g'(x_n) &=&  \frac{d x_0}{d x_n}=\tilde g'(\tilde x_n)= \frac{d\tilde x_0}{d\tilde x_n}\\
 &=& \frac{1}{e^{-2\tilde  An}\sin^2 \frac{\tilde  x_n}2 + e^{2\tilde  An}\cos^2 \frac{\tilde  x_n}2 }\\
 &=& \frac{1}{ \cosh (2\tilde  An) + \sinh (2\tilde  An) \cos\tilde x_n}.
\label{eq:gp}
 \eea 
 The derivative $g'$ corresponds to spatial contraction between time 0 and time $nT$, and hence is related to the Doppler shift in Eq. (\ref{eq:kDop}). Indeed, recognizing that $v = \Omega A$, the exponent in Eq. (\ref{eq:kDop}) is $2vt/cT = 2 \Omega A n /c = 2 \pi A n /L_0$, identical to $g'$ near the fixed point in the long time limit.
 
The energy density and energy can now be computed using Eq. (\ref{eq:edxt}) for arbitrary initial $\cA_0(x_0)$. The energy density is controlled by $[g'(x_n)]^2$.  This function has a peak of height $e^{4\tilde A n}$ near the stable fixed point. The width of the peak is exponentially small,  $\sim e^{-2\tilde A n}$.  

The full energy can be explicitly calculated for initially uniform energy density case; it grows exponentially starting from its initial value as 
\beq
E_n = E_0 \cosh 2\tilde A n\label{eq:E_n},
\eeq
again consistent with the qualitative picture presented above.

\section{Dynamical Casimir effect} \label{sec:Casi}

So far we've been focusing on the modification of the {\em classical} cavity field by modulating the mirror position. In this section we will consider the effect of the drive on  electromagnetic vacuum.  Even if mirrors are static, the vacuum is modified due to the boundary conditions imposed by the mirrors. The eignemodes have a discrete spectrum inside the cavity and continuous outside. While the total zero-point energy of any finite volume of space is infinite both with and without cavity, the energy {\em difference} between the two vacua is finite \cite{Casimir1948}. In 1D, this difference, the Casimir energy, is
\beq
E_C = -\frac{1}{24 \pi L}.\label{eq:CasiE}
\eeq
It is negative and is causing an observable attraction between mirrors separated by the distance $L_0$. The static Casimir effect has been experimentally measured \cite{Lamo1997, Mohi1998, Bressi2002}.

A dynamical counterpart of the static Casimir effect corresponds precisely to the modulated cavity problem that we have been studying, however, starting from the initial vacuum state. Modulation of the cavity has been predicted to lead to a squeezed vacuum state, and generation of detectable photons \cite{Moore70, Law94, Dodonov93, Dalvit98}. The signatures of the dynamical Casimir effect have been detected experimentally in superconducting resonator circuits \cite{Wilson2011}.
In this section we show how the earlier results can be obtained using the Floquet map, and point out some general features that follow directly from the Floquet picture.

As in Section \ref{sec:quali}, we examine the time-dependent electromagnetic energy density, given by the $\langle T_{00}(x,t)\rangle$ component  of the stress energy tensor (Eq. (\ref{eq:edxt})). For the initially static electromagnetic vacuum, upon regularization, it has been expressed in terms of the derivatives of the $R(x)$ function \cite{Fulling76, Dalvit98},
\beq
\langle T_{00}(x,t)\rangle = -\frac{1}{24\pi}\left[ \frac{R'''}{R'} - \frac32\left(\frac{R''}{R'}\right)^2  + \frac12\pi^2(R')^2 \right],
\eeq 
adding the contributions of the left and the right movers.  
For static mirrors, the first two terms vanish identically, and the last term gives rise to the static Casimir effect.
With the help of the relationship (\ref{eq:gR}), the energy density can be expressed in terms of the inverse map function, 
\beq
\langle T_{00}(x,t)\rangle = -\frac{\pi}{24L^2}\left[ \frac{\tilde g'''}{\tilde g'} - \frac32\left(\frac{\tilde g''}{\tilde g'}\right)^2  + \frac12(\tilde g')^2 \right],
\eeq 
where $\tilde g$ is the scaled map function [see, e.g., (\ref{eq:g})]. 
This is a general expression that can be used for any weak periodic drive protocol. In the case of harmonic resonant drive at the fundamental resonance, we can use the results of  Eqs. (\ref{eq:g}) and (\ref{eq:gp}). 
Direct substitution reveals that the spatial derivatives -- surprisingly -- conspire in such a way that the vacuum energy density remains uniform in space and constant in time, equal to the static Casimir energy density,
\beq
\langle T_{00}(x,t)\rangle = -\frac{\pi}{48L^2}.
\eeq 
Integration over $x\in[-L, L]$ that amounts to including both right and left movers gives the standard Casimir energy  (\ref{eq:CasiE}). The lack of spatial dependence is a consequence of a subtle cancellation effect and is special to the drive at the fundamental resonator frequency. For a drive at a higher harmonic of the fundamental, using Eq. (\ref{eq:gq}),   we find instead 
\beq
\langle T_{00}(x,t)\rangle = -\frac{\pi q^2}{48L^2} + \frac{\pi (q^2-1)}{48L^2} (\tilde g')^2,\label{eq:Casiq}
\eeq 
where
\bea
 g'(x) =\frac{1}{ \cosh (2qn \tilde  A) + \sinh (2qn \tilde  A) \cos q \tilde  x},
 \eea 
with tilde still indicating normalization by the cavity size, $\tilde x = \pi x/L$,  $n= t/T = ct/2L$, and $L$ is the size of the cavity. 

This result has interesting implications. As discussed earlier, and can also be readily seen here, $\tilde g'$ is  a strongly peaked function whose width exponentially decreases with time, and the height exponentially increases. Away from this peak (there are $q$ of them in fact), the vacuum energy density deficit that is responsible for the Casimir attraction is amplified by the factor $q^2$, compared to the stationary cavity. 
In a qualitative analogy with the classical case of Section \ref{sec:EC}, it appears as if $\frac{\pi (q^2-1)}{24L^2} $ of zero-point energy density is taken from the vacuum and moved into the sharp pulses, whose energy is amplified by pumping. Away from the peaks, the energy density approaches the value $-\frac{\pi q^2}{24L^2} $ exponentially rapidly, just as it approached zero energy density in the classical case. The Casimir energy density amplification by factor of four was previously obtained by Law in  an exactly solvable model of (unharmonic) drive at twice the fundamental frequency \cite{Law94}.

The total energy inside the cavity can be obtained by spatial integration, in analogy to Eq. (\ref{eq:E_n}),
\bea
E^q(t) &=& \int_{-L}^{L}dx\langle T_{00}(x,t)\rangle \\
&=& -\frac{\pi q^2}{24L} + \frac{\pi (q^2-1)}{24L} \cosh \frac{\pi q A ct}{L^2}\label{eq:E_nq}\\
&=&  - \frac{\pi }{24L} + \frac{\pi q^2(\cosh \frac{\pi q A ct}{L^2} - 1)}{24L} .
\eea 
It starts from the static Casimir value (\ref{eq:CasiE}), and then increases exponentially with time. When the argument of $\cosh$ becomes order 1, the Casimir attraction becomes the Casimir repulsion. The light pressure, however, becomes strongly time-dependent: periods of stronger than static Casimir attraction are interspersed by the short repulsion pulses from the peaks in the energy density.

\section{Cavity sweeping}\label{sec:sweep}
As we have seen,  parametrically driving a cavity near its resonances leads to spatio-temporal concentration of the initial classical field into short pulses with well defined trajectories. One application of this effect is  cooling, or ``sweeping" a cavity of unwanted electromagnetic radiation. The protocol is straightworward: First drive the cavity on resonance for a period of time needed to reduce the energy density to desired level everywhere except for the fixed-point peaks, and then ``open" the mirror at the moment when the pulse is about to arrive, to let it out. Though superficially similar to the Maxwell's demon, there is no need to ``observe" the pulse, since its location (fixed point trajectory) is a deterministic function of the drive. While the cavity is open, external radiation can enter; however, since the pulse -- ideally -- can be made exponentially short, the in-flow of energy can be made arbitrarily small.

In  Section \ref{sec:Casi}, we saw that similar energy concentration also occurs when the initial state of the cavity is  vacuum. Pumping at any higher resonance except for the fundamental ($q > 1$) leads to the formation of the spatially concentrated  energy density peaks riding on top of reduced vacuum energy density background, Eq. (\ref{eq:Casiq}). Again, it appears that the energy pulses can be removed from the cavity by briefly opening the mirror for time intervals exponentially short in the pumping time, seemingly allowing to make the energy in-flow arbitrarily small. After all the pulses are let out and the cavity is frozen in a static configuration, a paradoxical situation seems to arise: the energy deficit in the cavity is increased by a factor $q^2$ compared to the static Casimir energy! This has to be impossible, indicating a failure of the semiclassical reasoning  and the likely need to include the effects of quantum entanglement between the interior (``tails") and the exterior (``peaks") parts of the electromagnetic modes. The situation appears to be analogous to the entanglement between the interior and exterior of the black hole horizon, responsible for Hawking radiation, or entanglement between the right and left Rindler wedges responsble for Unruh effect \cite{Blenco2012} (see  Appendix \ref{sec:unruh} for connection between the pumped cavity problem and Unruh effect). 

\subsection{Practical considerations}
Under ideal conditions, the pulses become infinitely compressed. That would require ideal mirrors, perfectly reflecting in the infinite band of frequencies, and very stable modulation pattern of the moving mirror. Here we estimate the practical limits due to inevtable imperfections.
\subsubsection{Finite finesse}
The finesse, or quality factor, $Q$, roughly counts the number of times the waves bounce inside the cavity before escaping. This determines the number of times the map function is applied to the pulse. In optical cavities, it can be anywhere from 100 to $10^6$. The corresponding typical maximum compression factor (Eq. (\ref{eq:kDop})) is $\sim e^{2Qv/c}$, which can be quite large.

The second criterion considers the energy balance in the cavity. On every round trip, there is a probability $1/Q$ for light to leave the cavity. Therefore, in a static cavity energy decreases with time as  $E(t)= E_0 e^{-t/(QT)}$. On the other hand, in a driven cavity, there is an energy amplification effect due to pumping is $E(t)= E_0 e^{2vt/(cT)}$. Hence, as long as $1/Q < 2v/c$, pumping will dominate, producing net exponential gain in energy. This situation is similar to the single oscillator parametric resonance: In the presence of dissipation, even perfectly resonant pumping has to exceed a threshold set by dissipation.

\subsubsection{Mirror noise}
If the mirror position modulation not perfectly periodic, then after every round trip to the moving mirror, the pulse finds itself at a slightly different phase of the modulation.  Due to the fact that the stable fixed points of iteration persist over a range of parameters (cavity length, modulation frequency, etc), the pulse compression is expected to be quite stable. A conservative estimate for the effect of phase fluctuations can be obtained in the following way. Suppose the phase of the modulation jumps by order of $\pi$ every $T_\phi$. Right after the jump, the pulse finds itself outside the fixed point, and has time $T_\phi$ to drift and compress toward the new fixed point. Most conservatively, if it becomes completely decompressed during this time, the 
final compression will be $\sim e^{2T_\phi v/(cT)} \sim e^{2Q_M v/c}$, that is related to the quality factor of the mirror modulation.

\section{Stroboscopic General Relativity}\label{sec:GR}

In this section we will build an analogy between the stroboscopically observed light in a modulated cavity, and the light propagation in curved spaces, characteristic of the General Relativity (GR).  In GR, the gravitational field is encoded in the curvature of  space-time.   Among the consequences of the space-time curvature is the light deflection (lensing) by gravitating bodies, and even light trapping by the black holes. The light propagation itself can be used as a means to map the structure of space-time.

The space inside the modulated cavities that we study here is, of course, flat, and thus -- in continuous time -- the light propagation is trivial. The situation is changed, however, by introducing the mirror modulation and coarsegraining over the modulation period. Observed stroboscopically at the integer multiples of the modulation period, the cavity appears static. However, the light propagation, as represented by the stroboscopic light cones, encodes the mirror motion and can have non-trivial  curvature.  The effective -- stroboscopic -- speed of light can be arbitrarily slow, since the stroboscopic shift is smaller than the actual light ray path length inside the cavity. In particular, the light speed (one of them, to be precise) vanishes at the fixed points of the Floquet map. This suggests a connection between the fixed points and the black hole/white hole horizons.  An obvious caveat is that the map only applies to massless fields such as light; a massive neutral classical particle can be at rest inside the cavity, oblivious to any possible photonic black holes nearby.  Since the time step of the stroboscopic observation is at least the cavity period, the limit of continuous observation corresponds to sending the cavity size to zero.

Let us consider two representative cases: (a) a trivial map generated by observing a {\em stationary} cavity stroboscopically with some sampling frequency and (b) a cavity driven and observed with the same period, near one of the cavity resonances. In the former case, as expected, stroboscopic evolution corresponds to a space with a flat (Minkowski) metric. In the latter case, the metric becomes position dependent, and the fixed points of the map translate into the black hole and white hole event horizons.

\begin{figure}[htb]
     a \includegraphics[width=.4\columnwidth]{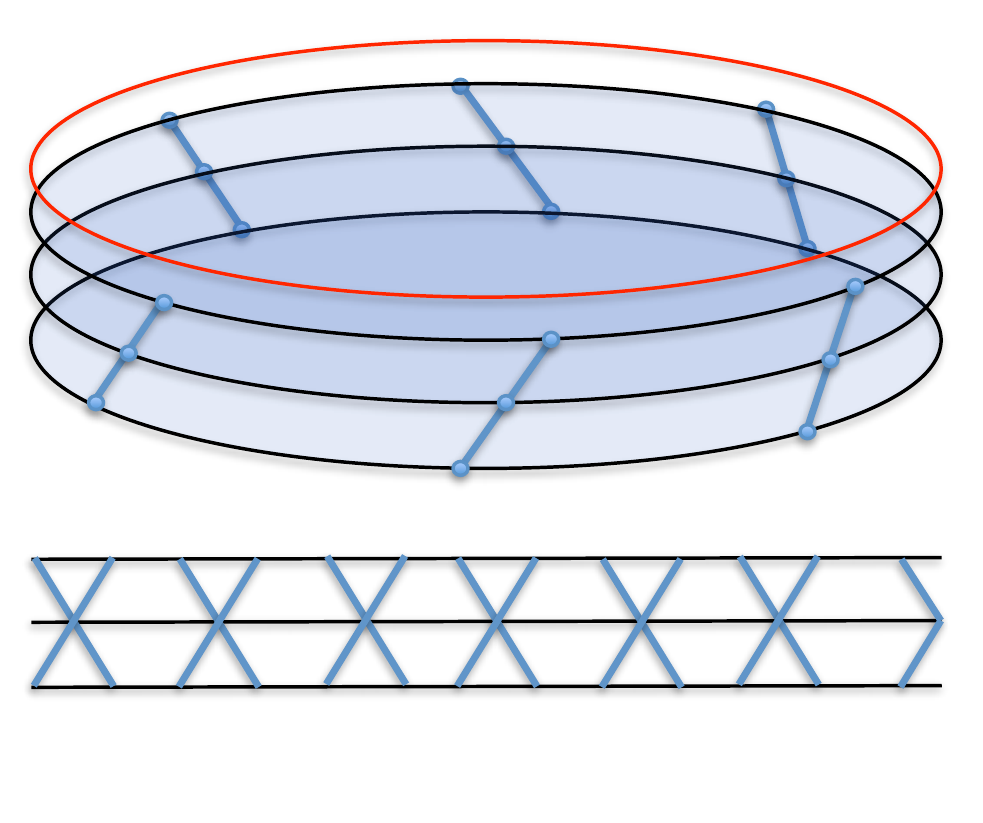} 
     \hspace{5mm}
     b \includegraphics[width=.4\columnwidth]{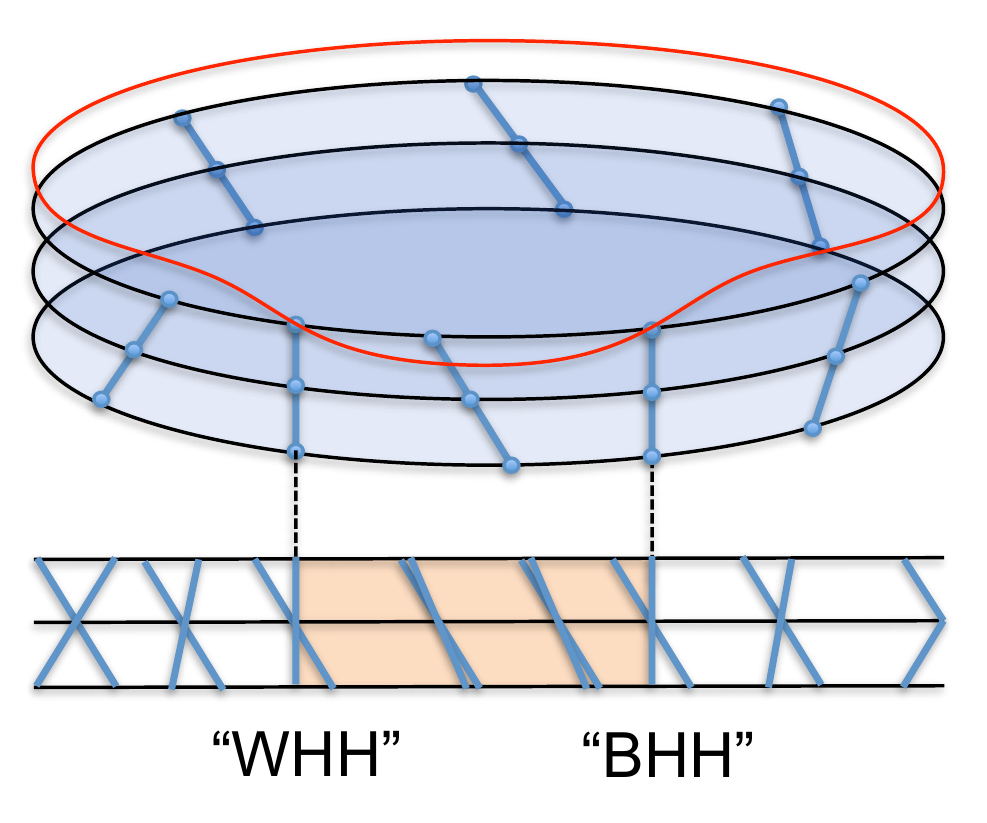}
	\caption{Stroboscopic observation of static (a) and driven (b) cavity. Black ellipses correspond to the cavity in a ring representation (Section \ref{sec:unfold}). The map function plotted as a function of position on the ring (red line). It indicates the distance by which the the light emitted from a point on a ring is sifted along the ring (counter-clock-wise, modulo the ring circumference) after one period of observation. The blue points on the ellipses are the corresponding stroboscopic ``null points" (connected by lines for clarity, so simulate ``light cones").
The lower row plots show the same information but in the linear cavity format.  For undriven cavity, which has a trivial (constant) map function, the ``light cones" are the same in every point of the cavity (panel a). For a driven cavity (panel b), the light cones become position dependent. Moreover, for a  map with fixed points (vertical dashed lines), one side of the light cone becomes vertical, corresponding to zero propagation velocity. The anomalous (shaded) region between the fixed points has only one direction of light propagation (to the left in this figure). Accordingly, the right fixed point is associated with the Black Hole horizon -- nothing can escape black hole interior, and the right fixed point with the White Hole horizon -- nothing can enter the white hole interior. 
}
	\label{fig:GR}
\end{figure}

In the case (a), the map function $x_{n+1} = f(x_n)$ transports all points  by the same distance and in the same direction along the ring (Figure  \ref{fig:GR}a). In the special case of  observation frequency equal to the cavity fundamental resonance all  points remain fixed, $x_{n+1} = x_n$. We are more interested however in a general ``detuned" case.  In the linear cavity representation, which reinstates left and right movers, we see that stroboscopic sampling induces a trivial metric with the same (stroboscopic) light cones at each position. Notice that the stroboscopic ``speed of light" is controlled by detuning between the sampling rate and the cavity resonance. 

In the case (b), we consider a map function that is position dependent and has fixed points (Figure  \ref{fig:GR}b). Now the left and right speeds of light are in general different. At the fixed points of the map function one of the speeds of light vanishes. This is indeed what happens at the black hole event horizon: the incoming light can cross the horizon, while the ``outgoing" light speed changes sign across the horizon. While not obvious in the Schwarzschild coordinates \cite{Misner2017, Landau2}, the Eddington-Finkelstein coordinates \cite{Eddington1924, Fink58, Wald1984} that parametrize space in terms of the incoming null lines and the radius reveal precisely that.

The fixed points separate regions of space where the speeds of light have the opposite or the same directions. The latter  is clearly anomalous, usually associated with the interior of the black holes, where light propagates away from the horizon in the direction of the $r = 0$ singularity. In the GR realized by the driven cavity, however, there is no singularity. The anomalous region is terminated instead by another fixed point, through which all light exits. This is the white hole horizon (WHH), a time reversed partner of the black hole horizon (BHH).

Notice that the BHH is associated with an unstable fixed point, while the WHH with the stable fixed point of the map, even though naively one could expect that stable fixed point, which concentrates all classical energy from the cavity, should be associated with a black hole. This is not so; the cosmological black hole horizon is indeed ``unstable": light just outside the horizon can escape, and just inside begins for fall onto singularity, again away from the horizon; this is the qualitative origin of the Hawking radiation.  On the other hand, the WHH concentrates the fluxes both from the interior and the incoming light from the exterior space. Because of the compact geometry in our cavity case, in the long time limit, all classical energy is concentrated on the WHH. 

Thus driven cavity, stroboscopically observed, can be used to simulate curved 1+1D spaces, including black and white holes.  The position of the ``fixed points" depends of course on the choice of the initial phase of the drive (or, equivalently, the choice of  the observation point within the period).  Observed continuously, instead of stroboscopically, the BHH and WHH propagate inside the cavity at true vacuum speed of light.  

It is worth noting that both black and white holes  appear simultaneously in the Kruskal-Szekeres coordinates \cite{Kruskal1960, Sze1960, Wald1984}, which join two time-reversed copies of Schwarzschild spaces. This parallels the fact that the stable and unstable fixed points of the Floquet map are the time reversed partners of each other: any non-trivial trajectory must start near an unstable fixed point and end near a stable fixed point. There are, however, no obvious singularities in the cavity model: the light that enters through the BHH, instead of falling onto singularity, approaches WHH. And vice versa, instead of having a singularity source inside the white hole, we have BHH.

Finally, it is tempting to explore if the stroboscopic GR toy model can be applied to the case of quantum fields.
We saw in Section \ref{sec:Casi} that driving the cavity vacuum at any resonance higher than the fundamental  would cause exponential energy and energy density growth in the vicinity of the stable fixed points (WHH) while simultaneously depleting the vacuum energy density everywhere else.  Can there be a quantum gravity/cosmological analog of this effect as well?

The GR model described here relies on a finely-tuned external driving of the mirror movement. However, as mentioned in the Introduction, the need for an external coherent drive is superfluous, since even time-independent energy source such as a battery can generate an AC output that could drive our ``toy universe''.

\section{Numerical illustrations}

In this  section we provide numerical illustrations for Section \ref{sec:FMap}, including single and multiperiod map functions for a cavity driven near fundamental resonance (Section \ref{sec:Nmap}); higher order fixed points that occur when the cavity is driven at a higher than fundamental resonance (Section \ref{sec:hr}); compression/decompression protocols that utilize the discrete time reversal technique presented in  Section \ref{sec:TR} (Section \ref{sec:CompDecomp}). 

We also illustrate the exponential energy concentration effect described in Section \ref{sec:En} that pumping has on the classical cavity field (Section \ref{sec:NEdens}). In this section we also compare the numerically exact results with the analytics derived in Section \ref{sec:whm} for weak drives.

\subsection{Single and multiperiod map function}\label{sec:Nmap}
Let us assume that a cavity mirror follows a  harmonic trajectory [Eq. (\ref{eq:ex1})],
\beq
z(t) = L + A \sin(\Omega t). \label{eq:ex11}
\eeq
We will keep the frequency and the drive amplitude fixed at $\Omega = \pi$ and $A = 0.1$ and only vary $L$ to scan through the resonance. The perfect resonance condition is  $L = L_0 = \pi/\Omega = 1$ (see Section \ref{sec:whm}); however, the fixed points are expected to persist in a range of cavity lengths,  $L\in [L_0 -A, L_0 + A] $.

\begin{figure*}[htb]
 
     \includegraphics[width=.5\columnwidth]{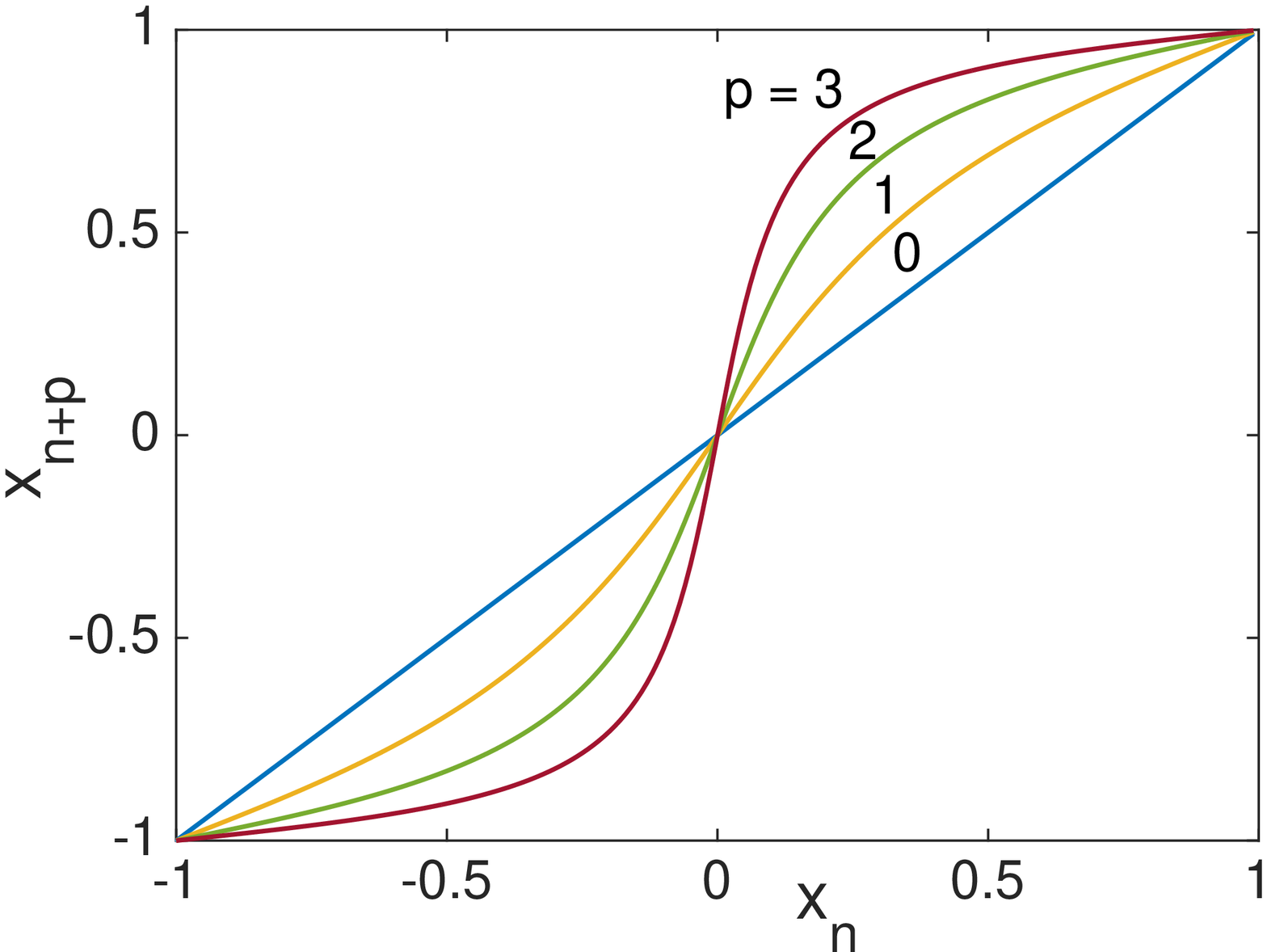}  
     \includegraphics[width=.5\columnwidth]{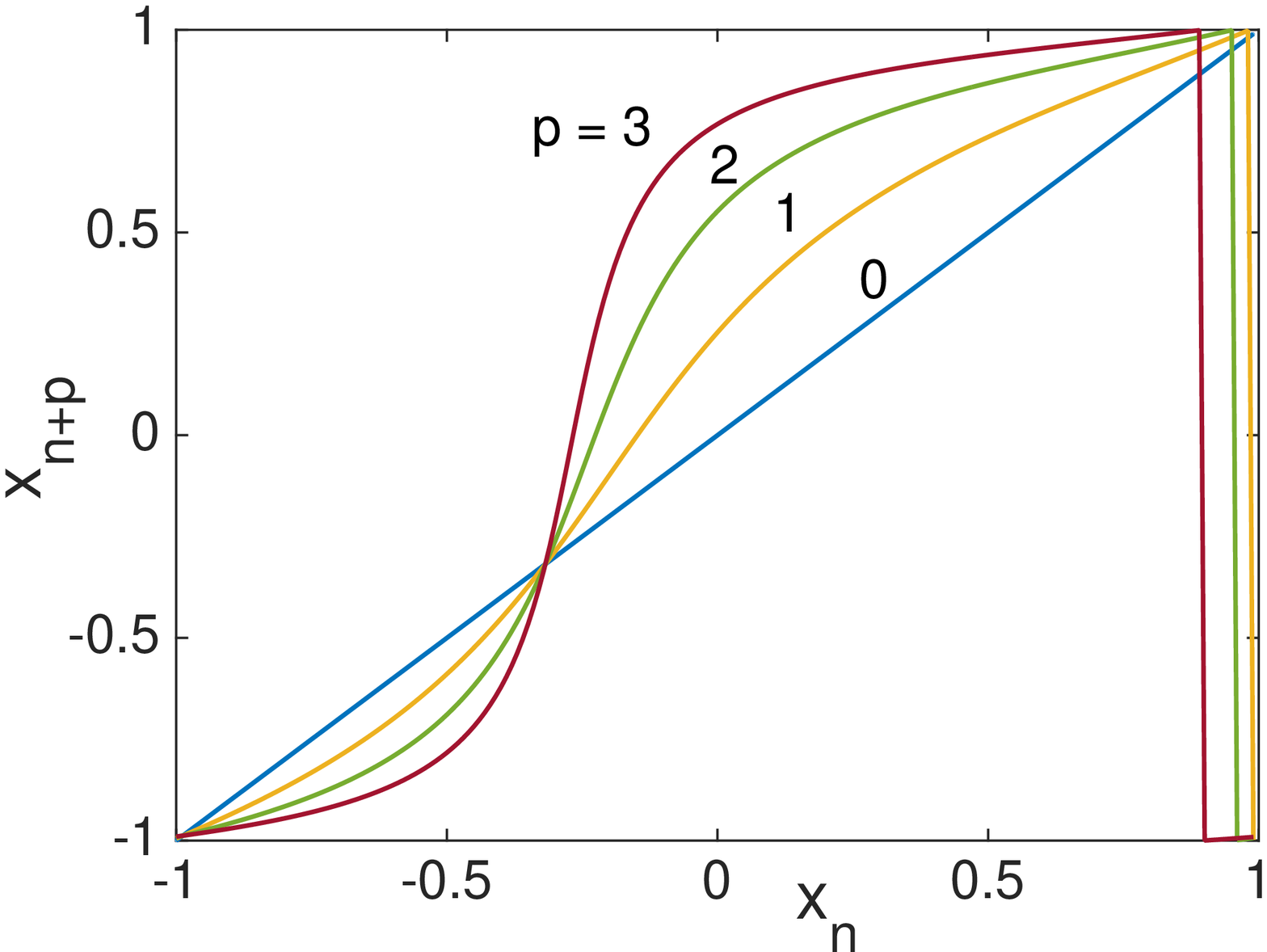}  
     \includegraphics[width=.5\columnwidth]{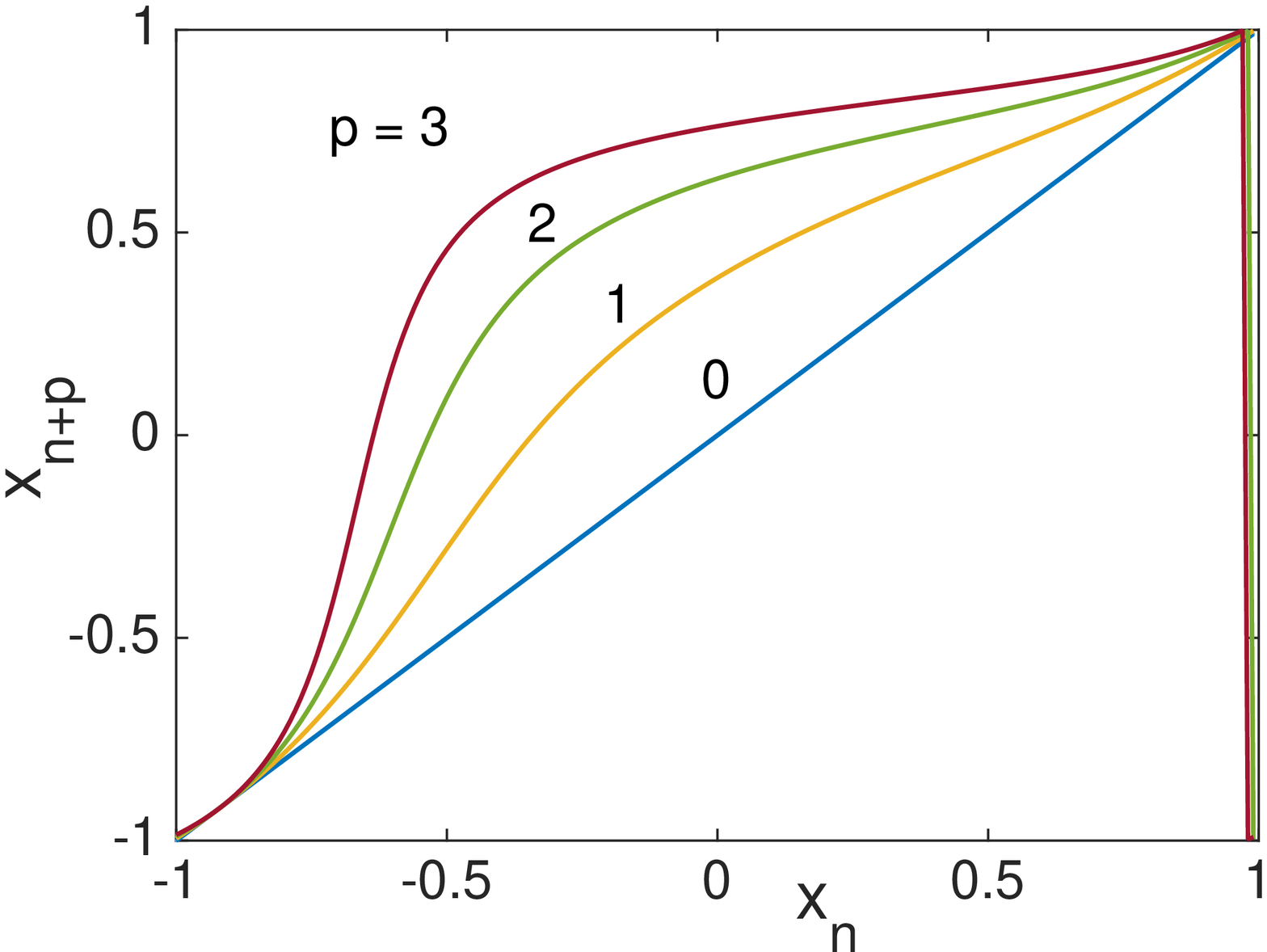}  
     \includegraphics[width=.5\columnwidth]{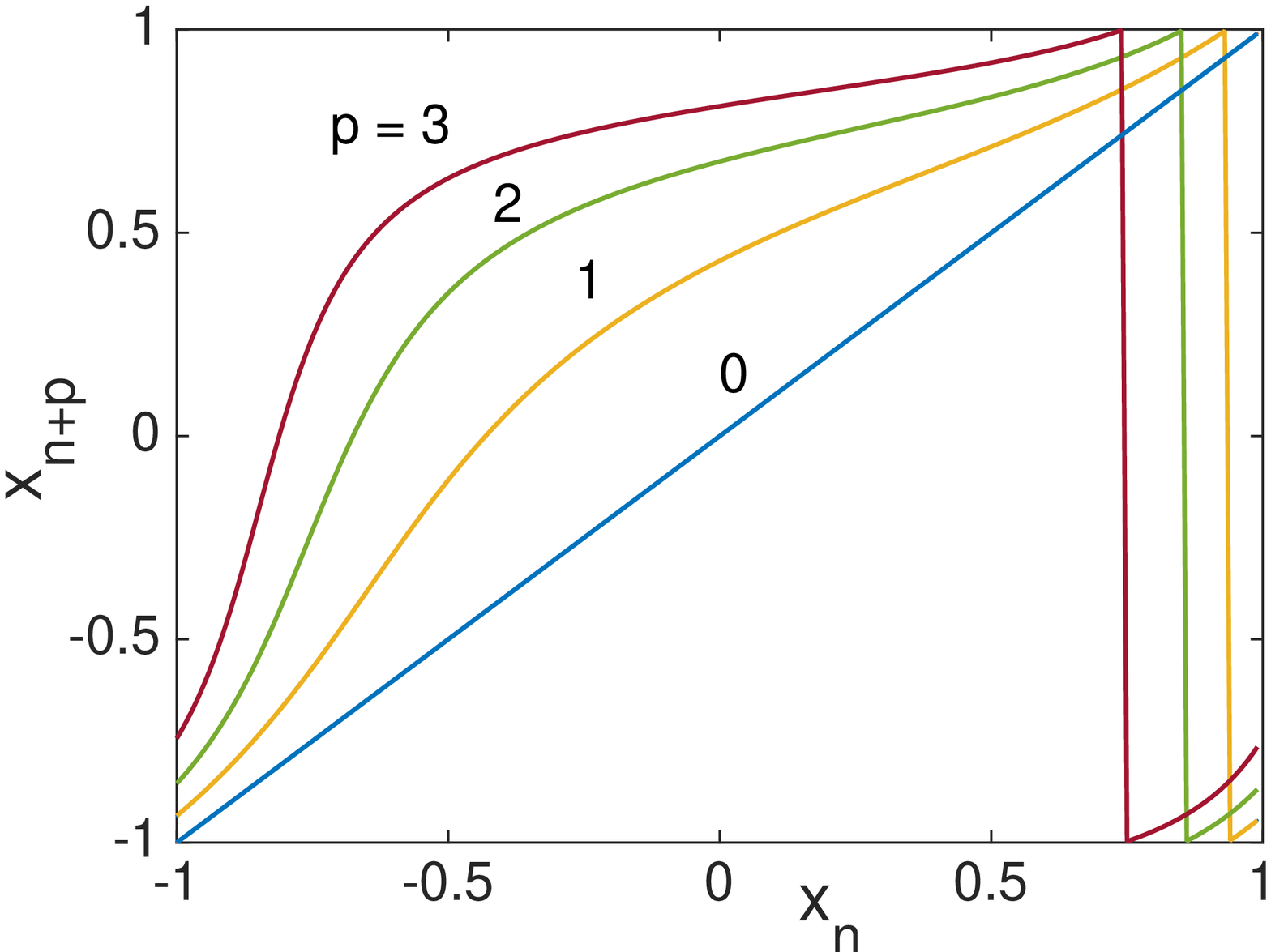}  
     \includegraphics[width=.5\columnwidth]{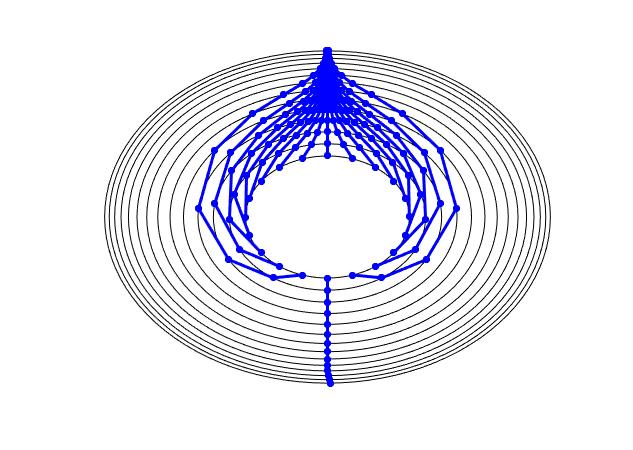}
     \includegraphics[width=.5\columnwidth]{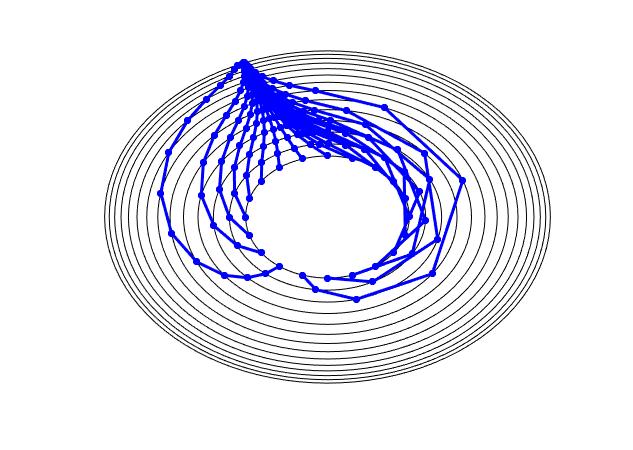}
     \includegraphics[width=.5\columnwidth]{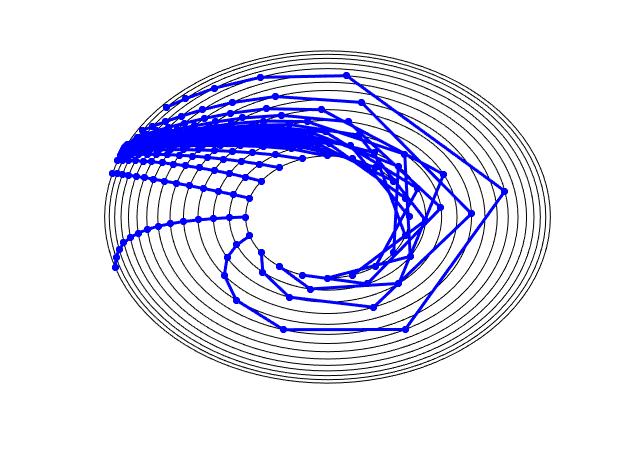}
     \includegraphics[width=.5\columnwidth]{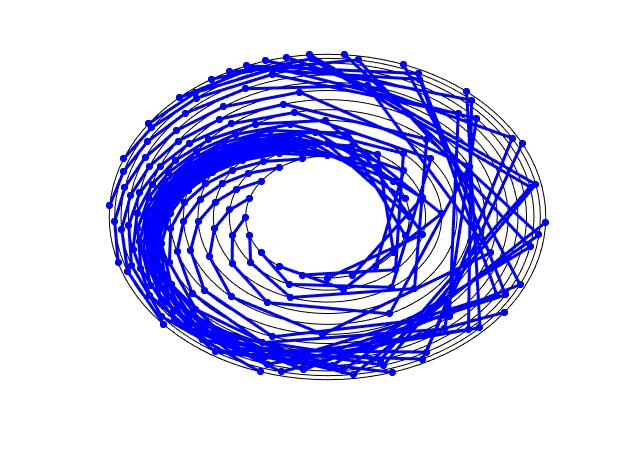}
	\caption{Top row: map functions $x_n \to x_{n+p}$ for Eq. (\ref{eq:ex1}). $\Omega = \pi$, $A = 0.1$.  Left to right: $L = 1, \ 0.95, \ 0.9, \ 0.87$. The horizontal and vertical axes have been rescaled to $[-1,+1]$. Periodicity in the vertical direction is implied. The crossing of the smooth part of the map with the diagonal are the fixed points. The vertical discontinuities are artifacts of representing a periodic function on a linear graph.
Bottom row: Iterative application of the map function (same parameters as the top row) to a set of initial positions  on $[-z(t_0), z(t_0))$ ring. Concentric rings count the numbers of iterations. 
}
	\label{fig:Map}
\end{figure*}

Figure \ref{fig:Map} shows the maps $f^{(p)}  = f(f(...f(x))) : x_n\to x_{n+p}$ for $L = 1, \ 0.95, \ 0.9, \ 0.87$. All of these cases are at or near the fundamental resonance; however, for $L = 0.87$, the period 1 map has no fixed points. 

The top row of figures should be interpreted with the periodic boundary conditions in mind, (Fig. \ref{fig:circle}b).  For clarity, the interval $(-z(0), z(0)]$ is scaled to $x\in(-1,1]$, with $x=1$ and $x=-1$ identified. The  intersections between the map function $f(x)$ and the $y = x$ line  are the fixed points of the iteration. At one of the crossings the slope is larger than one (unstable fixed point),  and at the other less than one (stable fixed point).
The map function is smooth; however, the plot sometimes suffers from artificial discontinuities since it presents periodic structure of the non-periodic domain.  

The lower row shows the results of repeated application of the map $f^{(p)}(x)$ to a set of points  $x$, initially uniformly distributed on the circle $(-z(t_0), z(t_0)]$. Those could be, for instance, the nodes of the cavity eigenmodes of the initially (for $t < 0$) static cavity. Since $x_n$ is defined on a ring, it is represented by the polar angle, $\phi_n = \pi x_n/z(t_0)$ (since $z(t)$ comes back to $z(0)$ after every period, the denominator does not change from iteration to iteration).  The discrete time flows in the outward radial direction, with the concentric circles counting the periods of the drive. The radial plots show the expected exponential convergence to fixed points when $L_0\in (L_0 -A, L_0 + A)$, $\delta_n \sim \exp -n$.
 At the critical value of the cavity length,  $L= L_0 -A = 0.9$, the stable and unstable fixed points merge. It is easy to see that the map function yields neither stable nor unstable fixed point, since $f' = 1$. Inclusion of the curvature $f''$
leads to ``one-sided" algebraic convergence $\delta_n \sim 1/n$  on the flatter side of $f$ near the touching point, and instability on the other
 \footnote{Near the touching point $x_0$, $\delta_{n+1} = \delta_{n} + f''(x_0) \delta_{n}^2/2$. For $\delta_n \ll 1$,  $n$ can be treated as a continuous variable: $d\delta/dn =  f'' \delta^2$/2.  Integration gives $\delta_n \sim \delta_0/(1 - \delta_0 f''/2)$ which yields $\delta_n \sim 1/n$ convergence for $\delta_0 f''(x_0) < 0 $ and an instability otherwise}.  For $L< L_0 -A $, $f(x)$ has no fixed points, and the map function gradually moves $x_n$ around the circle.

\subsection{Higher period fixed points} \label{sec:hr}
When cavity is driven at a frequency significantly higher than the fundamental,  a single application of the map function, which corresponds to time evolution over one period of the drive, does not have fixed points, that is, $x = f(x)$ does not have solutions. However, when the drive frequency approaches one of the higher harmonics of the fundamental frequency, $\Omega \approx p \Omega_0$, higher order fixed points become possible. Those correspond to nested applications of the single period map functions, $x =  f^{(p)}(x)$, which evolve over multiple periods of the drive needed to complete a round trip between the mirrors.
On general grounds (Section \ref{sec:FPevo}), fixed points should  come in sets of $p$ equivalent ones (their number can in fact be any multiple of $p$, depending on the details of function $z(t)$ and $L$; however, in general, different multiplets are going to be inequivalent). If there is a fixed point at $x^*$, then $f(x^*)$ is a distinct but equivalent fixed point.

\begin{figure}[htb]
     a\includegraphics[width=.7\columnwidth]{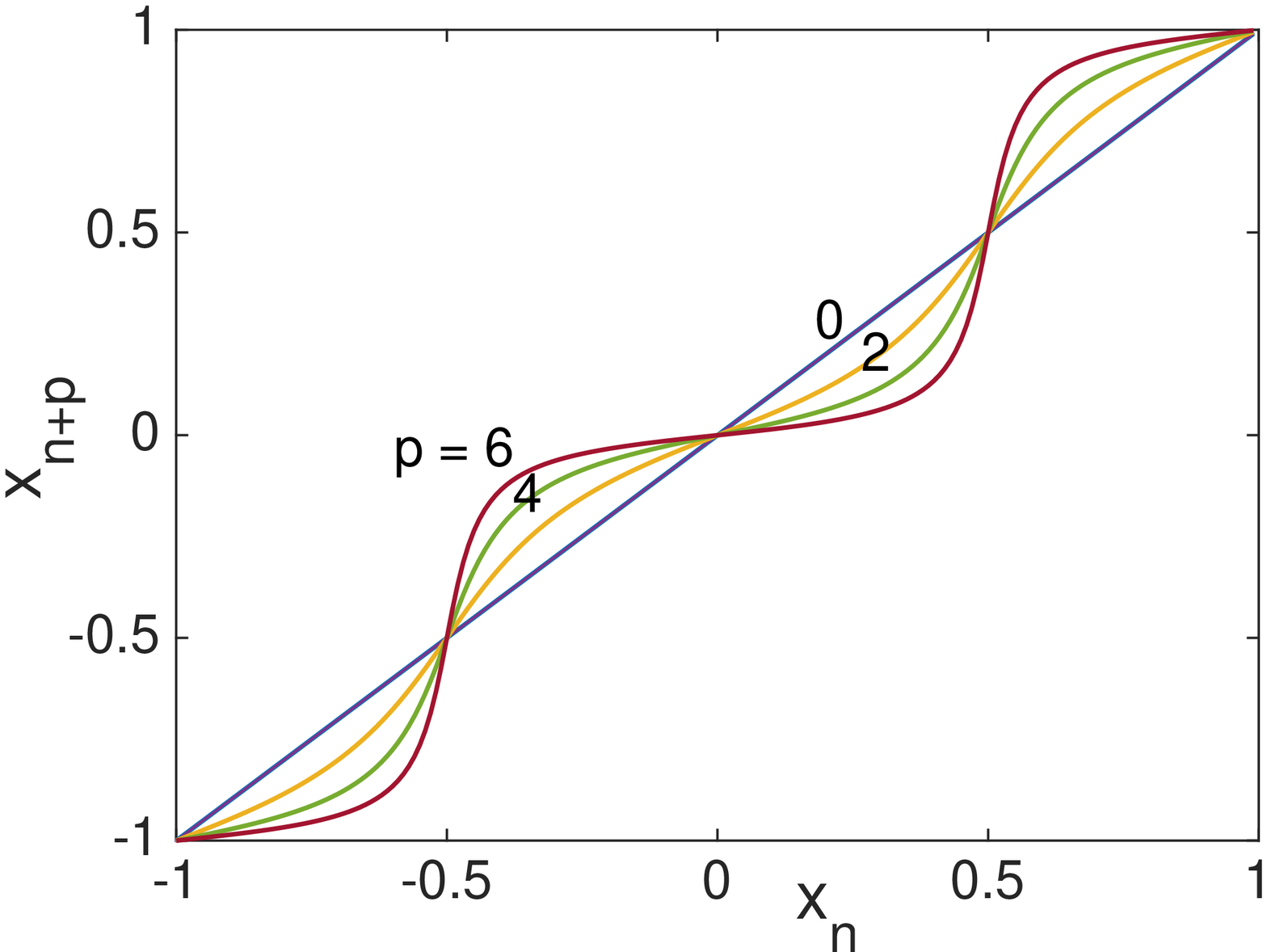}  
     b\includegraphics[width=.7\columnwidth]{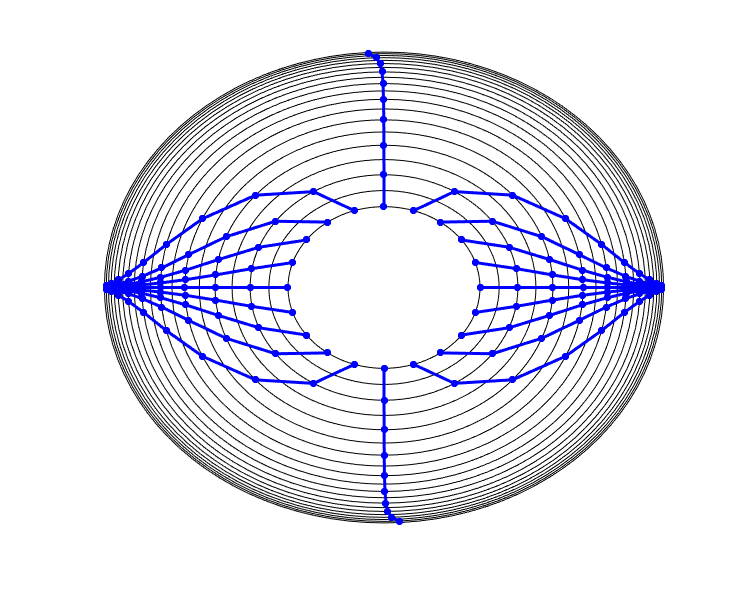}  
     c\includegraphics[width=.7\columnwidth]{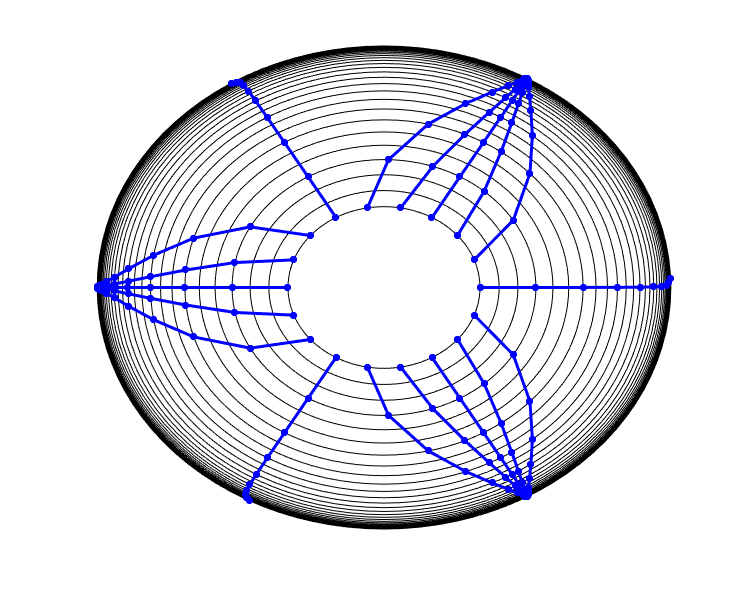}  
\caption{The model of Eq. (\ref{eq:ex1}) for $\Omega = \pi$, $A = 0.1$. (a) Period-two map function, $x_{n+1} = f(f(x_n))$ for $L_0 =2$. Notice two pairs of fixed points, two stable and two unstable. Similar to Figure \ref{fig:Map} the horizontal and vertical axes have been rescaled to $[-1,+1]$. (b) Iterative application of the map function from (a) to a set of initial positions on $[-z(t_0), z(t_0))$ circle. Black concentric rings count the numbers of drive periods. Note that the map iteration proceed in steps of 2.  (c) Same as (b), but for the case $L = 3$, and period-3 map function, $x_{n+1} = f(f(f(x_n)))$.}\label{fig:higherfp}
\end{figure}

Figure \ref{fig:higherfp}  shows numerical results for the model Eq. (\ref{eq:ex1}) for $\Omega = p\Omega_0$, for $p =2$ (a and b), and $p = 3$ (c).  

The higher order map function for the case $p = 2$ (Fig. \ref{fig:higherfp}a) shows twice the number of crossings  
as in the case $p=1$ (Fig. \ref{fig:Map}), two stable and two unstable. The iterative application of $f^{(2)}(x)$ map (Fig. \ref{fig:higherfp}b) shows convergence to two fixed points (black rings mark the individual drive periods; the iterations proceed in steps of 2). The almost straight "rays" coming out vertically correspond to unstable fixed points: arbitrarily tiny deviation from the exact fixed point grows exponentially with the discrete time.
Figure \ref{fig:higherfp}c shows analogous plot for $p = 3$.

\subsection{Time-reversal: Signal compression and decompression}\label{sec:CompDecomp}

In this section we illustrate how the discrete time reversal protocol described in Section \ref{sec:TR} can be applied for signal compression, transmission, and reconstruction.

As we have already seen, if a (multi-period) map function has fixed points, e.g.,  $x = f^{(p)}(x)$, then there is an exponential compression of the initial field $\cA(t_0,x)$  around a few fixed point trajectories. That is, the (near-resonant) pumping automatically leads to  compression, which may be useful, e.g., for signal transmission (shifting signal into frequency range to minimize losses, to avoid interception, or for multiplexing purposes).
The recovery of the original signal can be achieved with the help of the discrete time-reversal of Section \ref{sec:TR}, in particular utilizing the protocol (\ref{eq:TR}).

\begin{figure*}[htb]
     a\includegraphics[width=.66\columnwidth]{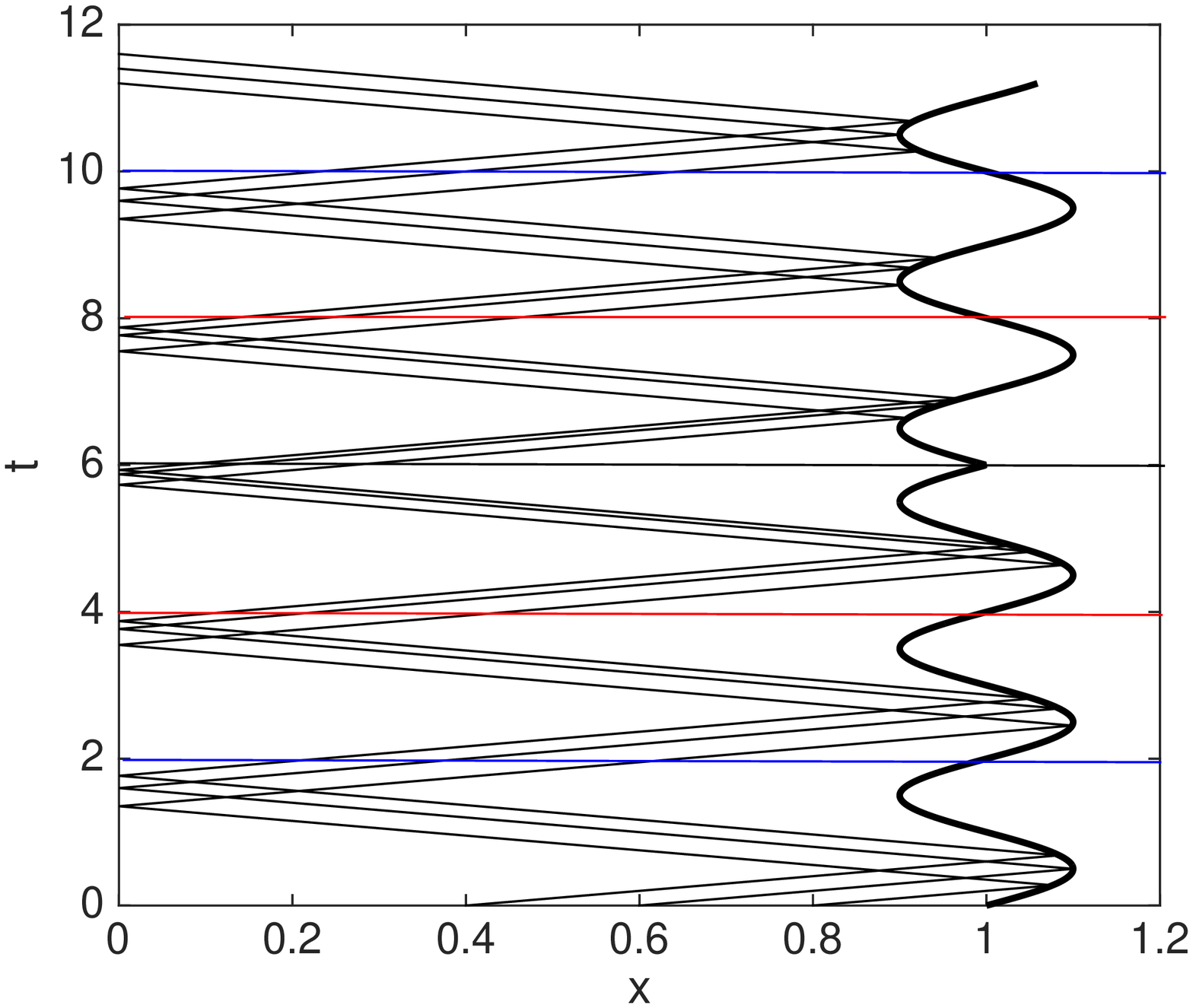}  
     b\includegraphics[width=.66\columnwidth]{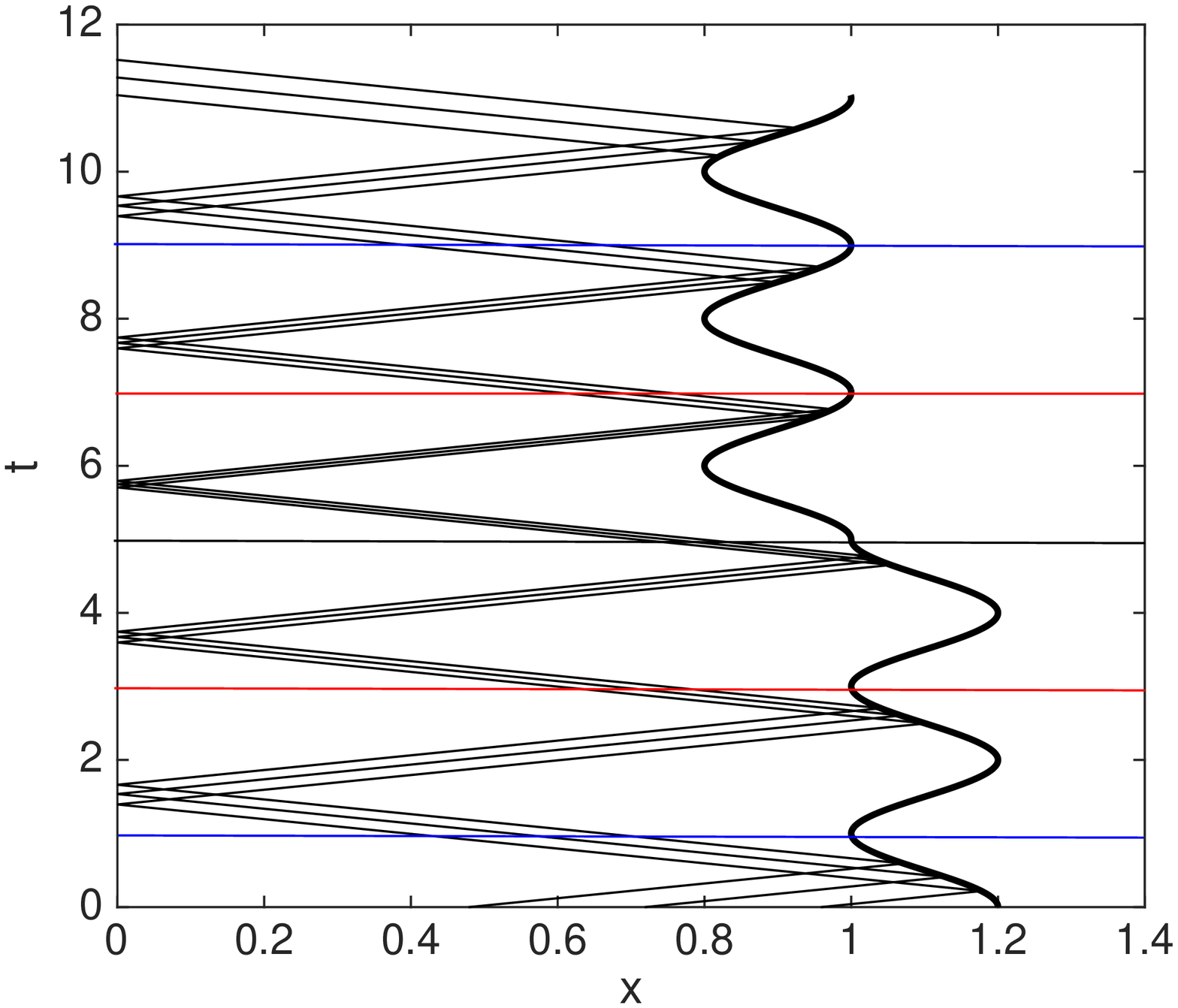}  
     c\includegraphics[width=.66\columnwidth]{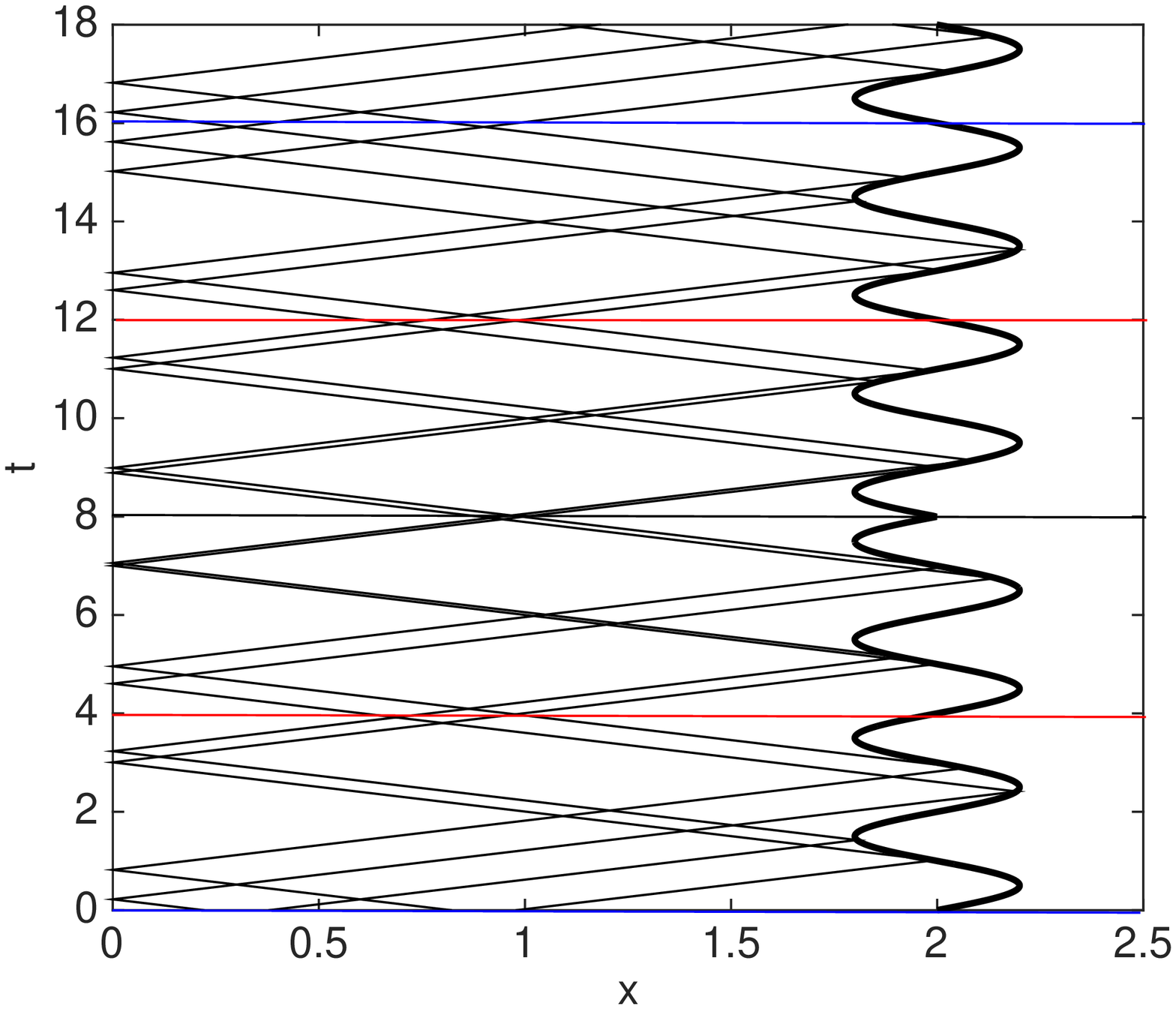}  
	\caption{Time-reversal operation (\ref{eq:TR}) applied to a model for Eq. (\ref{eq:ex1}); $\Omega = \pi$; thick wavy line on the right represents world line of the moving mirror. The second (static) mirror is located at $x = 0$. 
The moment of protocol time-reversal is marked by horizontal black line. Same color line below and above mark states that are time-reversed partners. (a) $L = 1$, $A = 0.1$; (b) $L = 1.1$, $A = 0.1$;  (c) $L = 2$, $A = 0.2$.
}
	\label{fig:TR}
\end{figure*}

In Figure \ref{fig:TR}, the time reversal protocol relative to some time $t^*$ is illustrated in the world-line picture. We used the model Eq. (\ref{eq:ex1}) for the movement of the mirror at $t< t^*$.  The thick black line on the right is the trajectory of the moving mirror.  The other mirror is located at $x = 0$. The break in the first (a and c) and the second (b) derivatives of the mirror world line indicates the time inversion boundary $t^*$ (also marked by horizontal black line). The map function relates the coordinate of null-line at time $t$ and $t+ T$. In the panels a and b, the horizontal thin lines are spaced vertically  by $T$.  The intersections of null lines with the black line map onto intersections with the upper red line, etc. The application of the time-reversal protocol (\ref{eq:TR}) makes the values at the intersections with the upper red line identical to the ones with the lower red line. Similarly for the blue lines.  The correspondence is not affected if the whole set of lines is shifted rigidly vertically. This simply corresponds to change in the reference time $t_0$ of the map.

Figure \ref{fig:TR}a is at the perfect resonance condition, $L = L_0$. The time reversed protocol $z(t>t^*)$ amounts to a mere $\pi$ shift relative to $z(t<t^*)$, which could have been naively guessed. Figure \ref{fig:TR}b, on the other hand, is at the boundary of the fixed point region, $L = L_0 + A$, with the fixed point world line reflecting from the mirror at its extreme left position. The time-reversed protocol $z(t>t^*)$ corresponds to the reflection around $x = L_0$ line and even has a different average cavity length, $L = L_0 -A$. Again, inspection of the intersections of the null-lines with the horizontal colored guides spaced by $T$ shows that the time-inversion procedure works. Finally, Fig.  \ref{fig:TR}c applies the generalized time-reversal protocol (\ref{eq:TRp}) to the case of a higher resonance, $p = 2$,  for $L = 2L_0$.  The time reversal now applies to the period-two map, $f^{(2)}$; therefore the horizontal reference (observation) lines are now spaced by $2T$ instead of $T$, and $z(t>t^*)$ corresponds to the reflection of $z(t<t^*)$ around $x = 2L_0$ line.
Again, the accurate mapping is verified.

\begin{figure}[htb]
     a\includegraphics[width=.99\columnwidth]{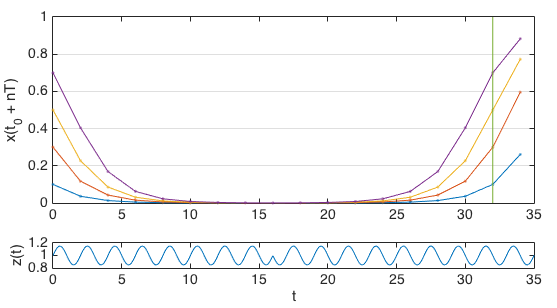}  
     b\includegraphics[width=.99\columnwidth]{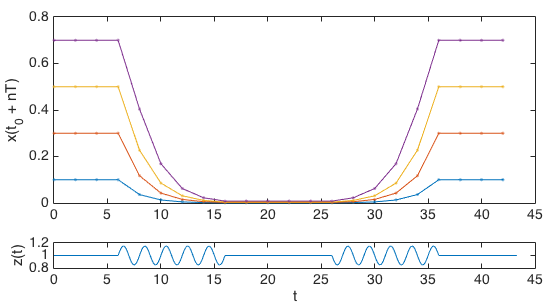}  
	\caption{Discrete time iterated map starting form a set of initial values of $x$ for $L = L_0 = 1$, $A = 0.15$.  (a) Compression followed by immediate decompression. Time reversal is applied at $t^* = 16$. The lines are symmetric with respect to this time. The vertical green line marks the return to the initial conditions. Lower subplot shows the mirror modulation protocol, $z(t)$.  (b) Compression followed by free propagation before decompression. Lower subplot shows the mirror modulation protocol, $z(t)$. 
}
	\label{fig:TRstrob}
\end{figure}

The stroboscopic view of the time-reversal procedure is shown in Figure \ref{fig:TRstrob}. Here, only the values of the iterated map $x_n$ (intersections of the null lines and horizontal guide lines from Fig. \ref{fig:TR}) are shown starting from several initial conditions for the null lines. Figure \ref{fig:TRstrob}a shows compression followed by immediate decompression. Vertical green line marks the return to the initial values $x_0$.
In Figure \ref{fig:TRstrob}b, a more complex protocol is employed: an undriven propagation ($z(t) = {\rm const}$), followed by compression, then another interval of free propagation, followed by decompression. It is important for the compression and decompression protocols ($z(t) \ne {\rm const}$) to be relatively synchronized; however, the free propagation need not even be inside the cavity.

\subsection{Classical energy density: Analytical vs. Numerical results}\label{sec:NEdens}

We now illustrate the exponential energy concentration phenomenon described in Section \ref{sec:En}. We consider here the classical fields, and assume that the initial energy density is uniform. 

In Section \ref{sec:Weak} we gave analytical results for the multiperiod map function $f^{(p)}$ that are expected to apply in the weak near-resonant drive limit, $A/L \ll 1$, $L - pL_0 \ll L_0$. Here we test the quality of the analytical solutions by comparing them to the numerically exact results at various drive amplitudes.

\begin{figure}[htb]
     a\includegraphics[width=.9\columnwidth]{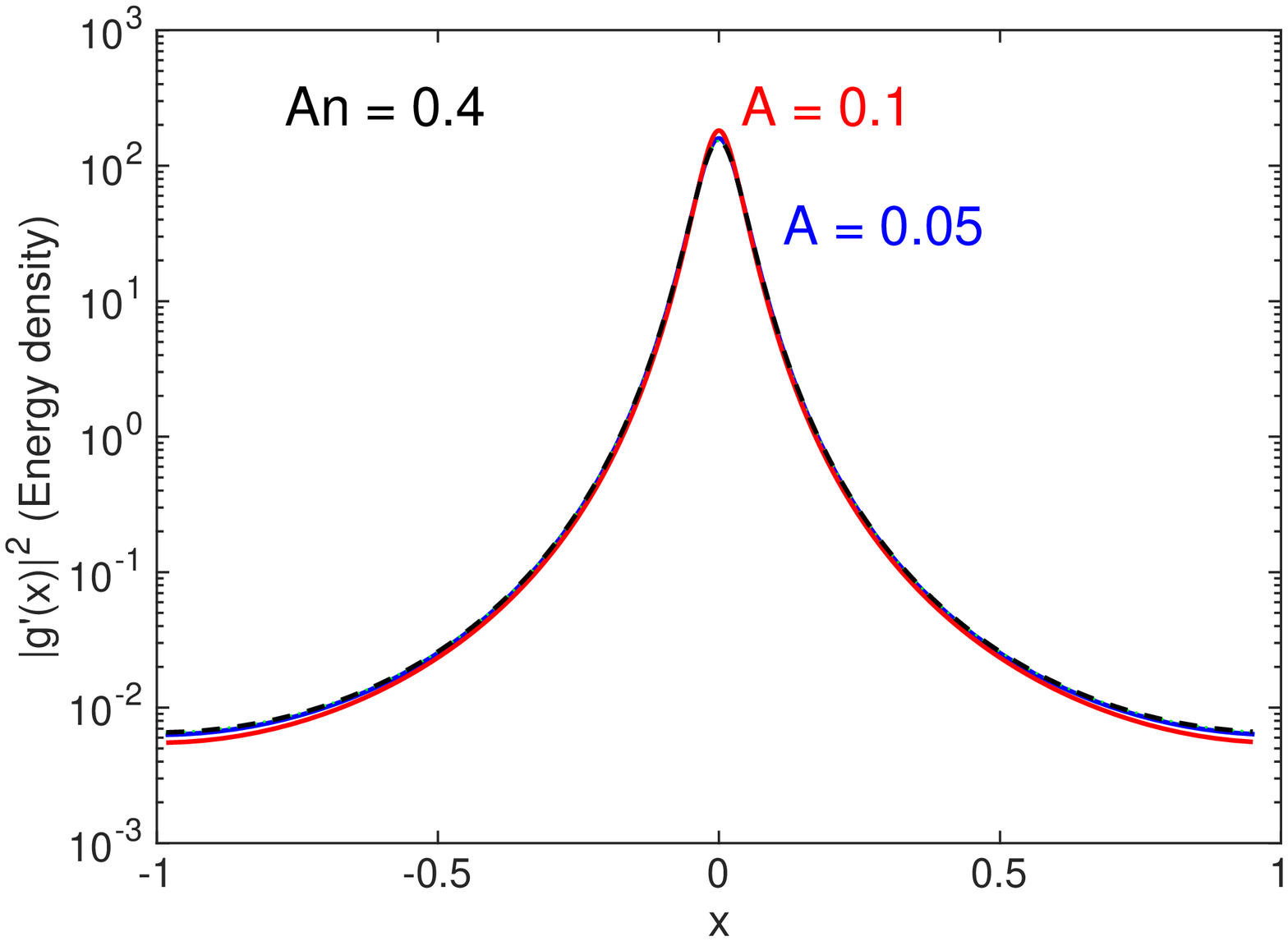}  
     b\includegraphics[width=.9\columnwidth]{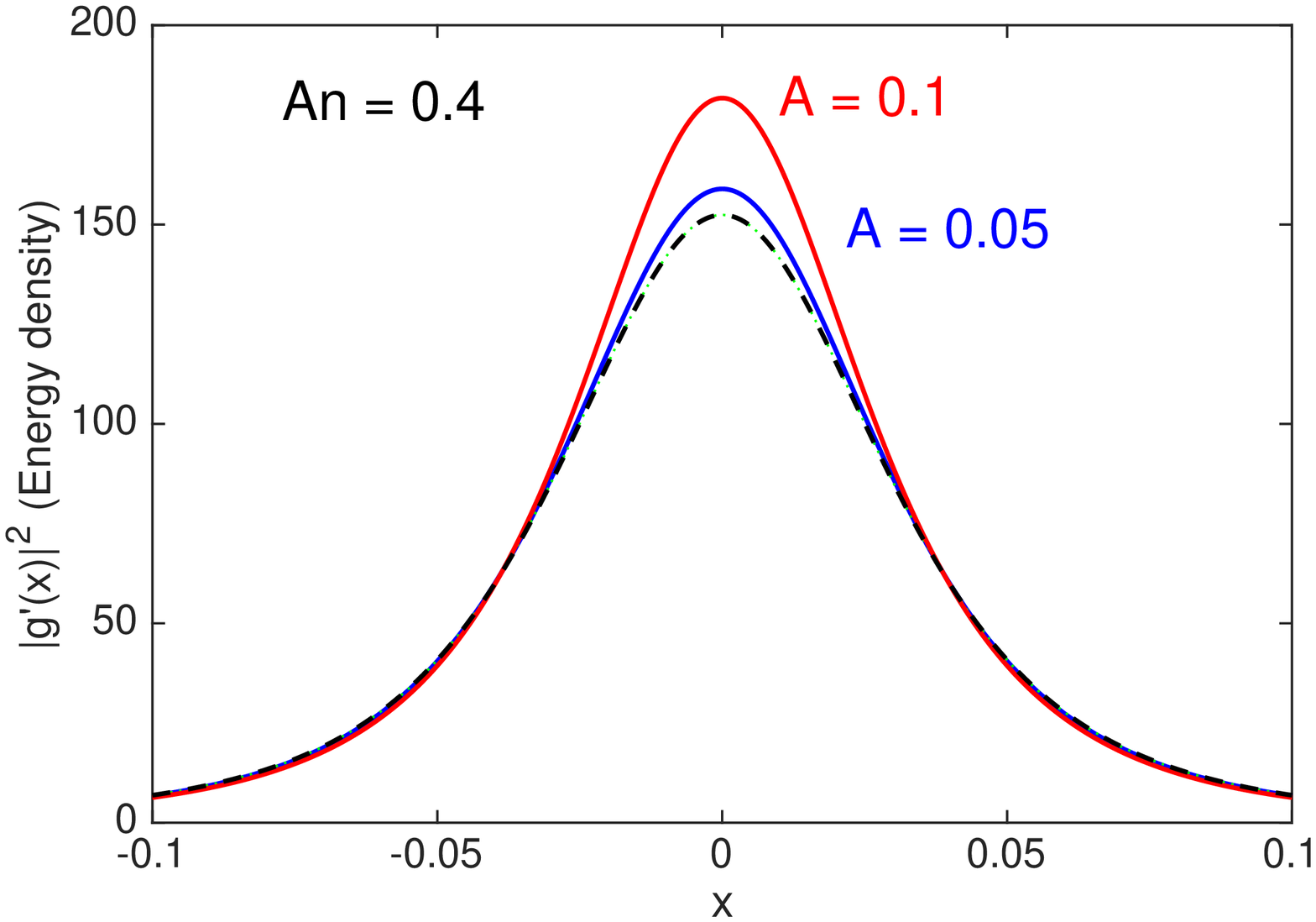}  
	\caption{Snapshot of the energy density, starting from the initially uniform for  $A = 0.1$ (red) and $A = 0.05$ (blue), after 4 and 8 periods of cavity modulation (on resonance $L = L_0 = 1$). (a) Log scale, (b) zoomed-in linear scale.
Black dashed line is the theoretical result (\ref{eq:edxt}) and  (\ref{eq:gp})  that corresponds to $An = 0.4$. The fit improves for weaker modulation, as expected.
}
	\label{fig:EnDens}
\end{figure}

Figure {\ref{fig:EnDens}} shows the snapshot of energy density after several periods of resonant drive ($L = L_0=1$) 
for two drive amplitudes, $A = 0.1$ and $A = 0.05$. The analytical result (\ref{eq:gp}) fits well in both cases in the full range of positions $x$, predictably improving for smaller $A$.

\begin{figure}[htb]
      a\includegraphics[width=.9\columnwidth]{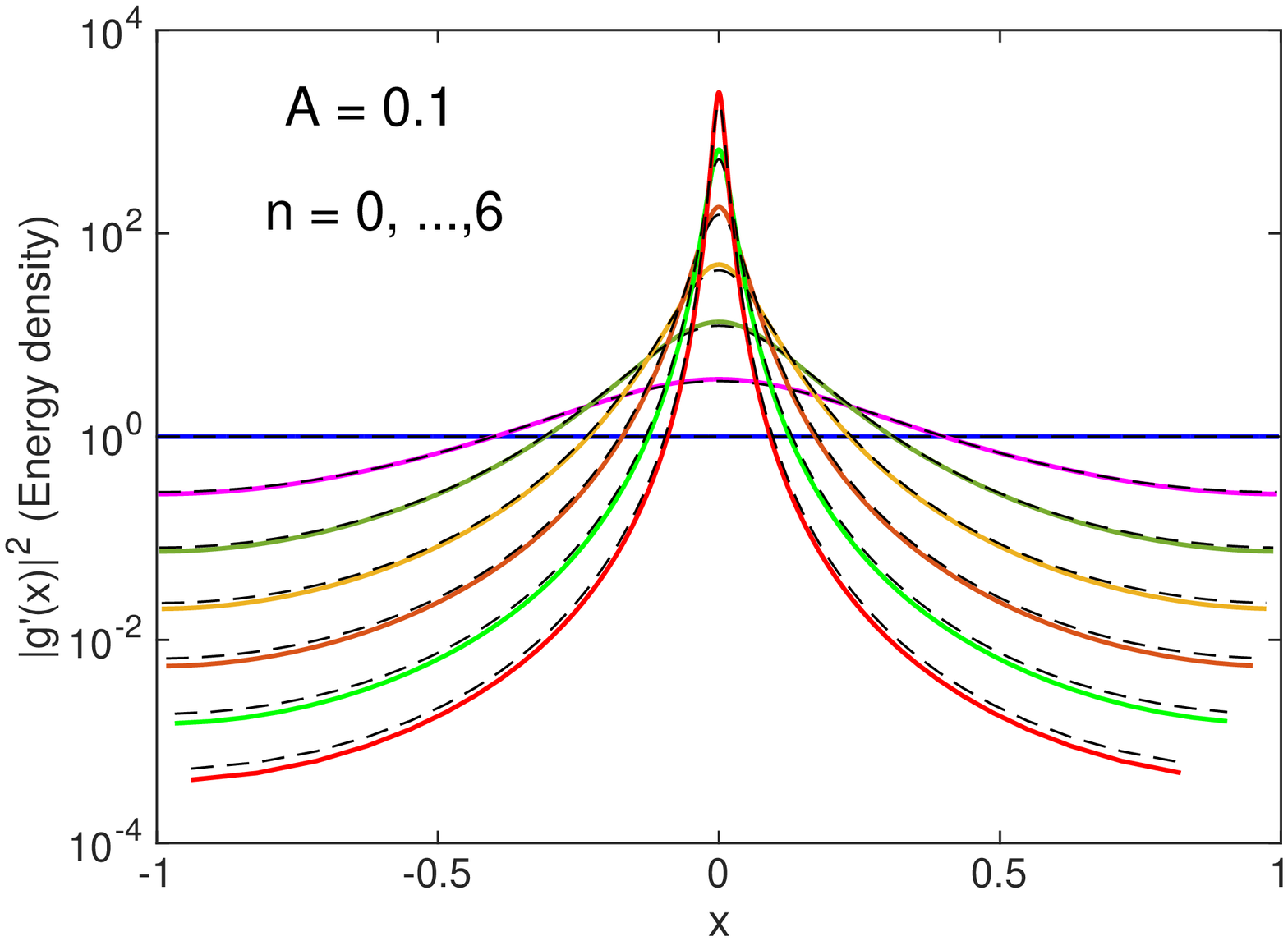}  
      b\includegraphics[width=.9\columnwidth]{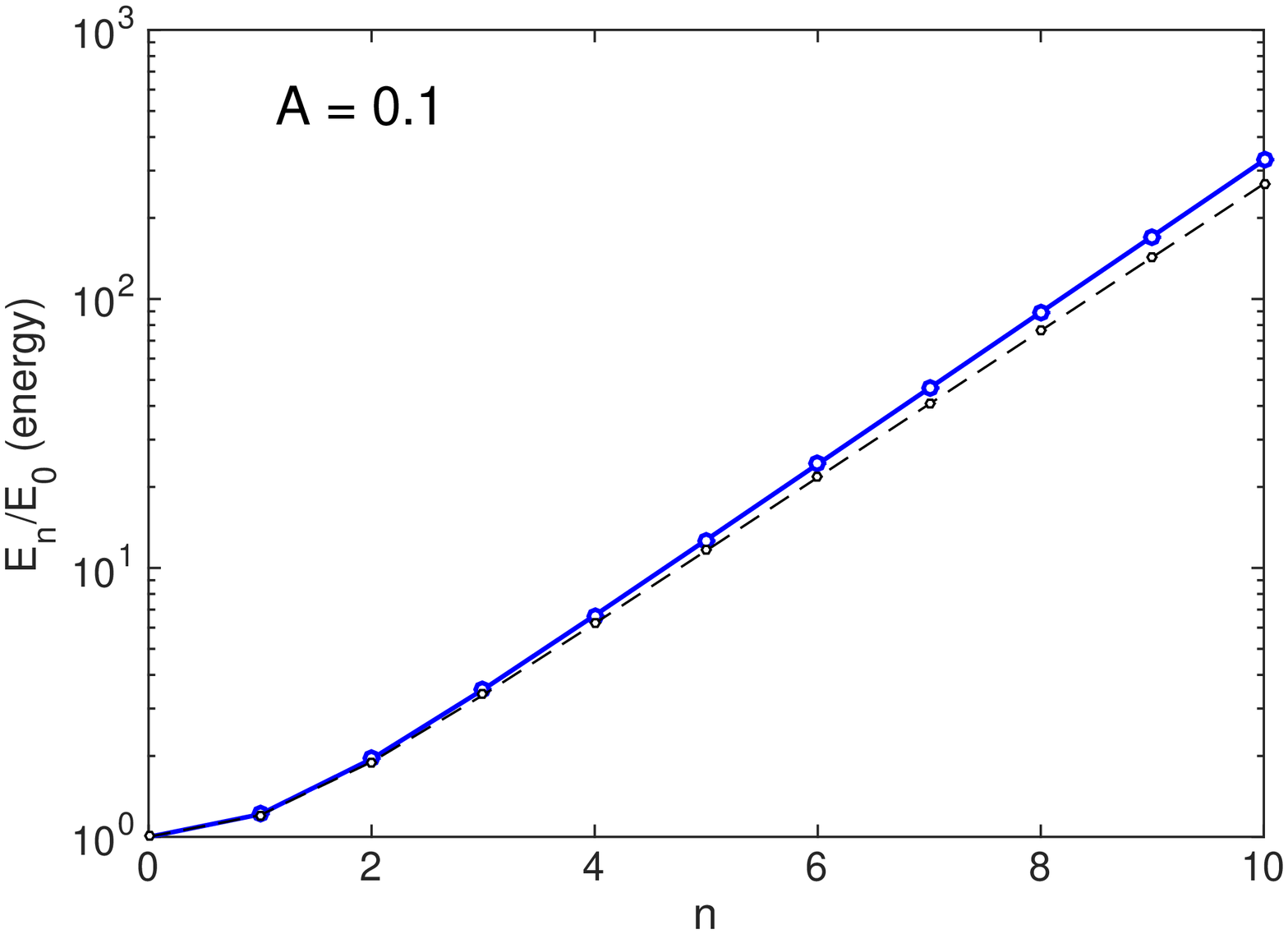}  
	\caption{(a) Energy density and (b) energy after $n$ pumping periods; $A = 0.1$, on resonance $L = L_0 = 1$.
Black dashed lines are the theoretical result from (\ref{eq:edxt}) and  (\ref{eq:gp}).
}
	\label{fig:EnDensN}
\end{figure}

Figure \ref{fig:EnDensN} shows the energy density and the total energy as a function of the number of drive periods (that is, discrete time), for drive amplitude $A = 0.1$.  Both the energy density and the total energy increase exponentially with time. Again, there is a good agreement with the analytical results (\ref{eq:edxt}) and  (\ref{eq:gp}), as well as the qualitative considerations based on the multiplicative Doppler shifts near the stable fixed point, Eq. (\ref{eq:kDop}).

\section{Summary and discussion}

The  principal result of this work is the connection between the problem of a parametrically driven electromagnetic cavity and dynamical systems.  The dynamical systems -- specifically, {\em maps} -- are characterized by their attractors. In 1D, these are fixed points, which can be stable and unstable, and, as a function of control parameters, can undergo bifurcations  and other transformations, including heralding chaos \cite{feigenbaum1978quantitative}. Interestingly, a similar --Floquet -- map account for the evolution of the electromagnetic field in a periodically modulated cavity. The fixed points of the Floquet map are the loci of the energy concentration in the long time limit. On the practical side,  Floquet maps can be tuned to perform signal compression and decompression. On the conceptual side, the maps realize spatial transformations that can be recast into the general relativity language, with the fixed points corresponding to the black hole and white hole horizons.
In this section, we point out connections to some well known phenomena,  discuss extensions and applications, as well as  expand on some nuances of the presented results. 

\subsection{Relation to other phenomena}

{\em Relation to mode-locked lasers.}
Resonant modulation of electromagnetic cavity properties has been the key invention that enabled the ultra-short pulse generation in mode locked lasers \cite{rp2008}.
In the modern solid state lasers, the mode locking effect is typically achieved by active electro-optical or acousto-optical modulation or via ``passive'' saturable absorbers. However, the original realizations of mode locking in the gas lasers indeed involved the spatial modulation of the mirror separation \cite{Hen1966, Smith1967, Baran1967}. In the case of lasers, the primary source of energy is electrical or optical one, which creates population inversion; the emitted light occupies a number of the cavity modes, with generally random relative phases. The standard picture of mode-locked lasing is that the cavity modulation locks all modes to be (very large) integer multiples of the modulation frequency, with synchronized phases. The result is very short and intense pulses of energy bouncing back and forth inside the cavity.
Clearly, there is a close relationship between this standard (Fourier space) picture and the real space description in this paper, which provides interpretation of the mode locking in terms of fixed points of a real space map.

{\em Relation to frequency modulation.}
Waves are a universal carrier of energy and information.  The standard technique for information transmission is to take a monochromatic {\em carrier wave} and encode information by modulating either the amplitude (AM) of the frequency/phase (FM) of the wave with the signal. A possible way to produce phase modulation is to reflect the carrier wave from a moving mirror -- changing the mirror position changes the path length, thus introducing the phase shift directly  proportional to the mirror displacement (and frequency shift proportional to the mirror velocity).  The cavity setup with one moving and one stationary mirror that's been the focus of the  present work is therefore a way to create a resonant phase modulation loop that shifts frequency to zero everywhere, except for the fixed point trajectory where it becomes infinite.

\subsection{Further directions and applications}

{\em Applicability to individual photon ``wavefunctions."}
The cavity modulation techniques can significantly change the properties of the classical cavity fields, and even modify the vacuum energy.  For quantum information applications it could be useful to be able to change the properties of individual photons, e.g. to shift their frequency to a more convenient range for transmission. Whether it is possible to perform such a manipulation while preserving quantum information \cite{klm2001} is an interesting question.  Moreover, the parametric pumping should allow creation of spatially multi-partite single and multi photon quantum states (cat stats), that should have unusual quantum correlations.

{\em Non-1D cavities.}
In this work we focused on the dynamics of driven 1D cavities for the reasons of simplicity. Such cavities indeed exist and can have high quality factor, as is the case for superconducting strip-line resonators. It is however interesting to explore whether it is possible to implement a synchronous driving protocol with 2D circular or 3D spherical cavities. The resonant locking conditions should remain essentially the same as in 1D for the symmetric ($s$)-waves in circular and spherical cavities; however, one can expect a dramatic energy density amplification  when the energy pulses pass through in the cavity center. For instance, for a spherical cavity of radius $R$, and the pulse width $\xi$, the amplification of energy density at the center, within a volume $\xi^3$ is by about a factor $(R/\xi)^2$ compared to when the pulse is near the resonator surface. Similar focusing phenomenon occurs in the acoustic waves in spherical cavities and can lead to sonoluminescence \cite{Putt1991}.

{\em Beyond simple resonances.}
Here we studied  the cases when the cavity modulation  is performed at the frequencies near the unperturbed cavity resonances, $\Omega = p \Omega_0$. However, we cannot rule out additional interesting features, including stable fixed points,  at other drive frequencies, e.g. rational $\Omega = (p/q) \Omega_0$ ($p$ and $q$ integers).  A possibility of a non-trivial dynamics follows from the close relationship between the Floquet maps and the Circle map that shows the Devil's staircase of phase lockings (Arnold tongues) at rational detunings between the resonant  and the drive frequencies \cite{Boyland1986}.

{\em Beyond periodic drives.}
The fixed point trajectories are expected to be stable to slow variation of parameters, such as frequency or the drive amplitude. Thus, for the applications such as the cavity ``sweeping" (parametric cooling), the drive can be turned off gradually, all the way down to zero amplitude, if driven precisely on resonance. 

{\em Classical refrigeration.}
The cavity sweeping protocol described in Section \ref{sec:sweep} can be equally well applied to mechanical resonators, as basic as a high quality violin string under tension. The corresponding modulation protocol comprises of (1) modulating the string length by moving one of the support points in resonance with one of the string modes, e.g. main resonance; this will concentrate most of the string energy in a short pulse; (2) as the pulse approaches one of the ends of the string, add another support point (node), so that the energy pulse becomes outside the ``redefined" string; (3) dissipate the pulsed energy into the environment; (5) Use the cold string as a mechanical refrigerant; (4) move the support (node) the the original position (reset the string); go back to step (1). Advantage of such a parametric cooling protocol compared to more standard side-band cooling techniques is that {\em all} vibrational modes are cooled simultaneously.

\subsection{Nuances}

{\em Stroboscopic General Relativity.}
We have argued in  Section \ref{sec:GR} that the Floquet map in a modulated cavity introduces non-trivial space-time curvature, with the fixed points corresponding to the event horizons. This may seem surprising, since the modulation is only affecting the boundary conditions for the field. There is an analogous situation that occurs in chaotic billiards. There, the interior of the billiard is trivial, and it is only the reflections at the boundary (or between particles) that may cause chaotic dynamics. A connection between this problem and kinematics in curved spaces was pointed out by I. Arnol'd \cite{Arnold63, Sinai70}. Indeed, the (two-sided) billiard can be thought of as a flattened ellipsoid (a specular reflection can then be thought of as a smooth transition between the ``upper" and the ``lower" surfaces of the ellipsoid-billiard). The integrated Gaussian curvature of an ellipsoid is $4\pi$; all of it becomes concentrated at the edge as the ellipsoid is flattened into a billiard.  Similarly, if there are two particles in the billiard, one of them  static, the effective topology becomes equivalent to a torus with two holes (genus two, ``kringel") that has integrated curvature $-4\pi$. 
Then, using the result by Hopf \cite{Hopf49} on geodesics on surfaces with negative curvature, Arnol'd argued qualitatively that the two particle case is chaotic, while for one particle may or may not be. Similarly, even though in the case of the dynamical mirrors studies here, the field is only affected at the time of reflection from a mirror, there coarse-grained effect is analogous to light propagation in a curved space-time.

\acknowledgements

I'd like to acknowledge useful discussions with B. I. Halperin, E. Demler, D. Orgad, H. Kapteyn, V. Balasubramanian, K. Schwab, V. Rosenhaus,  D. Dalvit, and K. Agarwal.
This work was initiated at the Kavli Institute to Theoretical Physics, supported in part by the National Science Foundation under Grant No. NSF PHY-1748958. Part of this work was performed at the Aspen Center for Physics, which is supported by National Science Foundation grant PHY-1607611. Work at Argonne National Laboratory was supported by the Department of Energy, Office of Science,  Materials Science and Engineering Division.

\appendix

\section{Connection between resonantly modulated cavity and Unruh effect}\label{sec:unruh}

According to the Unruh effect \cite{FullingUn73, DaviesUn75, Unruh76}, when an observer moves through vacuum with acceleration, she will see a thermal electromagnetic field. One way to obtain this result is to start with a monochromatic wave, $\cA \sim e^{ik(x - ct)}$ and transform into accelerating frame using the Rindler coordinates, 
\bea
x &=& \xi \cosh(\alpha \tau),\\
ct &=& \xi \sinh(\alpha \tau),
\eea
where scaled acceleration $\alpha = a/c$, $\tau$ is the observer proper time, and $\xi = c^2/a$ (here we reinstated the dimensionfull speed of light). Thus, in the moving frame,  
\beq
\cA \sim e^{ik\xi e^{\alpha\tau}}\label{eq:Unruh}.
\eeq
The spectrum of this highly unharmonic wave  yields the Plank distribution with the Unruh temperature, $T_U = \hbar \alpha/(2\pi k_B)$ \cite{PADMANABHAN200549, Blenco2012}.  

The form of the wave-vector renormalization obtained for cavity due to the repeated Doppler shifting (\ref{eq:kDop}) near a fixed point is the same as (\ref{eq:Unruh}), after identifying the Unruh acceleration $a$ with $2v/T$. This is to be expected, since on the fixed point trajectory, the moving mirror provides (delta-function) acceleration kicks relative to the incoming light. The factor of 2 is natural. Let's start with a light pulse with wavevector $\sim k$ in the lab frame, pointed towards moving mirror. To treat the reflection in stationary frame, let us change into the frame co-moving with mirror, so that at the moment of encounter the mirror is stationary. This is the first boost by $v$, with the velocity jump directed opposite to $k$. The pulse reflects from the mirror and starts traveling towards the other mirror (one stationary in the lab frame). To go back to the lab frame, we need to perform another boost, again in the direction against $k$. This is the second jump by $v$.  Two boosts by $v$ over the time period of $T$ yield an effective acceleration $a = 2v/T$ {\em relative} to the light pulse. However, in the lab frame, the boosts cancel each other. Thus we reach a seemingly paradoxical conclusion that the lab frame behaves as if it is accelerating relative to the fixed point trajectories.
Since in the long time limit, {\em all} classical radiation becomes concentrated near the stable fixed point trajectories, the {\em average} behavior of the driven cavity as a whole maps onto the uniformly accelerating system.

The presented reasoning applies to a classical cavity field. It would be interesting to see if it remains valid in the case of quantum fluctuation of vacuum, making the dynamical Casimir effect equivalent to the Unruh radiation in a reference frame accelerating with the fixed-point acceleration $a$.

\section{Spatio-temporal modulation of refraction index}\label{sec:nxt}
An alternative way to parametrically modulate a cavity is to change the  refraction index of the medium inside the cavity $n(t,x)$, without changing its shape or size. The resulting phenomenology is similar to the mirror resonator; however, instead of varying the physical length, modulation is implemented by {\em locally} varying the optical length, $dx_{opt} = n(t,x) dx$. As we will see, there is a close correspondence with the moving mirror case.  However, it holds only when the spatial and temporal variation of $n(t,x)$ is much slower and smoother than the modulated waves themselves (semiclassical limit). Otherwise, one has to include  back-scattering on the spatio-temporal ``boundaries", i.e., the regions of rapid change of $n$. The strong resonant compression effect around fixed point trajectories that we found in the main part of this paper appears when modulation is preformed around the fundamental or higher cavity mode frequencies. Thus, due to the slowness constraint in the case of modulated medium, the effect becomes limited to the waves whose Fourier spectrum contains only the high harmonics of the cavity. This is in contrast to the moving mirror case, where fixed point compression affects all possible modes equally well and no semiclassical limit is necessary.

\begin{figure}[htb]
      a\includegraphics[width=.7\columnwidth]{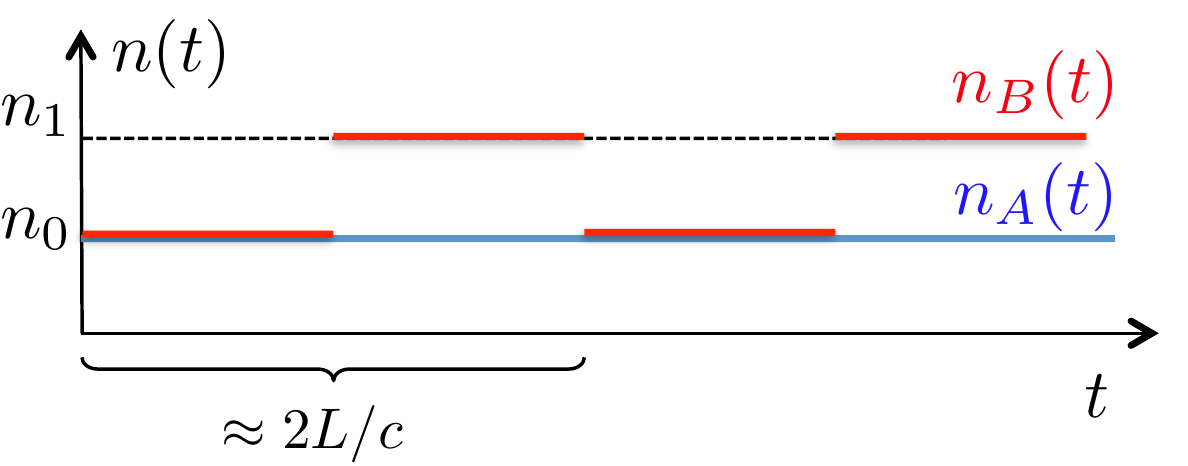}  
      
      \vspace{9mm}
      b\includegraphics[width=.9\columnwidth]{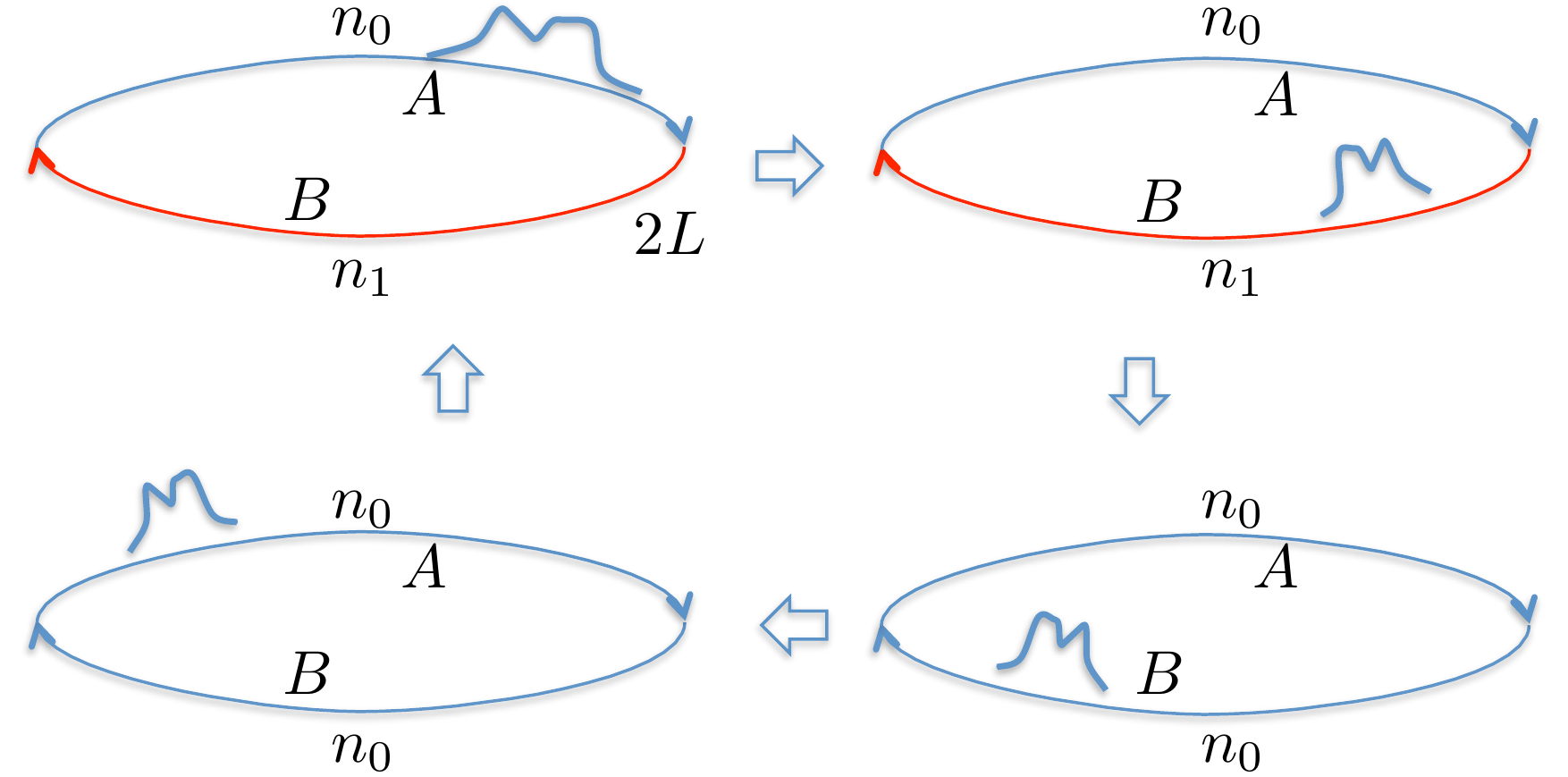}  
	\caption{ Wave packet contraction by spatiotemporally modulating refraction index. In one half of the ring cavity (A) refraction index is kept constant at $n_0$, in the other (B) it alternates between two values, $n_0$ and $n_1$ (panel a). Panel b: A wave packet (whose envelop is shown schematically), enters the B region when it is in the state $n_1$ ($> n_0$, color of the B segment indicates the value of $n$), and becomes spatially contracted, while the temporal frequencies remain unaffected. Then the refraction index drops to $n_0$ in B, while packets is still inside. This increases the temporal frequency, without affecting the spatial (wavevector) composition. If the modulation in region B is synchronized with the packet round trip, the contraction repeats multiplicatively every cycle.
	}
	\label{fig:nxt1}
\end{figure}

Qualitatively, the possibility of resonant wave compression by modulating $n(x,t)$ can be illustrated by the following example. As before, we will  represent the cavity as a ring of length $2L$ (see Section \ref{sec:unfold}; the actual cavity may  either be a ring  or a linear cavity of length $L$).
Suppose $n(t,x)$ is modulated in a step-like fashion, both in space and in time (assuming that the ``steps" are still smooth on the typical wave oscillation scale, making back-scattering unimportant). Namely, in half of the cavity $L<x< 2L$ (region B), the index alternates between values $n_0$ (same as the rest of the cavity, region A) and $n_1$ ($n_1> n_0$), with the frequency near the fundamental cavity mode (if $n_1 - n_0 \ll n_0$), see Figure \ref{fig:nxt1}. Now, let us prepare an initial wave packet in the region A. Let's choose the phase of the  drive such that when the packet arrives at  $x = L$, the refraction index in region B is at $n_1$. As the packet enters the region B with larger but static $n_1$, it spatially contracts, without changing its temporal frequency composition. As the next step of the protocol, while the packet is still inside B, the refraction index is reduced down to value $n_0$. Since the change is spatially uniform from the view point of the packet, the wave-vector composition remains unchanged in this process, while the frequency has to increase. The result of this two step process is therefore both spatial and temporal contraction of the incoming wavepacket. Now, arranging the drive frequency to be synchronous with the packet round-trip cycle (which is going to be close to the fundamental if $|n_1 - n_0| \ll n_0$), the packet can become -- in the ideal case -- infinitely contracted.  In the non-dispersive limit, when the phase velocity is constant, the drive frequency can be kept constant. But even if there is some dispersion, it can be compensated for by gradually adjusting the drive frequency (the same principle as used in synchrotrons). 

Clearly, this simple picture is valid only to the extent that the back-reflection can be neglected. For the part of the wave that is getting contracted, the semiclassical approximation becomes progressively better as the time goes by. Still, it would be interesting to go beyond the semiclassical limit, and systematically study the corrections due to the finiteness of the carrier frequency.

In the next subsections we derive the semiclassical equations for the wave propagation in the medium with spatiotemporally varying refraction, and derive the analytical map function for the weakly (and almost locally) driven medium, and see that it corresponds exactly to the case of the weakly modulated mirror described in Section \ref{sec:Weak}.

\subsection{Wave propagation in smoothly varying $n(t,x)$}

The electromagnetic wave equation in a nonstationary 1D medium with permittivity $\epsilon(t,x)$ and permeability $\mu(t,x)$ is \cite{Dodonov93}
\beq
\frac \partial {\partial t} \epsilon\frac \partial {\partial t} \cA - \frac \partial {\partial x} \frac1{\mu} \frac \partial {\partial x} \cA  = 0, \label{eq:WEM}
\eeq
which for uniform stationary medium reduces to Eq. (\ref{eq:WE}). [A mechanical analog is a balls-and-springs chain with space-time varying ball masses (analog of permittivity $\epsilon$) and spring stiffnesses (analog of the inverse  permeability, $1/\mu$.)] The refraction index in terms of these functions is $n(t,x) = \sqrt{\epsilon(t,x)\mu(t,x)}$.

For smooth spatiotemporal variations (relative to the ``carrier" wavevector and frequency), there is no back scattering, and we can look for an approximate solution in the form
\beq
\cA(t,x) = \tilde \cA(t,x) e^{i\phi(t,x)},
\eeq
where $\tilde A(t,x)$ varies on the same scale and the medium properties, and the exponential encodes the fast evolution of the ``carrier'' wave (a typical frequency in the packet).

After substituting this ansatz into Eq. (\ref{eq:WEM}), we  collect the terms that correspond to different powers of time and space derivatives of $\phi$. The first equation, 
\beq
\epsilon \dot \phi^2 - \frac {\phi'^2}{\mu} = 0,
\eeq
accounts for the evolution of the ``fast" phase.  It has two branches, 
\beq
\dot \phi \pm \frac {\phi'}{\sqrt{\epsilon \mu}} = 0,\label{eq:WEphi}
\eeq
corresponding to the two directions of propagation. The physical meaning is clear: for a given branch, the phase remains constant on the characteristics of the wave equation (\ref{eq:WEphi}) that has the space and time dependent speed of light, $c^* = c/\sqrt{\epsilon \mu}$. In the semiclassical limit, the two branches decouple; for concreteness, we pick the right movers, $\dot \phi + \frac {\phi'}{\sqrt{\epsilon \mu}} = 0$.

The other equation that we need involves the first powers of derivatives of $\phi$,
\beq
2(\epsilon\dot {\tilde \cA}\dot \phi - {\tilde \cA}'\phi'/\mu) = - [\dot\epsilon\dot\phi + \epsilon\ddot\phi - (1/\mu)'\phi' - \phi''/\mu]\tilde \cA. 
\eeq

Now, expressing all spatial derivatives of $\phi$ in terms of time derivatives using (\ref{eq:WEphi}), we obtain
\beq
2\left(\dot {\tilde \cA} + \frac{1}{\sqrt{\epsilon\mu}}{\tilde \cA}'\right) = - \frac12\left[\frac{\dot\epsilon}{\epsilon}+\frac{1}{\sqrt{\epsilon\mu}}\frac{{\epsilon}'}{\epsilon} - \frac{\dot\mu}{\mu} - \frac{1}{\sqrt{\epsilon\mu}}\frac{{\mu}'}{\mu}\right]\tilde \cA. 
\eeq
Note that this equation only contains derivatives in combinations that correspond to differentiation along the characteristics; therefore it describes evolution of $\tilde A$ on characteristics. It can be easily integrated, and implies that along the characteristics the following combination is conserved:
\beq
\sqrt{\frac{\epsilon}{\mu}} \tilde \cA^2 = {\rm const}.\label{eq:conserv}
\eeq

For a static inhomogeneous medium, it is equivalent to the spatial independence of the energy current (Poynting vector).

Since the ``fast" phase is also conserved on characteristics, one consequence of (\ref{eq:conserv})  is that if there is only an isolated ``active" region that is being modulated and the rest of the medium is static and uniform (``passive"), the values of the full (not only the slow part $\tilde \cA$!)   vector potential $\cA(t,x)$ inside the passive medium can be read off by tracing characteristics back to the initial conditions, in the same way as in the problem with a moving mirror (Figure \ref{fig:circle}). 

\begin{figure}[htb]
      \includegraphics[width=.7\columnwidth]{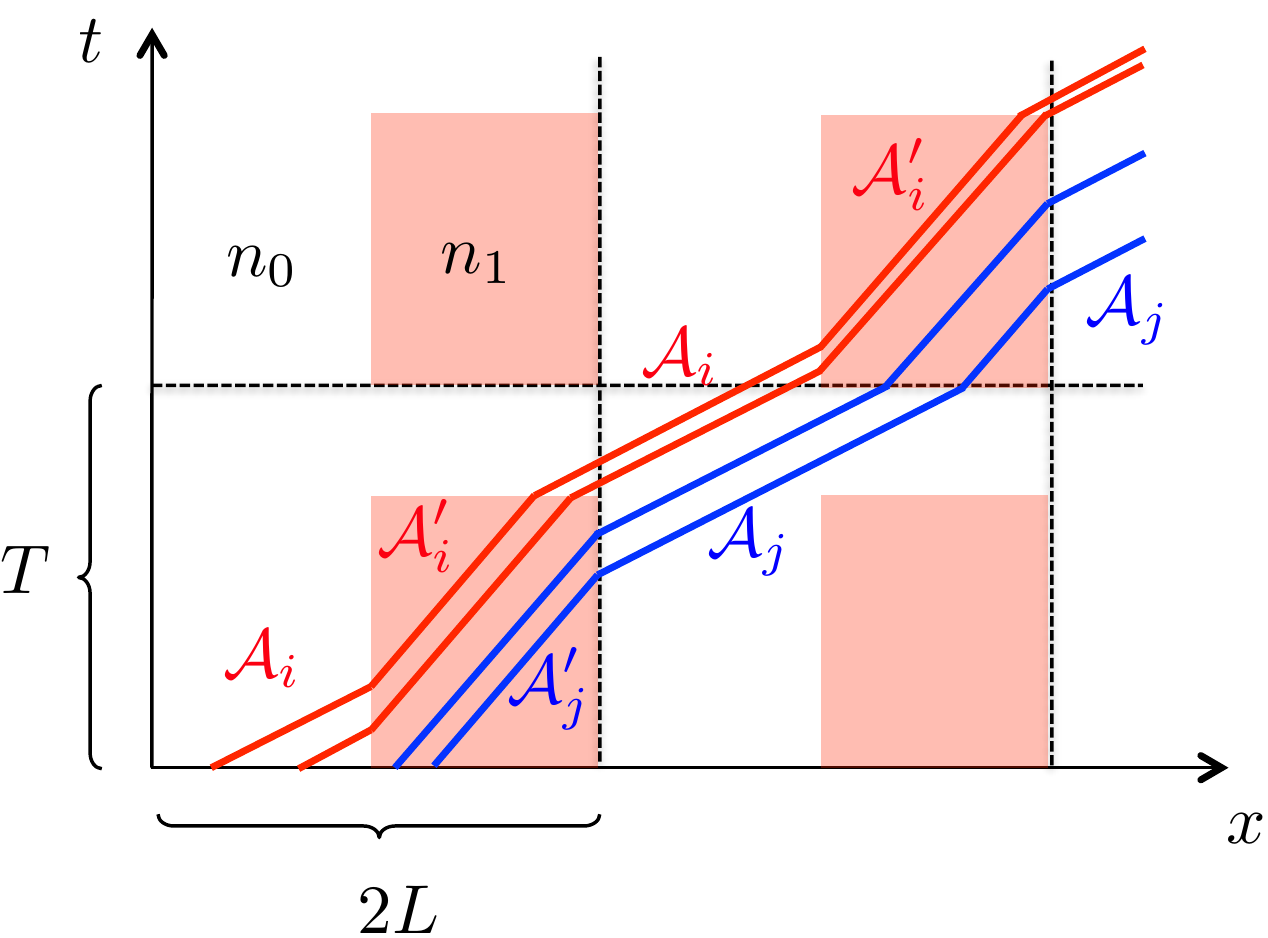}  
	\caption{Graphical solution of the semiclassical wave equation for spatiotemporally modulated medium specified in Figure \ref{fig:nxt1}. Shaded (pink) regions of space-time have refraction index  $n_1$ (spatial region B); the rest of space-time has refraction index $n_0$,   $n_1 > n_0$. The refraction index pattern is periodic both in space (period $2L$, the physical length of the ring cavity of twice the length of linear cavity) and time (period $T$ of modulation). The boundaries of one space-time unit cell are indicated by the dashed lines; however, the behavior of characteristics (tilted lines) is clearer in the extended space-time picture. 
The vector potential $\cA$ on a given characteristic is only a function of local material properties [Eq. (\ref{eq:conserv})]; it therefore remains constant within same color regions. The families of nearby characteristics show convergence and divergence properties, in the same way as in the case of cavities with moving mirrors (see Figure \ref{fig:circle}a).}
	\label{fig:nxt_char}
\end{figure}

We can now refine the qualitative picture of the field contraction described above and in Figure \ref{fig:nxt1}. For that, let us schematically construct the characteristics of the wave equation (\ref{eq:WEphi}),  Figure \ref{fig:nxt_char}.  The vector potential is transported along them, only changing value when the parameters of the medium change. Since we assume in this example that the parameters are piece-wise constant, there are only two values of $\cA$ along each characteristic. We see that the nearby characteristics can converge and diverge with time, depending on the drive phase, as we argued qualitatively above.

\subsection{Analytic result for the weak modulation limit}
We now consider the case when the properties of medium are  periodically modulated only in a narrow spatial region. This case is closely related to the one we studied in Section \ref{sec:Weak}, and similarly to it, can be solved analytically.

The length of the ring cavity is $2L$ (of $L$ for a linear resonator). We will take the reference dielectric permittivity in the cavity as 1, and will allow it to change within a spatial region of width $A \ll L$ (or $A/2$ near one of the mirrors in the case of linear cavity). Within this region, $n(t)$ can be an arbitrary periodic function. 

As described in the previous section, to find the field everywhere outside the modulated region, we only need to determine the characteristics, since $\cA$ is transported along them.

In precise analogy with the Section \ref{sec:Weak}, we can construct the Floquet map for the case of a drive near the fundamental resonance ($\Omega =\pi c/L_0\approx \pi c/L$ -- we reinstated here the vacuum speed of light for clarity). The equation relating spatial coordinates at time $t_0$ ($x_0$) and time $T = 2\pi/\Omega$ later ($x_1$), for spatially narrow modulation of small integrated strength, $|\Delta n| A \ll L$, is
\beq
x_1 + 2L - x_0 + A\left[n\left(t_0 + \frac{L-x_0}{c}\right) - 1\right] = cT, 
\eeq
which has the same form as Eq. (\ref{eq:map}) in Section \ref{sec:Weak}.

For example, for a harmonic modulation of depth $\nu$, for $x_n \to x_{n+1}$, we find
\beq
x_{n+1}  = x_n+ 2(L_0 - L) + A\nu \cos \pi\frac {x_n}{L_0}. 
\eeq

\bibliographystyle{apsrev4-1}
\bibliography{refs}

\begin{thebibliography}{59}%
\makeatletter
\providecommand \@ifxundefined [1]{%
 \@ifx{#1\undefined}
}%
\providecommand \@ifnum [1]{%
 \ifnum #1\expandafter \@firstoftwo
 \else \expandafter \@secondoftwo
 \fi
}%
\providecommand \@ifx [1]{%
 \ifx #1\expandafter \@firstoftwo
 \else \expandafter \@secondoftwo
 \fi
}%
\providecommand \natexlab [1]{#1}%
\providecommand \enquote  [1]{``#1''}%
\providecommand \bibnamefont  [1]{#1}%
\providecommand \bibfnamefont [1]{#1}%
\providecommand \citenamefont [1]{#1}%
\providecommand \href@noop [0]{\@secondoftwo}%
\providecommand \href [0]{\begingroup \@sanitize@url \@href}%
\providecommand \@href[1]{\@@startlink{#1}\@@href}%
\providecommand \@@href[1]{\endgroup#1\@@endlink}%
\providecommand \@sanitize@url [0]{\catcode `\\12\catcode `\$12\catcode
  `\&12\catcode `\#12\catcode `\^12\catcode `\_12\catcode `\%12\relax}%
\providecommand \@@startlink[1]{}%
\providecommand \@@endlink[0]{}%
\providecommand \url  [0]{\begingroup\@sanitize@url \@url }%
\providecommand \@url [1]{\endgroup\@href {#1}{\urlprefix }}%
\providecommand \urlprefix  [0]{URL }%
\providecommand \Eprint [0]{\href }%
\providecommand \doibase [0]{http://dx.doi.org/}%
\providecommand \selectlanguage [0]{\@gobble}%
\providecommand \bibinfo  [0]{\@secondoftwo}%
\providecommand \bibfield  [0]{\@secondoftwo}%
\providecommand \translation [1]{[#1]}%
\providecommand \BibitemOpen [0]{}%
\providecommand \bibitemStop [0]{}%
\providecommand \bibitemNoStop [0]{.\EOS\space}%
\providecommand \EOS [0]{\spacefactor3000\relax}%
\providecommand \BibitemShut  [1]{\csname bibitem#1\endcsname}%
\let\auto@bib@innerbib\@empty
\bibitem [{\citenamefont {Chirikov}(1979)}]{Chirikov1979}%
  \BibitemOpen
  \bibfield  {author} {\bibinfo {author} {\bibfnamefont {B.~V.}\ \bibnamefont
  {Chirikov}},\ }\href@noop {} {\bibfield  {journal} {\bibinfo  {journal}
  {Physics Reports}\ }\textbf {\bibinfo {volume} {52}},\ \bibinfo {pages} {263}
  (\bibinfo {year} {1979})}\BibitemShut {NoStop}%
\bibitem [{\citenamefont {Gonzalez}\ and\ \citenamefont {Piro}(1983)}]{NLO83}%
  \BibitemOpen
  \bibfield  {author} {\bibinfo {author} {\bibfnamefont {D.~L.}\ \bibnamefont
  {Gonzalez}}\ and\ \bibinfo {author} {\bibfnamefont {O.}~\bibnamefont
  {Piro}},\ }\href {\doibase 10.1103/PhysRevLett.50.870} {\bibfield  {journal}
  {\bibinfo  {journal} {Phys. Rev. Lett.}\ }\textbf {\bibinfo {volume} {50}},\
  \bibinfo {pages} {870} (\bibinfo {year} {1983})}\BibitemShut {NoStop}%
\bibitem [{\citenamefont {Oka}\ and\ \citenamefont {Aoki}(2009)}]{Oka09}%
  \BibitemOpen
  \bibfield  {author} {\bibinfo {author} {\bibfnamefont {T.}~\bibnamefont
  {Oka}}\ and\ \bibinfo {author} {\bibfnamefont {H.}~\bibnamefont {Aoki}},\
  }\href {\doibase 10.1103/PhysRevB.79.081406} {\bibfield  {journal} {\bibinfo
  {journal} {Phys. Rev. B}\ }\textbf {\bibinfo {volume} {79}},\ \bibinfo
  {pages} {081406} (\bibinfo {year} {2009})}\BibitemShut {NoStop}%
\bibitem [{\citenamefont {Kitagawa}\ \emph {et~al.}(2011)\citenamefont
  {Kitagawa}, \citenamefont {Oka}, \citenamefont {Brataas}, \citenamefont
  {Fu},\ and\ \citenamefont {Demler}}]{Kitagawa2011}%
  \BibitemOpen
  \bibfield  {author} {\bibinfo {author} {\bibfnamefont {T.}~\bibnamefont
  {Kitagawa}}, \bibinfo {author} {\bibfnamefont {T.}~\bibnamefont {Oka}},
  \bibinfo {author} {\bibfnamefont {A.}~\bibnamefont {Brataas}}, \bibinfo
  {author} {\bibfnamefont {L.}~\bibnamefont {Fu}}, \ and\ \bibinfo {author}
  {\bibfnamefont {E.}~\bibnamefont {Demler}},\ }\href {\doibase
  10.1103/PhysRevB.84.235108} {\bibfield  {journal} {\bibinfo  {journal} {Phys.
  Rev. B}\ }\textbf {\bibinfo {volume} {84}},\ \bibinfo {pages} {235108}
  (\bibinfo {year} {2011})}\BibitemShut {NoStop}%
\bibitem [{\citenamefont {Rudner}\ \emph {et~al.}(2013)\citenamefont {Rudner},
  \citenamefont {Lindner}, \citenamefont {Berg},\ and\ \citenamefont
  {Levin}}]{AFAI-1}%
  \BibitemOpen
  \bibfield  {author} {\bibinfo {author} {\bibfnamefont {M.~S.}\ \bibnamefont
  {Rudner}}, \bibinfo {author} {\bibfnamefont {N.~H.}\ \bibnamefont {Lindner}},
  \bibinfo {author} {\bibfnamefont {E.}~\bibnamefont {Berg}}, \ and\ \bibinfo
  {author} {\bibfnamefont {M.}~\bibnamefont {Levin}},\ }\href {\doibase
  10.1103/PhysRevX.3.031005} {\bibfield  {journal} {\bibinfo  {journal} {Phys.
  Rev. X}\ }\textbf {\bibinfo {volume} {3}},\ \bibinfo {pages} {031005}
  (\bibinfo {year} {2013})}\BibitemShut {NoStop}%
\bibitem [{\citenamefont {{Titum}}\ \emph {et~al.}(2016)\citenamefont
  {{Titum}}, \citenamefont {{Berg}}, \citenamefont {{Rudner}}, \citenamefont
  {{Refael}},\ and\ \citenamefont {{Lindner}}}]{AFAI-2}%
  \BibitemOpen
  \bibfield  {author} {\bibinfo {author} {\bibfnamefont {P.}~\bibnamefont
  {{Titum}}}, \bibinfo {author} {\bibfnamefont {E.}~\bibnamefont {{Berg}}},
  \bibinfo {author} {\bibfnamefont {M.~S.}\ \bibnamefont {{Rudner}}}, \bibinfo
  {author} {\bibfnamefont {G.}~\bibnamefont {{Refael}}}, \ and\ \bibinfo
  {author} {\bibfnamefont {N.~H.}\ \bibnamefont {{Lindner}}},\ }\href {\doibase
  10.1103/PhysRevX.6.021013} {\bibfield  {journal} {\bibinfo  {journal}
  {Physical Review X}\ }\textbf {\bibinfo {volume} {6}},\ \bibinfo {eid}
  {021013} (\bibinfo {year} {2016})}\BibitemShut {NoStop}%
\bibitem [{\citenamefont {von Keyserlingk}\ and\ \citenamefont
  {Sondhi}(2016)}]{vonKeyserlingk2016a}%
  \BibitemOpen
  \bibfield  {author} {\bibinfo {author} {\bibfnamefont {C.~W.}\ \bibnamefont
  {von Keyserlingk}}\ and\ \bibinfo {author} {\bibfnamefont {S.~L.}\
  \bibnamefont {Sondhi}},\ }\href@noop {} {\bibfield  {journal} {\bibinfo
  {journal} {Phys. Rev. B}\ }\textbf {\bibinfo {volume} {93}},\ \bibinfo
  {pages} {245145} (\bibinfo {year} {2016})}\BibitemShut {NoStop}%
\bibitem [{\citenamefont {Yao}\ \emph {et~al.}(2017)\citenamefont {Yao},
  \citenamefont {Potter}, \citenamefont {Potirniche},\ and\ \citenamefont
  {Vishwanath}}]{Yao17}%
  \BibitemOpen
  \bibfield  {author} {\bibinfo {author} {\bibfnamefont {N.~Y.}\ \bibnamefont
  {Yao}}, \bibinfo {author} {\bibfnamefont {A.~C.}\ \bibnamefont {Potter}},
  \bibinfo {author} {\bibfnamefont {I.-D.}\ \bibnamefont {Potirniche}}, \ and\
  \bibinfo {author} {\bibfnamefont {A.}~\bibnamefont {Vishwanath}},\ }\href
  {\doibase 10.1103/PhysRevLett.118.030401} {\bibfield  {journal} {\bibinfo
  {journal} {Phys. Rev. Lett.}\ }\textbf {\bibinfo {volume} {118}},\ \bibinfo
  {pages} {030401} (\bibinfo {year} {2017})}\BibitemShut {NoStop}%
\bibitem [{\citenamefont {Landauer}(1961)}]{Landauer}%
  \BibitemOpen
  \bibfield  {author} {\bibinfo {author} {\bibfnamefont {R.}~\bibnamefont
  {Landauer}},\ }\href {\doibase 10.1147/rd.53.0183} {\bibfield  {journal}
  {\bibinfo  {journal} {IBM Journal of Research and Development}\ }\textbf
  {\bibinfo {volume} {5}},\ \bibinfo {pages} {183} (\bibinfo {year}
  {1961})}\BibitemShut {NoStop}%
\bibitem [{\citenamefont {Bennett}(2003)}]{Bennett}%
  \BibitemOpen
  \bibfield  {author} {\bibinfo {author} {\bibfnamefont {C.~H.}\ \bibnamefont
  {Bennett}},\ }\href {\doibase https://doi.org/10.1016/S1355-2198(03)00039-X}
  {\bibfield  {journal} {\bibinfo  {journal} {Studies in History and Philosophy
  of Science Part B: Studies in History and Philosophy of Modern Physics}\
  }\textbf {\bibinfo {volume} {34}},\ \bibinfo {pages} {501 } (\bibinfo {year}
  {2003})}\BibitemShut {NoStop}%
\bibitem [{\citenamefont {Minsky}(1967)}]{minsky1967computation}%
  \BibitemOpen
  \bibfield  {author} {\bibinfo {author} {\bibfnamefont {M.~L.}\ \bibnamefont
  {Minsky}},\ }\href@noop {} {\emph {\bibinfo {title} {Computation: finite and
  infinite machines}}}\ (\bibinfo  {publisher} {Prentice-Hall, Inc.},\ \bibinfo
  {year} {1967})\BibitemShut {NoStop}%
\bibitem [{\citenamefont {Wolfram}(2002)}]{wolfram2002new}%
  \BibitemOpen
  \bibfield  {author} {\bibinfo {author} {\bibfnamefont {S.}~\bibnamefont
  {Wolfram}},\ }\href@noop {} {\emph {\bibinfo {title} {A new kind of
  science}}}\ (\bibinfo  {publisher} {Wolfram Media, Champaign, IL},\ \bibinfo
  {year} {2002})\BibitemShut {NoStop}%
\bibitem [{Note1()}]{Note1}%
  \BibitemOpen
  \bibinfo {note} {It should be mentioned that recently it has been speculated
  that the continuous time-translational symmetry can break spontaneously even
  in the ground states of quantum systems \cite {Wilcz}. It has been
  subsequently shown to be impossible under rather general conditions \cite
  {Bruno, Oshi}. In contrast, it appears that the {\protect \em discrete} time
  translation symmetry of {\protect \em driven} quantum systems can indeed
  further break spontaneously, for instance making the system evolve with twice
  the period of the drive \cite {Sondhi2016, Nayak2016, Yao17,
  choi2017observation, zhang2017observation}.}\BibitemShut {Stop}%
\bibitem [{\citenamefont {Seleznyova}(1995)}]{Selez95}%
  \BibitemOpen
  \bibfield  {author} {\bibinfo {author} {\bibfnamefont {A.~N.}\ \bibnamefont
  {Seleznyova}},\ }\href {\doibase 10.1103/PhysRevA.51.950} {\bibfield
  {journal} {\bibinfo  {journal} {Phys. Rev. A}\ }\textbf {\bibinfo {volume}
  {51}},\ \bibinfo {pages} {950} (\bibinfo {year} {1995})}\BibitemShut
  {NoStop}%
\bibitem [{\citenamefont {Moore}(1970)}]{Moore70}%
  \BibitemOpen
  \bibfield  {author} {\bibinfo {author} {\bibfnamefont {G.~T.}\ \bibnamefont
  {Moore}},\ }\href {\doibase 10.1063/1.1665432} {\bibfield  {journal}
  {\bibinfo  {journal} {Journal of Mathematical Physics}\ }\textbf {\bibinfo
  {volume} {11}},\ \bibinfo {pages} {2679} (\bibinfo {year}
  {1970})}\BibitemShut {NoStop}%
\bibitem [{\citenamefont {Law}(1994)}]{Law94}%
  \BibitemOpen
  \bibfield  {author} {\bibinfo {author} {\bibfnamefont {C.~K.}\ \bibnamefont
  {Law}},\ }\href {\doibase 10.1103/PhysRevLett.73.1931} {\bibfield  {journal}
  {\bibinfo  {journal} {Phys. Rev. Lett.}\ }\textbf {\bibinfo {volume} {73}},\
  \bibinfo {pages} {1931} (\bibinfo {year} {1994})}\BibitemShut {NoStop}%
\bibitem [{\citenamefont {Fulling}\ and\ \citenamefont
  {Davies}(1976)}]{Fulling76}%
  \BibitemOpen
  \bibfield  {author} {\bibinfo {author} {\bibfnamefont {S.~A.}\ \bibnamefont
  {Fulling}}\ and\ \bibinfo {author} {\bibfnamefont {P.~C.}\ \bibnamefont
  {Davies}},\ }\href {\doibase 10.1098/rspa.1976.0045} {\bibfield  {journal}
  {\bibinfo  {journal} {Proceedings of the Royal Society of London A:
  Mathematical, Physical and Engineering Sciences}\ }\textbf {\bibinfo {volume}
  {348}},\ \bibinfo {pages} {393} (\bibinfo {year} {1976})}\BibitemShut
  {NoStop}%
\bibitem [{\citenamefont {Dalvit}\ and\ \citenamefont
  {Mazzitelli}(1998)}]{Dalvit98}%
  \BibitemOpen
  \bibfield  {author} {\bibinfo {author} {\bibfnamefont {D.~A.~R.}\
  \bibnamefont {Dalvit}}\ and\ \bibinfo {author} {\bibfnamefont {F.~D.}\
  \bibnamefont {Mazzitelli}},\ }\href {\doibase 10.1103/PhysRevA.57.2113}
  {\bibfield  {journal} {\bibinfo  {journal} {Phys. Rev. A}\ }\textbf {\bibinfo
  {volume} {57}},\ \bibinfo {pages} {2113} (\bibinfo {year}
  {1998})}\BibitemShut {NoStop}%
\bibitem [{\citenamefont {Yablonovitch}(1989)}]{Yablo1989}%
  \BibitemOpen
  \bibfield  {author} {\bibinfo {author} {\bibfnamefont {E.}~\bibnamefont
  {Yablonovitch}},\ }\href {\doibase 10.1103/PhysRevLett.62.1742} {\bibfield
  {journal} {\bibinfo  {journal} {Phys. Rev. Lett.}\ }\textbf {\bibinfo
  {volume} {62}},\ \bibinfo {pages} {1742} (\bibinfo {year}
  {1989})}\BibitemShut {NoStop}%
\bibitem [{\citenamefont {Dodonov}\ \emph
  {et~al.}(1993{\natexlab{a}})\citenamefont {Dodonov}, \citenamefont {Klimov},\
  and\ \citenamefont {Nikonov}}]{DodonovM93}%
  \BibitemOpen
  \bibfield  {author} {\bibinfo {author} {\bibfnamefont {V.~V.}\ \bibnamefont
  {Dodonov}}, \bibinfo {author} {\bibfnamefont {A.~B.}\ \bibnamefont {Klimov}},
  \ and\ \bibinfo {author} {\bibfnamefont {D.~E.}\ \bibnamefont {Nikonov}},\
  }\href {\doibase 10.1103/PhysRevA.47.4422} {\bibfield  {journal} {\bibinfo
  {journal} {Phys. Rev. A}\ }\textbf {\bibinfo {volume} {47}},\ \bibinfo
  {pages} {4422} (\bibinfo {year} {1993}{\natexlab{a}})}\BibitemShut {NoStop}%
\bibitem [{\citenamefont {Leonhardt}\ and\ \citenamefont
  {Philbin}(2006)}]{Leonhardt2006}%
  \BibitemOpen
  \bibfield  {author} {\bibinfo {author} {\bibfnamefont {U.}~\bibnamefont
  {Leonhardt}}\ and\ \bibinfo {author} {\bibfnamefont {T.~G.}\ \bibnamefont
  {Philbin}},\ }\href@noop {} {\bibfield  {journal} {\bibinfo  {journal} {New
  Journal of Physics}\ }\textbf {\bibinfo {volume} {8}},\ \bibinfo {pages}
  {247} (\bibinfo {year} {2006})}\BibitemShut {NoStop}%
\bibitem [{Note2()}]{Note2}%
  \BibitemOpen
  \bibinfo {note} {In principle, it is not impossible to``move'' the mirror
  superluminally, since no information needs to be transferred when the
  reflective properties of the medium are changed. A classic example of the
  same phenomenon is the ``propagation'' of the scissor cut}\BibitemShut
  {NoStop}%
\bibitem [{\citenamefont {Casimir}(1948)}]{Casimir1948}%
  \BibitemOpen
  \bibfield  {author} {\bibinfo {author} {\bibfnamefont {H.~B.}\ \bibnamefont
  {Casimir}},\ }in\ \href@noop {} {\emph {\bibinfo {booktitle} {Proc. Kon. Ned.
  Akad. Wet.}}},\ Vol.~\bibinfo {volume} {51}\ (\bibinfo {year} {1948})\ p.\
  \bibinfo {pages} {793}\BibitemShut {NoStop}%
\bibitem [{\citenamefont {Lamoreaux}(1997)}]{Lamo1997}%
  \BibitemOpen
  \bibfield  {author} {\bibinfo {author} {\bibfnamefont {S.~K.}\ \bibnamefont
  {Lamoreaux}},\ }\href {\doibase 10.1103/PhysRevLett.78.5} {\bibfield
  {journal} {\bibinfo  {journal} {Phys. Rev. Lett.}\ }\textbf {\bibinfo
  {volume} {78}},\ \bibinfo {pages} {5} (\bibinfo {year} {1997})}\BibitemShut
  {NoStop}%
\bibitem [{\citenamefont {Mohideen}\ and\ \citenamefont
  {Roy}(1998)}]{Mohi1998}%
  \BibitemOpen
  \bibfield  {author} {\bibinfo {author} {\bibfnamefont {U.}~\bibnamefont
  {Mohideen}}\ and\ \bibinfo {author} {\bibfnamefont {A.}~\bibnamefont {Roy}},\
  }\href {\doibase 10.1103/PhysRevLett.81.4549} {\bibfield  {journal} {\bibinfo
   {journal} {Phys. Rev. Lett.}\ }\textbf {\bibinfo {volume} {81}},\ \bibinfo
  {pages} {4549} (\bibinfo {year} {1998})}\BibitemShut {NoStop}%
\bibitem [{\citenamefont {Bressi}\ \emph {et~al.}(2002)\citenamefont {Bressi},
  \citenamefont {Carugno}, \citenamefont {Onofrio},\ and\ \citenamefont
  {Ruoso}}]{Bressi2002}%
  \BibitemOpen
  \bibfield  {author} {\bibinfo {author} {\bibfnamefont {G.}~\bibnamefont
  {Bressi}}, \bibinfo {author} {\bibfnamefont {G.}~\bibnamefont {Carugno}},
  \bibinfo {author} {\bibfnamefont {R.}~\bibnamefont {Onofrio}}, \ and\
  \bibinfo {author} {\bibfnamefont {G.}~\bibnamefont {Ruoso}},\ }\href
  {\doibase 10.1103/PhysRevLett.88.041804} {\bibfield  {journal} {\bibinfo
  {journal} {Phys. Rev. Lett.}\ }\textbf {\bibinfo {volume} {88}},\ \bibinfo
  {pages} {041804} (\bibinfo {year} {2002})}\BibitemShut {NoStop}%
\bibitem [{\citenamefont {Dodonov}\ \emph
  {et~al.}(1993{\natexlab{b}})\citenamefont {Dodonov}, \citenamefont {Klimov},\
  and\ \citenamefont {Nikonov}}]{Dodonov93}%
  \BibitemOpen
  \bibfield  {author} {\bibinfo {author} {\bibfnamefont {V.~V.}\ \bibnamefont
  {Dodonov}}, \bibinfo {author} {\bibfnamefont {A.~B.}\ \bibnamefont {Klimov}},
  \ and\ \bibinfo {author} {\bibfnamefont {D.~E.}\ \bibnamefont {Nikonov}},\
  }\href {\doibase 10.1063/1.530093} {\bibfield  {journal} {\bibinfo  {journal}
  {Journal of Mathematical Physics}\ }\textbf {\bibinfo {volume} {34}},\
  \bibinfo {pages} {2742} (\bibinfo {year} {1993}{\natexlab{b}})}\BibitemShut
  {NoStop}%
\bibitem [{\citenamefont {Wilson}\ \emph {et~al.}(2011)\citenamefont {Wilson},
  \citenamefont {Johansson}, \citenamefont {Pourkabirian}, \citenamefont
  {Simoen}, \citenamefont {Johansson}, \citenamefont {Duty}, \citenamefont
  {Nori},\ and\ \citenamefont {Delsing}}]{Wilson2011}%
  \BibitemOpen
  \bibfield  {author} {\bibinfo {author} {\bibfnamefont {C.}~\bibnamefont
  {Wilson}}, \bibinfo {author} {\bibfnamefont {G.}~\bibnamefont {Johansson}},
  \bibinfo {author} {\bibfnamefont {A.}~\bibnamefont {Pourkabirian}}, \bibinfo
  {author} {\bibfnamefont {M.}~\bibnamefont {Simoen}}, \bibinfo {author}
  {\bibfnamefont {J.}~\bibnamefont {Johansson}}, \bibinfo {author}
  {\bibfnamefont {T.}~\bibnamefont {Duty}}, \bibinfo {author} {\bibfnamefont
  {F.}~\bibnamefont {Nori}}, \ and\ \bibinfo {author} {\bibfnamefont
  {P.}~\bibnamefont {Delsing}},\ }\href@noop {} {\bibfield  {journal} {\bibinfo
   {journal} {Nature}\ }\textbf {\bibinfo {volume} {479}},\ \bibinfo {pages}
  {376} (\bibinfo {year} {2011})}\BibitemShut {NoStop}%
\bibitem [{\citenamefont {Nation}\ \emph {et~al.}(2012)\citenamefont {Nation},
  \citenamefont {Johansson}, \citenamefont {Blencowe},\ and\ \citenamefont
  {Nori}}]{Blenco2012}%
  \BibitemOpen
  \bibfield  {author} {\bibinfo {author} {\bibfnamefont {P.~D.}\ \bibnamefont
  {Nation}}, \bibinfo {author} {\bibfnamefont {J.~R.}\ \bibnamefont
  {Johansson}}, \bibinfo {author} {\bibfnamefont {M.~P.}\ \bibnamefont
  {Blencowe}}, \ and\ \bibinfo {author} {\bibfnamefont {F.}~\bibnamefont
  {Nori}},\ }\href {\doibase 10.1103/RevModPhys.84.1} {\bibfield  {journal}
  {\bibinfo  {journal} {Rev. Mod. Phys.}\ }\textbf {\bibinfo {volume} {84}},\
  \bibinfo {pages} {1} (\bibinfo {year} {2012})}\BibitemShut {NoStop}%
\bibitem [{\citenamefont {Misner}\ \emph {et~al.}(2017)\citenamefont {Misner},
  \citenamefont {Thorne},\ and\ \citenamefont {Wheeler}}]{Misner2017}%
  \BibitemOpen
  \bibfield  {author} {\bibinfo {author} {\bibfnamefont {C.~W.}\ \bibnamefont
  {Misner}}, \bibinfo {author} {\bibfnamefont {K.~S.}\ \bibnamefont {Thorne}},
  \ and\ \bibinfo {author} {\bibfnamefont {J.~A.}\ \bibnamefont {Wheeler}},\
  }\href@noop {} {\emph {\bibinfo {title} {Gravitation}}}\ (\bibinfo
  {publisher} {Princeton University Press},\ \bibinfo {year}
  {2017})\BibitemShut {NoStop}%
\bibitem [{\citenamefont {Landau}\ and\ \citenamefont
  {Lifshitz}(2000)}]{Landau2}%
  \BibitemOpen
  \bibfield  {author} {\bibinfo {author} {\bibfnamefont {L.~D.}\ \bibnamefont
  {Landau}}\ and\ \bibinfo {author} {\bibfnamefont {E.~M.}\ \bibnamefont
  {Lifshitz}},\ }\href@noop {} {\emph {\bibinfo {title} {The classical theory
  of fields: Volume 2 (course of theoretical physics series)}}}\ (\bibinfo
  {publisher} {Oxford Pergamon Press, Oxford},\ \bibinfo {year}
  {2000})\BibitemShut {NoStop}%
\bibitem [{\citenamefont {Eddington}(1924)}]{Eddington1924}%
  \BibitemOpen
  \bibfield  {author} {\bibinfo {author} {\bibfnamefont {A.~S.}\ \bibnamefont
  {Eddington}},\ }\href@noop {} {\bibfield  {journal} {\bibinfo  {journal}
  {Nature}\ }\textbf {\bibinfo {volume} {113}},\ \bibinfo {pages} {192}
  (\bibinfo {year} {1924})}\BibitemShut {NoStop}%
\bibitem [{\citenamefont {Finkelstein}(1958)}]{Fink58}%
  \BibitemOpen
  \bibfield  {author} {\bibinfo {author} {\bibfnamefont {D.}~\bibnamefont
  {Finkelstein}},\ }\href {\doibase 10.1103/PhysRev.110.965} {\bibfield
  {journal} {\bibinfo  {journal} {Phys. Rev.}\ }\textbf {\bibinfo {volume}
  {110}},\ \bibinfo {pages} {965} (\bibinfo {year} {1958})}\BibitemShut
  {NoStop}%
\bibitem [{\citenamefont {Wald}(1984)}]{Wald1984}%
  \BibitemOpen
  \bibfield  {author} {\bibinfo {author} {\bibfnamefont {R.~M.}\ \bibnamefont
  {Wald}},\ }\href@noop {} {\emph {\bibinfo {title} {General Relativity}}}\
  (\bibinfo  {publisher} {University of Chicago Press},\ \bibinfo {year}
  {1984})\BibitemShut {NoStop}%
\bibitem [{\citenamefont {Kruskal}(1960)}]{Kruskal1960}%
  \BibitemOpen
  \bibfield  {author} {\bibinfo {author} {\bibfnamefont {M.~D.}\ \bibnamefont
  {Kruskal}},\ }\href@noop {} {\bibfield  {journal} {\bibinfo  {journal}
  {Physical Review}\ }\textbf {\bibinfo {volume} {119}},\ \bibinfo {pages}
  {1743} (\bibinfo {year} {1960})}\BibitemShut {NoStop}%
\bibitem [{\citenamefont {Szekeres}(1960)}]{Sze1960}%
  \BibitemOpen
  \bibfield  {author} {\bibinfo {author} {\bibfnamefont {G.}~\bibnamefont
  {Szekeres}},\ }\href@noop {} {\bibfield  {journal} {\bibinfo  {journal}
  {Publicationes Mathematicae Debrecen}\ }\textbf {\bibinfo {volume} {7}},\
  \bibinfo {pages} {285} (\bibinfo {year} {1960})}\BibitemShut {NoStop}%
\bibitem [{Note3()}]{Note3}%
  \BibitemOpen
  \bibinfo {note} {Near the touching point $x_0$, $\delta _{n+1} = \delta _{n}
  + f''(x_0) \delta _{n}^2/2$. For $\delta _n \ll 1$, $n$ can be treated as a
  continuous variable: $d\delta /dn = f'' \delta ^2$/2. Integration gives
  $\delta _n \sim \delta _0/(1 - \delta _0 f''/2)$ which yields $\delta _n \sim
  1/n$ convergence for $\delta _0 f''(x_0) < 0 $ and an instability
  otherwise}\BibitemShut {NoStop}%
\bibitem [{\citenamefont {Feigenbaum}(1978)}]{feigenbaum1978quantitative}%
  \BibitemOpen
  \bibfield  {author} {\bibinfo {author} {\bibfnamefont {M.~J.}\ \bibnamefont
  {Feigenbaum}},\ }\href@noop {} {\bibfield  {journal} {\bibinfo  {journal}
  {Journal of statistical physics}\ }\textbf {\bibinfo {volume} {19}},\
  \bibinfo {pages} {25} (\bibinfo {year} {1978})}\BibitemShut {NoStop}%
\bibitem [{\citenamefont {Paschotta}\ \emph {et~al.}(2008)\citenamefont
  {Paschotta} \emph {et~al.}}]{rp2008}%
  \BibitemOpen
  \bibfield  {author} {\bibinfo {author} {\bibfnamefont {R.}~\bibnamefont
  {Paschotta}} \emph {et~al.},\ }\href@noop {} {\emph {\bibinfo {title}
  {Encyclopedia of laser physics and technology}}},\ Vol.~\bibinfo {volume}
  {1}\ (\bibinfo  {publisher} {Wiley-vch Berlin},\ \bibinfo {year}
  {2008})\BibitemShut {NoStop}%
\bibitem [{\citenamefont {Henneberger}\ and\ \citenamefont
  {Schulte}(1966)}]{Hen1966}%
  \BibitemOpen
  \bibfield  {author} {\bibinfo {author} {\bibfnamefont {W.}~\bibnamefont
  {Henneberger}}\ and\ \bibinfo {author} {\bibfnamefont {H.}~\bibnamefont
  {Schulte}},\ }\href@noop {} {\bibfield  {journal} {\bibinfo  {journal}
  {Journal of Applied Physics}\ }\textbf {\bibinfo {volume} {37}},\ \bibinfo
  {pages} {2189} (\bibinfo {year} {1966})}\BibitemShut {NoStop}%
\bibitem [{\citenamefont {Smith}(1967)}]{Smith1967}%
  \BibitemOpen
  \bibfield  {author} {\bibinfo {author} {\bibfnamefont {P.}~\bibnamefont
  {Smith}},\ }\href@noop {} {\bibfield  {journal} {\bibinfo  {journal} {Applied
  Physics Letters}\ }\textbf {\bibinfo {volume} {10}},\ \bibinfo {pages} {51}
  (\bibinfo {year} {1967})}\BibitemShut {NoStop}%
\bibitem [{\citenamefont {Baranov}\ and\ \citenamefont
  {Shirokov}(1967)}]{Baran1967}%
  \BibitemOpen
  \bibfield  {author} {\bibinfo {author} {\bibfnamefont {R.}~\bibnamefont
  {Baranov}}\ and\ \bibinfo {author} {\bibfnamefont {Y.~M.}\ \bibnamefont
  {Shirokov}},\ }\href@noop {} {\bibfield  {journal} {\bibinfo  {journal} {Sov.
  Phys. JETP}\ }\textbf {\bibinfo {volume} {53}},\ \bibinfo {pages} {2123}
  (\bibinfo {year} {1967})}\BibitemShut {NoStop}%
\bibitem [{\citenamefont {Knill}\ \emph {et~al.}(2001)\citenamefont {Knill},
  \citenamefont {Laflamme},\ and\ \citenamefont {Milburn}}]{klm2001}%
  \BibitemOpen
  \bibfield  {author} {\bibinfo {author} {\bibfnamefont {E.}~\bibnamefont
  {Knill}}, \bibinfo {author} {\bibfnamefont {R.}~\bibnamefont {Laflamme}}, \
  and\ \bibinfo {author} {\bibfnamefont {G.~J.}\ \bibnamefont {Milburn}},\
  }\href@noop {} {\bibfield  {journal} {\bibinfo  {journal} {Nature}\ }\textbf
  {\bibinfo {volume} {409}},\ \bibinfo {pages} {46} (\bibinfo {year}
  {2001})}\BibitemShut {NoStop}%
\bibitem [{\citenamefont {Barber}\ and\ \citenamefont
  {Putterman}(1991)}]{Putt1991}%
  \BibitemOpen
  \bibfield  {author} {\bibinfo {author} {\bibfnamefont {B.~P.}\ \bibnamefont
  {Barber}}\ and\ \bibinfo {author} {\bibfnamefont {S.~J.}\ \bibnamefont
  {Putterman}},\ }\href@noop {} {\bibfield  {journal} {\bibinfo  {journal}
  {Nature}\ }\textbf {\bibinfo {volume} {352}},\ \bibinfo {pages} {318}
  (\bibinfo {year} {1991})}\BibitemShut {NoStop}%
\bibitem [{\citenamefont {Boyland}(1986)}]{Boyland1986}%
  \BibitemOpen
  \bibfield  {author} {\bibinfo {author} {\bibfnamefont {P.~L.}\ \bibnamefont
  {Boyland}},\ }\href {https://projecteuclid.org:443/euclid.cmp/1104115779}
  {\bibfield  {journal} {\bibinfo  {journal} {Comm. Math. Phys.}\ }\textbf
  {\bibinfo {volume} {106}},\ \bibinfo {pages} {353} (\bibinfo {year}
  {1986})}\BibitemShut {NoStop}%
\bibitem [{\citenamefont {Arnol'd}(1963)}]{Arnold63}%
  \BibitemOpen
  \bibfield  {author} {\bibinfo {author} {\bibfnamefont {V.~I.}\ \bibnamefont
  {Arnol'd}},\ }\href@noop {} {\bibfield  {journal} {\bibinfo  {journal}
  {Uspekhi Mat. Nauk}\ }\textbf {\bibinfo {volume} {18}},\ \bibinfo {pages}
  {91} (\bibinfo {year} {1963})}\BibitemShut {NoStop}%
\bibitem [{\citenamefont {Sinai}(1970)}]{Sinai70}%
  \BibitemOpen
  \bibfield  {author} {\bibinfo {author} {\bibfnamefont {Y.~G.}\ \bibnamefont
  {Sinai}},\ }\href@noop {} {\bibfield  {journal} {\bibinfo  {journal} {Russ.
  Math. Surv.}\ }\textbf {\bibinfo {volume} {25}},\ \bibinfo {pages} {137}
  (\bibinfo {year} {1970})}\BibitemShut {NoStop}%
\bibitem [{\citenamefont {Hopf}(1949)}]{Hopf49}%
  \BibitemOpen
  \bibfield  {author} {\bibinfo {author} {\bibfnamefont {E.}~\bibnamefont
  {Hopf}},\ }\href@noop {} {\bibfield  {journal} {\bibinfo  {journal} {Uspekhi
  Mat. Nauk}\ }\textbf {\bibinfo {volume} {4}},\ \bibinfo {pages} {129Ð170}
  (\bibinfo {year} {1949})}\BibitemShut {NoStop}%
\bibitem [{\citenamefont {Fulling}(1973)}]{FullingUn73}%
  \BibitemOpen
  \bibfield  {author} {\bibinfo {author} {\bibfnamefont {S.~A.}\ \bibnamefont
  {Fulling}},\ }\href {\doibase 10.1103/PhysRevD.7.2850} {\bibfield  {journal}
  {\bibinfo  {journal} {Phys. Rev. D}\ }\textbf {\bibinfo {volume} {7}},\
  \bibinfo {pages} {2850} (\bibinfo {year} {1973})}\BibitemShut {NoStop}%
\bibitem [{\citenamefont {Davies}(1975)}]{DaviesUn75}%
  \BibitemOpen
  \bibfield  {author} {\bibinfo {author} {\bibfnamefont {P.~C.~W.}\
  \bibnamefont {Davies}},\ }\href {http://stacks.iop.org/0305-4470/8/i=4/a=022}
  {\bibfield  {journal} {\bibinfo  {journal} {Journal of Physics A:
  Mathematical and General}\ }\textbf {\bibinfo {volume} {8}},\ \bibinfo
  {pages} {609} (\bibinfo {year} {1975})}\BibitemShut {NoStop}%
\bibitem [{\citenamefont {Unruh}(1976)}]{Unruh76}%
  \BibitemOpen
  \bibfield  {author} {\bibinfo {author} {\bibfnamefont {W.~G.}\ \bibnamefont
  {Unruh}},\ }\href {\doibase 10.1103/PhysRevD.14.870} {\bibfield  {journal}
  {\bibinfo  {journal} {Phys. Rev. D}\ }\textbf {\bibinfo {volume} {14}},\
  \bibinfo {pages} {870} (\bibinfo {year} {1976})}\BibitemShut {NoStop}%
\bibitem [{\citenamefont {Padmanabhan}(2005)}]{PADMANABHAN200549}%
  \BibitemOpen
  \bibfield  {author} {\bibinfo {author} {\bibfnamefont {T.}~\bibnamefont
  {Padmanabhan}},\ }\href {\doibase
  https://doi.org/10.1016/j.physrep.2004.10.003} {\bibfield  {journal}
  {\bibinfo  {journal} {Physics Reports}\ }\textbf {\bibinfo {volume} {406}},\
  \bibinfo {pages} {49 } (\bibinfo {year} {2005})}\BibitemShut {NoStop}%
\bibitem [{\citenamefont {Wilczek}(2013)}]{Wilcz}%
  \BibitemOpen
  \bibfield  {author} {\bibinfo {author} {\bibfnamefont {F.}~\bibnamefont
  {Wilczek}},\ }\href {\doibase 10.1103/PhysRevLett.111.250402} {\bibfield
  {journal} {\bibinfo  {journal} {Phys. Rev. Lett.}\ }\textbf {\bibinfo
  {volume} {111}},\ \bibinfo {pages} {250402} (\bibinfo {year}
  {2013})}\BibitemShut {NoStop}%
\bibitem [{\citenamefont {Bruno}(2013)}]{Bruno}%
  \BibitemOpen
  \bibfield  {author} {\bibinfo {author} {\bibfnamefont {P.}~\bibnamefont
  {Bruno}},\ }\href {\doibase 10.1103/PhysRevLett.110.118901} {\bibfield
  {journal} {\bibinfo  {journal} {Phys. Rev. Lett.}\ }\textbf {\bibinfo
  {volume} {110}},\ \bibinfo {pages} {118901} (\bibinfo {year}
  {2013})}\BibitemShut {NoStop}%
\bibitem [{\citenamefont {Watanabe}\ and\ \citenamefont
  {Oshikawa}(2015)}]{Oshi}%
  \BibitemOpen
  \bibfield  {author} {\bibinfo {author} {\bibfnamefont {H.}~\bibnamefont
  {Watanabe}}\ and\ \bibinfo {author} {\bibfnamefont {M.}~\bibnamefont
  {Oshikawa}},\ }\href {\doibase 10.1103/PhysRevLett.114.251603} {\bibfield
  {journal} {\bibinfo  {journal} {Phys. Rev. Lett.}\ }\textbf {\bibinfo
  {volume} {114}},\ \bibinfo {pages} {251603} (\bibinfo {year}
  {2015})}\BibitemShut {NoStop}%
\bibitem [{\citenamefont {Khemani}\ \emph {et~al.}(2016)\citenamefont
  {Khemani}, \citenamefont {Lazarides}, \citenamefont {Moessner},\ and\
  \citenamefont {Sondhi}}]{Sondhi2016}%
  \BibitemOpen
  \bibfield  {author} {\bibinfo {author} {\bibfnamefont {V.}~\bibnamefont
  {Khemani}}, \bibinfo {author} {\bibfnamefont {A.}~\bibnamefont {Lazarides}},
  \bibinfo {author} {\bibfnamefont {R.}~\bibnamefont {Moessner}}, \ and\
  \bibinfo {author} {\bibfnamefont {S.~L.}\ \bibnamefont {Sondhi}},\ }\href
  {\doibase 10.1103/PhysRevLett.116.250401} {\bibfield  {journal} {\bibinfo
  {journal} {Phys. Rev. Lett.}\ }\textbf {\bibinfo {volume} {116}},\ \bibinfo
  {pages} {250401} (\bibinfo {year} {2016})}\BibitemShut {NoStop}%
\bibitem [{\citenamefont {Else}\ \emph {et~al.}(2016)\citenamefont {Else},
  \citenamefont {Bauer},\ and\ \citenamefont {Nayak}}]{Nayak2016}%
  \BibitemOpen
  \bibfield  {author} {\bibinfo {author} {\bibfnamefont {D.~V.}\ \bibnamefont
  {Else}}, \bibinfo {author} {\bibfnamefont {B.}~\bibnamefont {Bauer}}, \ and\
  \bibinfo {author} {\bibfnamefont {C.}~\bibnamefont {Nayak}},\ }\href
  {\doibase 10.1103/PhysRevLett.117.090402} {\bibfield  {journal} {\bibinfo
  {journal} {Phys. Rev. Lett.}\ }\textbf {\bibinfo {volume} {117}},\ \bibinfo
  {pages} {090402} (\bibinfo {year} {2016})}\BibitemShut {NoStop}%
\bibitem [{\citenamefont {Choi}\ \emph {et~al.}(2017)\citenamefont {Choi},
  \citenamefont {Choi}, \citenamefont {Landig}, \citenamefont {Kucsko},
  \citenamefont {Zhou}, \citenamefont {Isoya}, \citenamefont {Jelezko},
  \citenamefont {Onoda}, \citenamefont {Sumiya}, \citenamefont {Khemani} \emph
  {et~al.}}]{choi2017observation}%
  \BibitemOpen
  \bibfield  {author} {\bibinfo {author} {\bibfnamefont {S.}~\bibnamefont
  {Choi}}, \bibinfo {author} {\bibfnamefont {J.}~\bibnamefont {Choi}}, \bibinfo
  {author} {\bibfnamefont {R.}~\bibnamefont {Landig}}, \bibinfo {author}
  {\bibfnamefont {G.}~\bibnamefont {Kucsko}}, \bibinfo {author} {\bibfnamefont
  {H.}~\bibnamefont {Zhou}}, \bibinfo {author} {\bibfnamefont {J.}~\bibnamefont
  {Isoya}}, \bibinfo {author} {\bibfnamefont {F.}~\bibnamefont {Jelezko}},
  \bibinfo {author} {\bibfnamefont {S.}~\bibnamefont {Onoda}}, \bibinfo
  {author} {\bibfnamefont {H.}~\bibnamefont {Sumiya}}, \bibinfo {author}
  {\bibfnamefont {V.}~\bibnamefont {Khemani}},  \emph {et~al.},\ }\href@noop {}
  {\bibfield  {journal} {\bibinfo  {journal} {Nature}\ }\textbf {\bibinfo
  {volume} {543}},\ \bibinfo {pages} {221} (\bibinfo {year}
  {2017})}\BibitemShut {NoStop}%
\bibitem [{\citenamefont {Zhang}\ \emph {et~al.}(2017)\citenamefont {Zhang},
  \citenamefont {Hess}, \citenamefont {Kyprianidis}, \citenamefont {Becker},
  \citenamefont {Lee}, \citenamefont {Smith}, \citenamefont {Pagano},
  \citenamefont {Potirniche}, \citenamefont {Potter}, \citenamefont
  {Vishwanath} \emph {et~al.}}]{zhang2017observation}%
  \BibitemOpen
  \bibfield  {author} {\bibinfo {author} {\bibfnamefont {J.}~\bibnamefont
  {Zhang}}, \bibinfo {author} {\bibfnamefont {P.}~\bibnamefont {Hess}},
  \bibinfo {author} {\bibfnamefont {A.}~\bibnamefont {Kyprianidis}}, \bibinfo
  {author} {\bibfnamefont {P.}~\bibnamefont {Becker}}, \bibinfo {author}
  {\bibfnamefont {A.}~\bibnamefont {Lee}}, \bibinfo {author} {\bibfnamefont
  {J.}~\bibnamefont {Smith}}, \bibinfo {author} {\bibfnamefont
  {G.}~\bibnamefont {Pagano}}, \bibinfo {author} {\bibfnamefont {I.-D.}\
  \bibnamefont {Potirniche}}, \bibinfo {author} {\bibfnamefont {A.~C.}\
  \bibnamefont {Potter}}, \bibinfo {author} {\bibfnamefont {A.}~\bibnamefont
  {Vishwanath}},  \emph {et~al.},\ }\href@noop {} {\bibfield  {journal}
  {\bibinfo  {journal} {Nature}\ }\textbf {\bibinfo {volume} {543}},\ \bibinfo
  {pages} {217} (\bibinfo {year} {2017})}\BibitemShut {NoStop}%
\end{thebibliography}%

%

\end{document}